%% file: PrecisionMuonPhysics_final.tex
\pdfoutput=1
\documentclass[preprint]{elsarticle}

\usepackage{ifthen}
\usepackage{amsmath}
\usepackage{graphicx}
\newlength{\tabcont}
\journal{Progress in Nuclear and Particle Physics}
\graphicspath{{Figures/}}
\newcommand{\opera}{{\tt OPERA}}

\newcommand{\cernexp}{{CERN\,III}}
\newcommand{\bnlexp}{{BNL\,E821}}
\newcommand{\fnalexp}{{FNAL\,E989}}
\newcommand{\jparcexp}{{J-PARC\,E34}}
\newcommand{\gm}{\ensuremath{(g-2)}}
\newcommand{\wa}{\mbox{\ensuremath{\omega_a}}}
\renewcommand{\wp}{\mbox{\ensuremath{\omega_p}}}
\newcommand{\amu}{\ensuremath{a_{\mu}}}

\newcommand{\tab}[1]{\settowidth{\tabcont}{#1}
\ifthenelse{\lengthtest{\tabcont < .25\linewidth}}
{\makebox[.25\linewidth][l]{#1}\ignorespaces}
{\makebox[.5\linewidth][l]{#1}\ignorespaces}
}

\begin{document}
\begin{frontmatter}

\title{Precision Muon Physics}

\author[uky]{T.\,P.~Gorringe}
\author[uw]{D.\,W.~Hertzog}
\address[uky]{University of Kentucky, Lexington, KY 40506, USA}
\address[uw]{University of Washington, Seattle, WA 98195, USA}

\date{today}

\input{abstract_final}

\end{frontmatter}
\tableofcontents

\input{Introduction_final}
\input{MuonDecay_final}

\input{MuonMichel_final}

\input{MuonCLFV_final}
\input{Muon-g2_final}

\input{MuonEDM_final}
\input{Muonium_final}
\input{MuonProtonRadius_final}

\input{MuonCapture_final}
\input{Summary_final}
\input{Acknowledgments}

\bibliography{PrecisionMuonPhysics_final}

\end{document}

%% file: abstract_final.tex
\begin{abstract}

The muon is playing a unique role in sub-atomic physics.
Studies of muon decay both determine the overall strength 
and establish the chiral structure of weak interactions,
as well as setting extraordinary limits 
on charged-lepton-flavor-violating processes.
Measurements of the muon's anomalous magnetic moment
offer singular sensitivity to the completeness of the standard model
and the predictions of many speculative theories.
Spectroscopy of muonium and muonic atoms gives
unmatched determinations of fundamental quantities including
the magnetic moment ratio $\mu_\mu / \mu_p$, 
lepton mass ratio $m_{\mu} / m_e$, and proton charge radius $r_p$. 
Also, muon capture experiments  
are exploring elusive features 
of weak interactions involving nucleons and nuclei.

We will review the experimental landscape of contemporary high-precision
and high-sensitivity experiments with muons. One focus 
is the novel methods and ingenious techniques 
that achieve such precision and sensitivity in 
recent, present, and planned experiments. Another focus 
is the uncommonly broad and topical range of questions in atomic, nuclear 
and particle physics that such experiments explore.   

\end{abstract}

%% file: Introduction_final.tex
\section{Introduction}

The muon is not a building block of ordinary matter.
It's much heavier than the electron but much lighter than the proton.
It interacts through its electric charge and magnetic moment
and its weak charged and neutral currents---but not the strong force.
Positive muons can form hydrogen-like atoms with electrons 
while negative muons can form hydrogen-like atoms with nuclei. 
The muon is unstable,  but sufficiently
long-lived to precisely study its properties
and sufficiently short-lived to precisely study its decays.
By a quirk of nature---parity non-conservation---muons
are produced fully polarized
and when they decay they are self-analyzing.

Since its discovery the muon has played
a rather unique and versatile role in physics.
In this review we discuss recent, current and
near-future efforts involving precision measurements
of properties and decays of free muons and muonic atoms.
The physics topics---which
range from fundamental constants and basic symmetries,
to weak nucleonic and nuclear interactions,
and standard model tests and new physics searches---are quite diverse.
These unique experiments are generally designed
to do one thing and do it well.
We aim to provide an experimentalist's perspective
into how these measurements are performed and their
physics impact.

This article is organized as follows. Section\ \ref{ssc:muonrole}
introduces the historical evolution of muon experiments in sub-atomic
physics.
Some general comments on muon facilities,
experimental technologies and blind-analysis procedures
are made in Sec.\ \ref{ssc:beams}.
Section\ \ref{sc:muonlifetime} discusses the measurement
of the muon lifetime and determination of the Fermi Constant.
The measurements of the muon decay parameters
and tests of the $V$-$A$ theory
are described in Sec.\ \ref{sc:decayparameters}.
Searches for charged lepton flavor violating muon decays
are described in Sec.\ \ref{sc:cLFV}.
Section \ref{sc:muondipole} covers
the measurement of the muon's magnetic and electric dipole moments,
and their sensitivities to new particles and unknown forces.
Lastly, a number of precision measurements
involving muonium ($\mu^+ e$) atoms
and muonic ($\mu^- Z$) atoms, which determine
fundamental constants and elementary interactions, are
discussed in Secs.\ \ref{sc:muonium}, \ref{sc:mup} and \ref{sc:muz}.


\subsection{Historical overview of muon experiments in sub-atomic physics}
\label{ssc:muonrole}


Table~\ref{tb:muonproperties} summarizes the important properties
and decay modes of the muon. The discovery of
muons---new particles about 200 times more massive than electrons---as
cosmic-ray constituents was famously made by Anderson and Neddermeyer \cite{Anderson:1936zz}
at Caltech in 1936  and Street and Stevenson \cite{Street:1937me} at Harvard in 1937. Their work
was actually the culmination of numerous experiments
conducted over many years,
which eventually demonstrated that the highly-penetrating component of cosmic radiation
was neither electrons nor protons.\footnote{The existence
of highly-penetrating cosmic rays was
known since the work of Bothe and Kolhorster in 1929.
In 1933, Kunze \cite{Kunze:1933} noted a ``particle of uncertain nature'' in his
investigations of cosmic rays using ionization chambers.}
Ultimately, the discovery became our first evidence for generations
of elementary particles and the hierarchical structure of the standard model.

\begin{table}
\centering
  \caption{Summary of Measured Muon Properties and Selected Decay Rates and Limits}\label{tb:muonproperties}
{\small
\begin{tabular}{lcccc}
  \hline
  Property & Symbol & Value & Precision & Ref.\  \\ \hline
  Mass & $m_\mu$ & 105.658\,3715(35) MeV & 34 ppb & \cite{Mohr:2012tt} \\
  Mean Lifetime & $\tau_\mu$& $2.196\,9811(22)\times10^{-6}$~s  & 1.0 ppm & \cite{Tishchenko:2012ie} \\
  Anom.\ Mag.\ Moment & $a_\mu$ & $116\,592\,091(63)\times10^{-11}$ & 0.54 ppm & \cite{Mohr:2012tt,Bennett:2006fi} \\
  Elec.\ Dipole Moment & $d_\mu$ & $< 1.9 \times 10^{-19} e\cdot$cm & 95\% C.L. & \cite{Bennett:2008dy} \\ \hline\hline
  Branching Ratios & PDG average & B.R. Limits & 90\% C.L. & Ref.\\ \hline
  $\mu^- \rightarrow e^-{\bar{\nu}}_{e}\nu_\mu$ & $\approx 100\%$ & $\mu^- \rightarrow e^- \gamma$ & $5.7\times10^{-13}$ &\cite{Adam:2013mnn} \\
  $\mu^- \rightarrow e^-{\bar{\nu}}_{e}\nu_\mu \gamma$ & $1.4(4)\%$ & $\mu^- \rightarrow e^-e^+e^-$ & $1.0\times10^{-12}$ &\cite{Bellgardt:1987du} \\
  $\mu^- \rightarrow e^-{\bar{\nu}}_{e}\nu_\mu e^+e^-$ & $3.4(4)\times10^{-5}$ & $\mu^- \rightarrow e^-$ conversion & $7\times10^{-13}$ &\cite{Bertl:2006up} \\
\hline
\end{tabular}}
\end{table}

The first cloud chamber photograph of the decay of a muon 
was taken in 1940 \cite{Williams:1940}.
The earliest determination of the muon lifetime
was made by interpreting the ``anomalous absorption''
of cosmic-ray muons with decreasing altitude as muon disintegration \cite{Rossi:1941zz}.
The direct measurement of the muon lifetime
was made shortly afterwards
by recording the time intervals between stopping muons
and decay electrons using cosmic rays \cite{Rasetti:1941ie}.
These early measurements---involving in-flight and stopped
muons---afforded a decisive test of time dilation for moving particles.
Nowadays the precision measurement of the muon lifetime $\tau_{\mu}$
provides the best determination of the Fermi constant $G_F$,
the quantity governing the universal strength of weak interactions.

The $\mu \rightarrow e \nu \bar{\nu}$ decay scheme
was established through cosmic-ray measurements
of the decay electron energy spectrum.
It was demonstrated---via the continuum distribution
and the energy endpoint---that
muons decay into three or more particles with small or zero masses \cite{PhysRev.75.1432}.
Fermi's theory of nuclear $\beta$-decay
was consequently expanded to accommodate
$\mu \rightarrow e \nu \bar{\nu}$ decay
thus affording the first glimpse of weak universality.
These early experiments---along with theoretical work
on the tensor structure
of the current-current interaction \cite{Michel:1950,PhysRev.93.354}---were
the beginnings of our modern precision studies
of the muon decay parameters
as a valued probe of the weak force.



In a profound paper in 1956, Lee and Yang suggested that the discrete symmetry
of parity might be violated in the weak interaction~\cite{Lee:1956qn}.
One important prediction of parity non-conservation concerned the by-then well-known
$\pi \rightarrow \mu \rightarrow e$ weak decay chain.
First, the non-conservation of parity in $\pi \rightarrow \mu \bar{\nu}$ decay
would cause muons to be polarized along the muon momentum axis.
Second, the non-conservation of parity in $\mu \rightarrow e \nu \bar{\nu}$  decay
would cause electrons to be emitted anisotropically about the muon polarization axis.
Soon afterwards the non-conservation of parity in $\pi \rightarrow \mu \rightarrow e$
decay was reported by Garwin {\it et al.}\ \cite{Garwin:1957hc}, and Friedman and Telegdi \cite{Friedman:1957mz},
following their observation of large decay-electron anisotropies.
Parity non-conservation is now codified
in the  $V$-$A$ structure of the weak currents \cite{Sudarshan:1958vf,Feynman:1958ty}
which imparts a distinctive angle-energy correlation
on the decay electrons with the muon polarization.
Since this early work, 
the precision measurement of angle-energy correlations
has enabled increasingly precise tests of $V$-$A$ theory.

The hypothesis of a weak interaction mediated by a force-carrying boson
was originally introduced in 1940 \cite{Lee:1949qk}.
The early theories involving weak bosons predicted
the existence of $\mu \rightarrow e \gamma$ decay
at the level of about $10^{-4}$ (the mechanism for
$\mu \rightarrow e \gamma$ involved the neutrino emitted
by the muon being absorbed by the electron).
By the late 1950s the $\mu \rightarrow e \gamma$ experimental limit
was  orders-of-magnitude below this theoretical prediction.
The crisis concerning $\mu \rightarrow e \gamma$ decay
led Pontecorvo \cite{PhysRev.72.246} and others to
postulate the existence of two distinct neutrino types; an electron-flavored neutrino and a muon-flavored neutrino.
Since the genesis of lepton flavor in $\mu \rightarrow e \gamma$,
increasingly delicate searches for rare processes including
$\mu \rightarrow e \gamma$ and $\mu \rightarrow e$ conversion
have continued to shape our understanding of flavor.

The aforementioned Garwin {\it et al.}\ experiment on parity non-conservation
in muon decay also yielded  the first measurement of the magnetic moment of the positive muon.
Their result for the gyromagnetic ratio $g = +2.00\pm0.10$
was a demonstration that the muon was a structureless, spin-1/2 Dirac particle.
Subsequently---through increasingly sophisticated measurements
that utilize the possibilities of polarization and polarimetry of muons in
$\pi \rightarrow \mu \rightarrow e$ decay---the determination of the anomalous part $( g-2 )$
of the muon magnetic moment has been measured
to an astonishing sub-part-per-million level.
The anomalous moment arises
through quantum vacuum fluctuations that
accrue from all particles of
nature---both known and unknown---and thereby affords a
unique window on new physics at high-energy scales

The simplest atom involving muons is muonium,
a pure QED bound state of $\mu^+e^-$, sometimes dubbed the ``perfect atom.''
Because both constituents are point-like leptons
the muonium  energy levels are completely free
of perturbations arising from nuclear size effects.
Its formation was first identified
by detecting the characteristic
Larmor precession frequency of polarized muonium formed when muons stop in certain gases.
Following its discovery  the measurement
of the hyperfine splitting of the muonium ground state---by inducing microwave transitions
between hyperfine states in external magnetic fields---was developed
by Hughes and co-workers (for details see \cite{PhysRevA.1.595}).
This work has provided the most precise determinations
of the muon-to-proton magnetic moment ratio $\mu_{\mu} / \mu_p$
and the muon-to-electron mass ratio $m_{\mu} / m_e$
as well as important tests of quantum electrodynamics.

When negative muons are brought to rest in matter
they undergo atomic capture and form muonic atoms.
These atoms are hydrogen-like systems where---unimpeded by the
exclusion principle---the muon cascades from an initially high principal quantum number state
to the $1S$ atomic ground state.
The existence of such atoms was first discussed
in Refs.\ \cite{PhysRev.71.320,PhysRev.72.399}
that demonstrated the timescale for formation
of muonic atoms was much shorter than the muon lifetime.
Compared to ordinary atoms,
the muonic atom radii are $( m_{\mu} / m_e )$ times smaller
and the energy levels are $( m_{\mu} / m_e )$ times greater.
Consequently, the overlap between the muon orbits
and the nucleus is much larger than in ordinary atoms
and the energy levels can be significantly perturbed
by the nuclear charge distribution.
Precision spectroscopy of muonic atoms thus became a
workhorse for studies of nuclear charge radii and electromagnetic moments.
In the simplest case of a $\mu p$ atom, the recent measurements of the
tiny energy splitting (Lamb shift) between the muonic $2S_{1/2}-2P_{1/2}$ orbitals
have provided the most precise determination of the proton charge radius.

After muonic atoms are formed they
disintegrate by either muon decay or nuclear capture
$\mu^- [A, Z] \rightarrow [A, Z-1] \nu$.
The first experimental evidence
for muon capture was observations of
differing electron yields
when stopping negative and positive muons
in carbon and iron \cite{PhysRev.68.232}.
This work demonstrated the weak nature
of muon capture on atomic nuclei thus dispelling the notion
that cosmic-ray muons were the force-carriers of the strong interaction,
and ultimately leading to early ideas of weak universality
in muon capture, muon decay and beta decay.
Modern measurements of muon capture
are investigating weak nucleonic and nuclear interactions
that address subjects which range from standard model symmetries to
fundamental astrophysical processes.

\subsection{Common features of precision muon experiments}
\label{ssc:beams}

Muon beams are derived from pion decays, the pions being
produced in the nuclear collisions between an accelerated beam and a fixed target.
There are three basic types of muon beam lines
involving so-called surface, cloud and decay muons.
In surface $\mu^+$ beams the muons originate from at-rest decay
of $\pi^+$ stops in the surface layer of the production target.
The resulting muons are  mono-energetic 29.8~MeV/c, 100\% longitudinally polarized,
and because of the localized source have a sharp focus.\footnote{Only surface  $\mu^+$ beams
are available  as $\pi^-$ stops form pionic atoms and rapidly undergo nuclear capture.}
In cloud beams the muons originate from in-flight decays
of parent pions in the region between production target  and the first bending magnet of the secondary beam line.
The resulting muons arise from ``forward-decays'' and ``backward-decays'' of pions
and therefore the resulting polarization is considerably lower than surface beams.
In decay beams an upstream section of the beam line selects the parent pion momentum
and a downstream section of the beam line selects the daughter muon momentum.
The resulting muons are typically highly polarized
and free of electron contamination.

Intense muon sources at the Paul Scherrer Institute (PSI)
in Switzerland and TRIUMF in Canada are examples of muon beams based
on high current, medium energy, proton cyclotrons
that provide an essentially continuous beam
(with the micro-time structure
of the cyclotron radio-frequency).
Facilities at Fermilab in the U.S.\ and J-PARC in Japan
are examples of muon beams based
on lower current, higher energy, proton synchrotrons
and provide a pulsed beam with typical
repetition rates of tens of Hertz.
Additionally, novel muon sources
for ultra-cold muons, based on formation and ionization of muonium atoms,
and ultra-intense muons, based on capture and transport by superconducting solenoids,
are under development for future muon experiments
(see Secs.\ \ref{ssc:e34} and \ref{ssc:mu2e} respectively).

Today's precision muon experiments---which require both
enormous statistics and extraordinary limits on possible biases
from systematic effects---are
benefiting from advances in areas including
radiation detectors, readout electronics,
computer hardware and software infrastructure.

Challenges for detector design include
stringent requirements on timing, tracking and calorimetry
in high-rate environments.
Ultra-low mass tracking chambers
and modern silicon detector technology
have been used or are being developed
for high-rate, high-precision tracking
applications. High-density electromagnetic
calorimeters (liquid xenon, lead fluoride,
lead tungstate) have been built
or are being constructed for
both good energy resolution and
fast-timing applications.
Modern silicon photomultipliers
are enabling readout
of scintillation and  Cherenkov light
in magnetic fields
and restricted geometries.
And sophisticated NMR-based field measurements
and finite-element field modeling
are being utilized for precision magnetometry.


Commercial applications for real-time measurement,
processing and distribution of massive data-streams
have driven hardware development including
high sampling-rate waveform digitizers,
field programmable gate arrays and
graphical processing units.
The digital capture of detector signals
offers invaluable opportunities
for systematics investigations
in precision measurements and waveform digitizers
have been used or are being deployed
in many recent, present and future experiments.
Field programmable gate arrays---{\it i.e.}\
high-performance user-customizable integrated circuits---are
now common for special purpose tasks
in timing, trigger and control logic.
A hybrid architecture of multicore CPUs
and teraflop-performance GPUs is being developed
for real-time readout and processing
in the Fermilab muon \gm\ experiment.

Recent precision muon experiments have stored datasets
of hundreds of terabytes of raw data and
future experiments will store datasets of
many petabytes of raw data.
This scale of data analysis and data simulations
has required grid computing facilities
such as WesGrid (Canada) and the
National Center for Supercomputing Applications (U.S.).
The extreme demands on high statistics and
understanding systematics are pushing applications
of GEANT (GEometry ANd Tracking software toolkit)
into new territories.

Increasingly,  modern measurements are carried out using
so-called ``blind-analysis'' techniques.
A simple example for measurements that use precision oscillators for timing is to
prescriptively de-tune the oscillator by a small offset from its nominal setting
during the data taking; the offset is unknown to anyone analyzing the data.
The data analysis time unit is then a somewhat arbitrarily defined  ``clock tick.''
When the analyses are complete, the clock ticks are converted into physical time units,
and the unblinded result is revealed. In efforts where rare events are investigated, blinded
regions must be established in which the rare events would be found.  Backgrounds and cuts are
determined from allowed events outside and near the forbidden box.  Once the analysis is complete,
the blinded box is opened to see if any events have survived. These disciplined procedures provide
a needed level of integrity to the experiments.

%% file: MuonDecay_final.tex
\section{Muon lifetime}
\label{sc:muonlifetime}

\subsection{Fermi constant $G_F$}


The strength of the weak interaction is governed
by the Fermi constant $G_F$.
The roots of $G_F$ are Fermi's theory---based on an analogy between the
emission of an electron-neutrino pair
by a radioactive nucleus and a photon
by a charged particle---of a current-current weak interaction.
Of course, since 1934, our modern understanding of weak interactions
has evolved  to incorporate parity-violating $V-A$ currents and
the massive $W$ and $Z$ gauge bosons.
However, the constant $G_F$ and Fermi interaction
have survived as a convenient, low energy, effective
theory of the weak sector
in the standard model (and presumably any successor).

\begin{figure}
\begin{centering}
  \includegraphics[width=0.8\columnwidth]{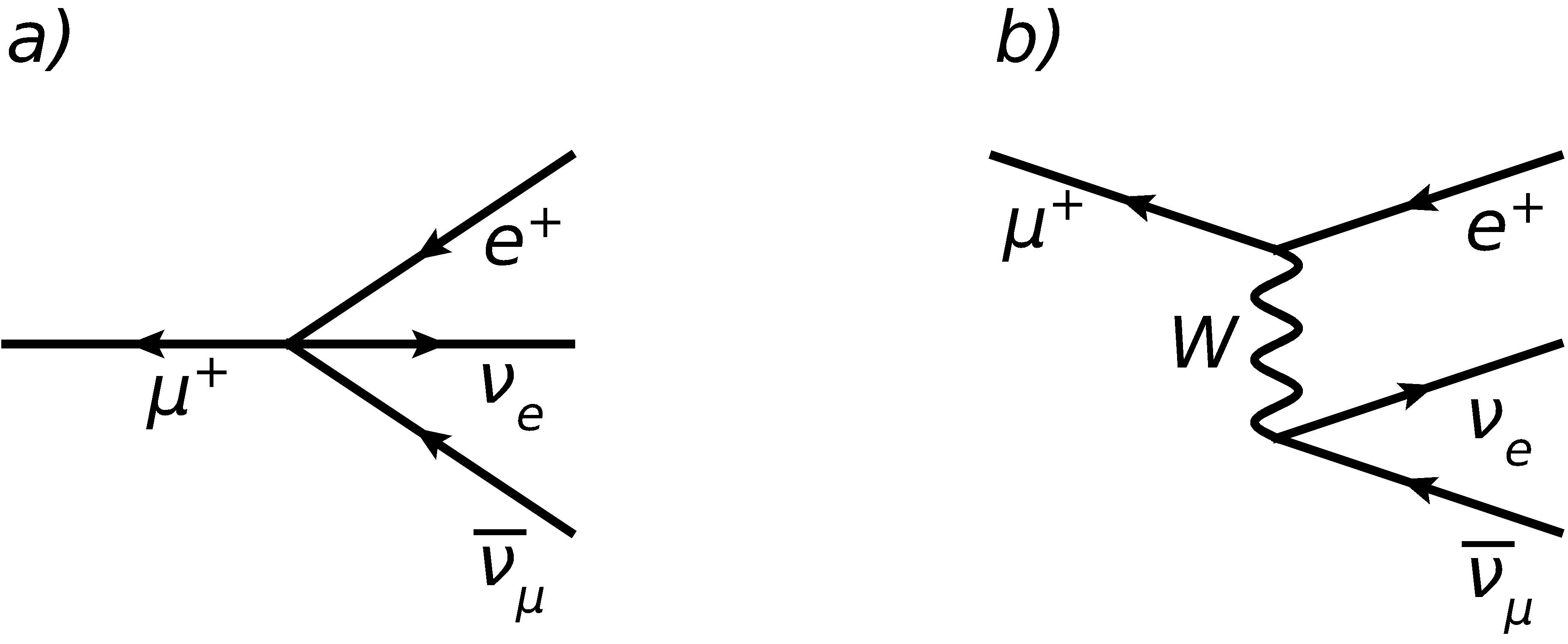}
  \caption{ \label{fg:GFfeynman}
a).~ Tree-level Feynman diagram for ordinary muon decay in Fermi's
current-current interaction.
b).~ Tree-level standard model diagram for ordinary muon decay
indicating the W-boson mediated weak interaction between the leptonic
currents.}
\end{centering}
\end{figure}

Within the standard model the Fermi constant (see Fig.\ \ref{fg:GFfeynman})
is given by
\begin{equation}
\frac{G_F}{\sqrt{2}} = \frac{g^2}{8M_W^2}\left( 1 + \sum_i r_i\right)\
\label{eqn:GFSM}
\end{equation}
where $1 / M^2_W$ represents  the tree-level propagator corresponding
to $W$-boson exchange and $g$ the weak coupling.
The term $\sum_i r_i$ incorporates the higher-order electroweak interaction
corrections \cite{Awramik:2003rn}.
The factors of $\sqrt{2}$ and $8$ in Eqn.\ \ref{eqn:GFSM} are reminders of the origins
of the Fermi constant in a vector current - vector current weak interaction.

%



By far the best determination of the Fermi constant is obtained
by the measurement of the positive muon lifetime, $\tau_{\mu}$.
Experimentally, intense beams of low-energy muons are nowadays available
and the 2.2~$\mu$s muon lifetime with its associated decay electrons
are nicely suited to precision measurements of time distributions.
Theoretically, because muon decays---$\mu \rightarrow e \nu \bar{\nu}$,
$\mu \rightarrow e \nu \bar{\nu} \gamma$
and $\mu \rightarrow e e e \nu \bar{\nu}$---are pure leptonic weak interactions,
their interpretation is unambiguous.

The determination of the Fermi constant $G_F$ from the muon lifetime $\tau_{\mu}$
represents a reference point for subatomic physics.
It permits the testing of weak universality, for example, through
precision measurements of leptonic tau-decays.
It enables the determination of weak mass-mixing angles, for example,
through precision measurements of neutron decay.
Moreover---together with the fine structure constant $\alpha$ and the
Z gauge boson mass $M_Z$---it completely determines the electroweak sector of the
standard model and enables searches for new forces and particles.

\subsection{Experimental approaches to measuring $\tau_{\mu}$}

As already discussed, after the discovery of the muon by
Anderson and Neddermeyer \cite{Anderson:1936zz} and Street and Stevenson \cite{Street:1937me}
the earliest measurements of the lifetime---for
stopped muons and in-flight muons---were
important in verifying the time dilation of moving particles.

By the beginning of this century the
Particle Data Group world average of the muon lifetime
was $\tau_{\mu} = 2.19703\pm0.00004$~$\mu$s or 18 parts-per-million (ppm) \cite{Caso:1998tx}.
The world average was largely determined
by three measurements: Giovanetti {\it et al.}\ \cite{Giovanetti:1984yw},
Bardin {\it et al.}\ \cite{Bardin:1984ie}
and Balandin {\it et al.}\ \cite{Balandin:1975fe},
that were conducted in the seventies and the eighties.
The experiments of Giovanetti {\it et al.}\ and Balandin {\it et al.}\
used low-rate continuous beams in order to insure
the arrival and decay of muons occur one-by-one,
{\it i.e.}  avoiding any incorrect assignment
of daughter electrons with parent muons.
At higher rates, if only the previous stop was associated with a particular electron
the measured lifetime would be distorted,
or, if all the neighboring stops were associated
with a particular electron a random background would be incurred.
Such one-by-one measurements therefore limit
the collection of decays to roughly 10$^{10}$
and the uncertainty on $\tau_{\mu}$ to roughly 10~ppm.



The two most recent measurements of the positive muon lifetime---the FAST experiment \cite{Barczyk:2007hp}
and MuLan experiment \cite{Tishchenko:2012ie}---were specifically designed
to circumvent the statistical limitations
of one-by-one measurements.
The FAST approach involved an active pixelated target in order
to reconstruct muon-electron vertices and thereby correctly
associate each decay electron with its parent muon.
In principle therefore,
multiple muons could be simultaneously stopped and
multiple decays could be simultaneously recorded,
without losing the parent-daughter association.
The MuLan approach involved a time-structured muon beam
in order to first prepare a ``radioactive source'' of muons
and afterwards measure  the ``emanating radiation'' of
electrons. In this scheme no association
of a particular daughter electron with a particular parent muon is necessary;
the observed lifetime of a radioactive source is independent of the source preparation.



\subsubsection{MuLan experiment}
\label{ssc:mulan}

\begin{figure}
\begin{centering}
  \includegraphics[width=0.8\columnwidth]{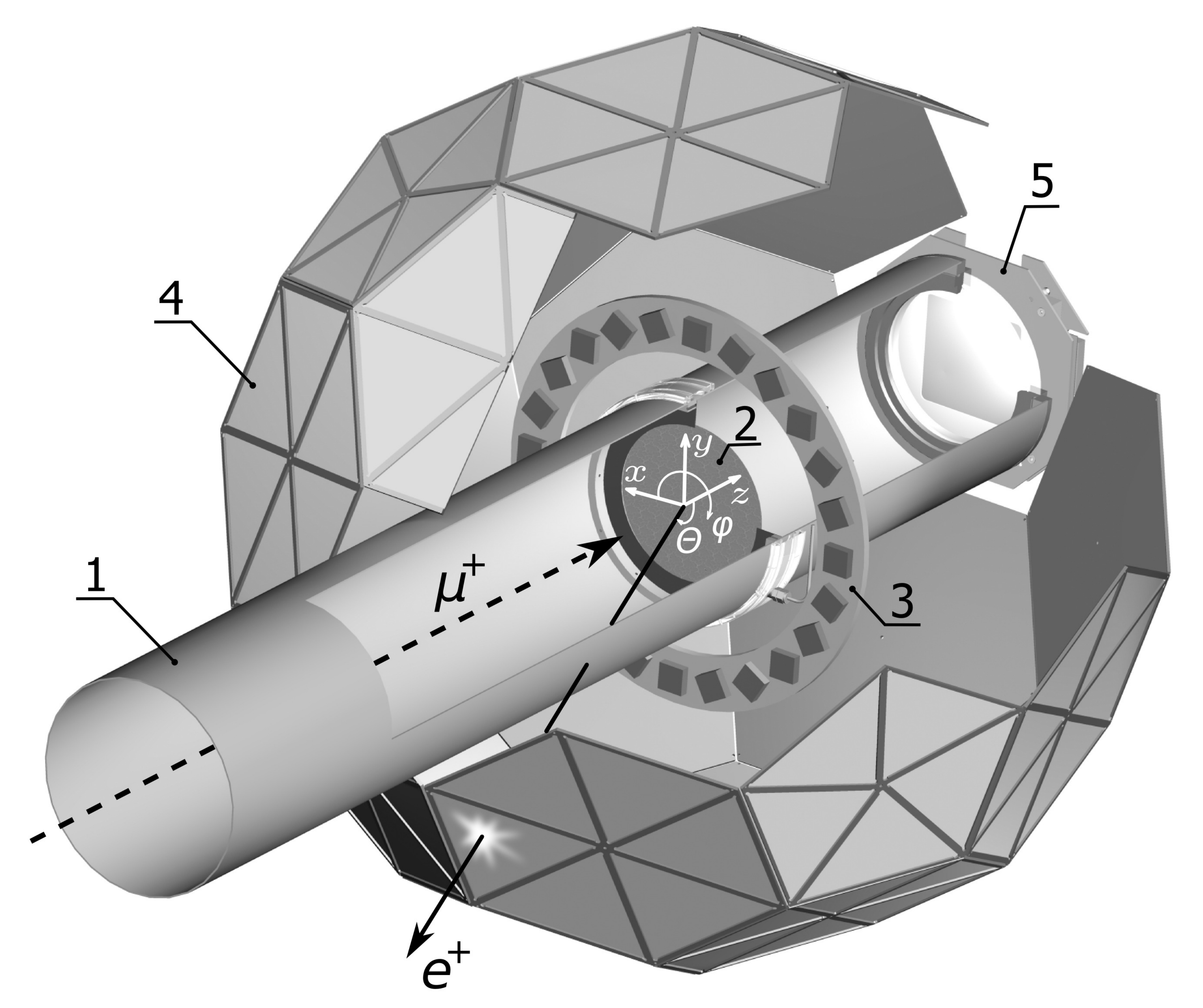}
  \caption{ \label{fg:MuLanSetup} Cutaway diagram of the MuLan experiment
indicating an incoming muon and outgoing positron and showing the
beam pipe (1), stopping target (2), optional magnet array (3),
scintillator detector (4) and beam monitor (5). Figure courtesy MuLan
collaboration.}
\end{centering}
\end{figure}

The MuLan experiment \cite{Tishchenko:2012ie} was conducted at PSI.
The setup---which comprised
an in-vacuum stopping target and
fast-timing finely-segmented positron detector---is shown in Fig.\ \ref{fg:MuLanSetup}.
The setup was designed to both accumulate the
necessary quantity of muon decays
and minimize distortions
arising from muon spin rotation in the stopping target
and electron pile-up in the detector array.

\begin{figure}
\begin{centering}
  \includegraphics[width=0.8\columnwidth]{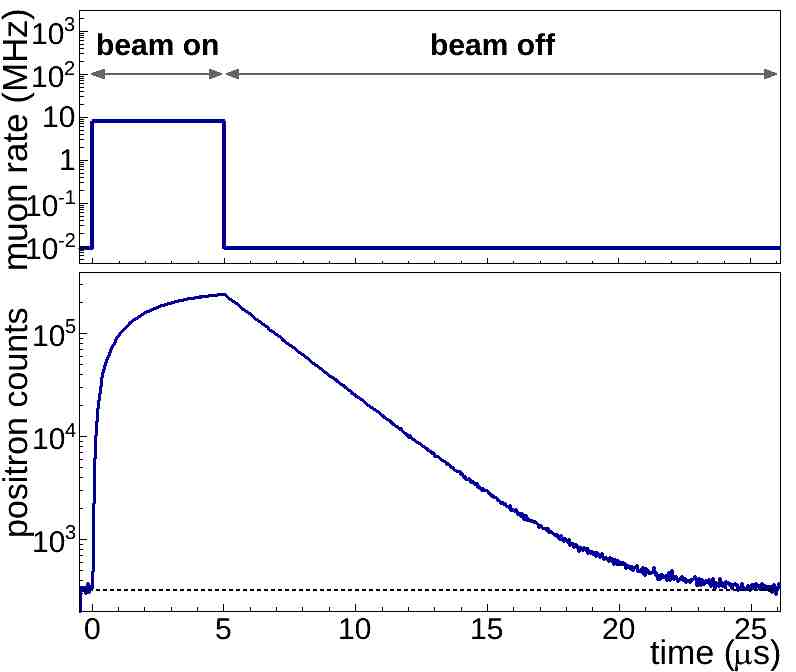}
  \caption{ \label{fg:mulanbeamonoff}
The 5-$\mu$s beam-on, 22-$\mu$s beam-off, time structure
of the MuLan experiment. The upper panel shows the muon arrival
time distribution and the lower panel shows the decay electron time
distribution.}
\end{centering}
\end{figure}

To reach the necessary statistics of 10$^{12}$ decays
the experiment relied on a time-structured surface-muon beam
(see Fig.\ \ref{fg:mulanbeamonoff}).
The experiment involved cycles of 5~$\mu$s duration beam-on periods
to accumulate stopped muons and 22~$\mu$s duration beam-off periods
to measure decay positrons.
The arrangement permitted an average rate of stopped muons
of $1$-$2 \times 10^{6}$~s$^{-1}$; much more than permissible
in a one-by-one lifetime measurement.

Surface muon beams are nearly 100\% longitudinally polarized.
The positrons emitted
in muon decay are distributed asymmetrically
about the $\mu$-spin axis with high-energy positrons
preferentially emitted in the spin direction
and low energy positrons preferentially emitted
opposite the spin direction.
The spin vectors of stopped muons both precess and relax
in the local magnetic field of the target material;
a phenomenon known as muon spin rotation.
$\mu$SR yields a time-dependent muon-ensemble polarization and
thereby a time-dependent decay-positron angular distribution.
It results---when detecting positrons in specific directions---in a geometry-dependent
modulation of the exponential decay curve
by the $\mu$SR signal.

Two combinations of stopping target material
and environmental magnetic
fields were used.
One configuration involved a magnetized ferromagnetic foil
with a roughly 0.5~T internal field orientated perpendicular
to the beam axis.
Another configuration involved a nonmagnetic quartz crystal
with a roughly 80~G external field orientated perpendicular
to the beam axis.
Muon stops in the ferromagnetic target were mostly
diamagnetic $\mu^+$ ions exhibiting a precession frequency
of 13.6~kHz per Gauss while
muon stops in the quartz target were mostly
paramagnetic $\mu^+ e^-$ atoms exhibiting a precession frequency
of 1.39~MHz per Gauss.
The targets were mounted in the beam vacuum
to avoid either multiple scattering or stopping
in upstream detectors or vacuum windows.

The 5~$\mu$s duration of muon accumulation
was important in dephasing the spins of the polarized
stopping muons in the transverse magnetic fields
of the two target configurations.
In both cases the precession frequencies of either
$\mu^+$ ions or $\mu^+ e^-$ atoms was sufficient to
reduce the transverse polarization by roughly a thousand-fold.
The residual polarization
was ultimately limited by the alignment accuracy
between the transverse field and the beam polarization,
{\it i.e.}\ the presence of a small longitudinal component
of the muon polarization.

The positron detector comprised a nearly 4$\pi$ array
of 170 fast-timing, double-layered, scintillator detectors arranged
in a truncated icosahedron (soccer ball) geometry.
The high granularity and fast timing characteristics
were important in minimizing the incidence of positron pile-up.
The forward-backward symmetry of the 3$\pi$ solid angle detector
also suppressed the imprint of $\mu$SR
on the time distribution of the decay positrons.\footnote{A detector
with perfectly forward-backward symmetry about the muon spin direction
would display no $\mu$SR signal in the positron time distribution.}
Therefore the MuLan $\mu$SR signal is strongly suppressed and
largely determined by the the non-uniformity of the detection efficiency.

Analog signals from detector elements were digitized
using fast-sampling ADCs and digitized ``islands'' of contiguous
samples of above-threshold signals were identified by
FPGAs. A distributed data acquisition enabled
the storage of all above-threshold signals from the
positron detector.

A temperature-stabilized crystal oscillator was used
as the timebase for the fast-sampling ADCs.
The collaboration was blinded
to the exact frequency of the timebase during the
data taking and the subsequent analysis.
Only after completing the entire analysis
was the frequency unblinded and the lifetime revealed.

In analyzing the data the digitized islands were first fit
to pulse templates to determine the times and energies of
individual pulses. A software-defined minimum amplitude was
applied to distinguish the minimally ionizing positrons
from low-energy backgrounds and a software-defined
minimum deadtime was applied to establish an explicit resolving
time between neighboring pulses. The time distribution
of coincident hits between inner-outer tile pairs
was then constructed.


After the application
of small data-driven corrections
for positron pileup and gain changes,
the time distributions of coincident hits
were fit to extract $\tau_{\mu}$.
The ferromagnetic target data showed
no evidence of $\mu$SR effects
and was fit to the function
\begin{equation*}
N(t) = N e^{-t/\tau_\mu} + B
\end{equation*}
where the time-independent background $B$ arose from both cosmic rays
and the imperfect beam extinction during the measurement period.
The quartz target data showed a clear $\mu$SR signal
and was therefore fit to a modified function that incorporated
both longitudinal- and transverse-field $\mu$SR effects.
The fits were performed for different software deadtime
and extrapolated to zero software deadtime to obtain $\tau_{\mu}$.


The values of $\tau_{\mu}$ obtained from the two targets
were in good agreement and yielded a combined result of
\begin{equation*}
\tau_{\mu} = 2~196~980.3 \pm  2.1 ({\rm stat}) \pm 0.7 ({\rm syst})~{\rm ps} ;
\end{equation*}
an overall uncertainty of 2.2~ps (1.0~ppm) and thirty-fold improvement
over earlier generations of experiments.

\subsubsection{FAST experiment}

\begin{figure}
\begin{centering}
  \includegraphics[width=1.0\columnwidth]{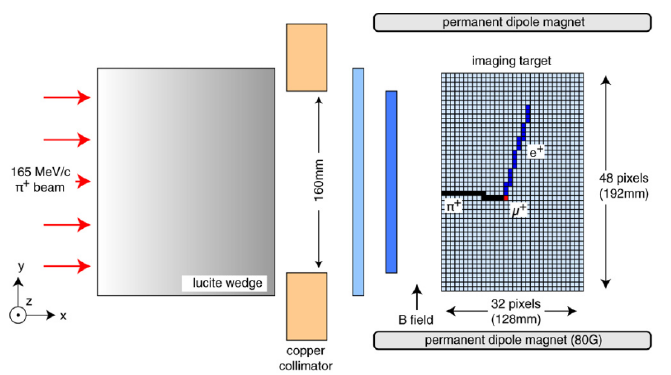}
  \caption{ \label{fg:FASTSetup} Horizontal cross section of the
FAST experiment showing (from left to right) the wedge-shaped degrader, beam defining scintillators
and pixelated stopping target. The pattern of hits from a $\pi \rightarrow \mu \rightarrow e$ decay chain are also indicated. Figure courtesy FAST
collaboration.}
\end{centering}
\end{figure}

The FAST experiment \cite{Barczyk:2007hp} was also conducted at PSI.
The setup---which consisted of beam defining counters
and a pixelated stopping
target---is shown in Fig \ref{fg:FASTSetup}.

The approaches of FAST and MuLan to measuring $\tau_{\mu}$ were very different.
In FAST the segmented target was used to associate
the daughter positrons with parent muons by identifying
the coordinates of the muon stop and the trajectory of the decay electron.

Additionally, the FAST experiment employed a stopping pion beam.
When pions were stopped the subsequent muons from $\pi \rightarrow \mu \nu$ decay
and subsequent positrons from $\mu \rightarrow e \nu \bar{\nu}$ decay were
detected, thereby identifying the entire $\pi \rightarrow \mu \rightarrow e$
decay chain.
Since the decay muons from stopped pions are emitted isotropically,
the overall ensemble of stopped muons had essentially zero polarization.
This minimizes the  $\mu$SR distortions of the decay-positron
time distribution.

To widely distribute the stopping pions over the segmented target
the beam was defocused at the target location and directed through a wedge-shaped degrader.
The degrader varied the penetration depth of pions according to
their vertical coordinate. Beam counters were situated both upstream
and downstream and provided the identification of stopping pions and
through going electrons. A permanent magnet comprising two planes of
ferrite blocks provided a 80~G transverse magnetic field.

The segmented stopping target comprised a $32 \times 48$ pixelated array
of $4 \times 4 \times 200$~mm$^3$, vertically orientated, plastic scintillator bars.
The light from individual bars
was readout by multi-anode photomultipliers. After amplification and
discrimination the hit times of individual pixels were recorded by multihit TDCs.
The setup employed two discriminator thresholds,
a higher-level threshold capable of identifying the heavily ionizing pions / muons
that was used in the trigger logic and a lower-level threshold capable of identifying
the minimally ionizing positrons that was used in the data analysis.

In normal data-taking the TDC information
was readout after the identification of a pion stop
and a corresponding decay muon in the pixelated target.
A level-one trigger distinguished
the incident pions from other beam particles.
A level-two trigger identified
the sequence involving the prompt pion signal
and the delayed muon signal
(it used an FPGA array to perform the trigger decision
within several microseconds).
The range of the 4.1~MeV muons from the $\pi \rightarrow \mu \nu$ decay
was approximately 1.5~mm and therefore the muon stop was
located in either the same pixel or an adjacent pixel to the pion stop.
On fulfilling the above trigger the pion stop pixel was
used to define a $7 \times 7$ pixel region and a $-8$ to $+22$~$\mu$s time window
for the selective readout of the TDC modules.


Data were collected at
incident beam rates of roughly 160~kHz and
stopping pion rates of roughly 80~kHz,
and yielded a sample of approximately $1.0 \times 10^{10}$
$\mu \rightarrow e \nu \bar{\nu}$ decay events.
To handle the large data volume
the TDC information was both readout and
also processed in real-time.
The real-time processing used lookup tables
to identify the characteristic topologies
of pixel hits corresponding
to a positron emanating from the muon stop
(the topologies allowed for pixel inefficiencies).

The identification of $\pi \rightarrow \mu \rightarrow e$
events yielded the corresponding times $t_{\pi}$, $t_{\mu}$ and $t_{e}$
of the pion, muon and positron and thereby the time differences
between the stopped pion and the decay electron,
$t_{e} - t_{\pi}$,
and between the stopped muon and the decay electron,
$t_{e} - t_{\mu}$.
This real-time processing accumulated both global
$t_{e} - t_{\pi}$ and $t_{e} - t_{\mu}$
histograms for all $\pi \rightarrow \mu \rightarrow e$
events as well as sub-sets for
different pion coordinates,
positron topologies, {\it etc.}
Both the measured $t_e - t_{\pi}$ and $t_e - t_{\mu}$ distributions
exhibit the muon lifetime.\footnote{The $t_e - t_{\pi}$ distribution is
a single decay curve with lifetime $\tau_{\mu}$ for time intervals much greater 
than $\tau_{\pi}$.}



The time distributions showed the exponential decay curve
as well as time-independent and cyclotron-RF correlated backgrounds.
The two backgrounds arose from random backgrounds and beam particles
that generated fake positron topologies and thereby
$\pi \rightarrow \mu \rightarrow e$ events.
The time distributions showed no evidence of $\mu$SR effects.

The RF-correlated background
was handled by rebinning data
with the cyclotron RF period.
After the rebinning procedure
a maximum likelihood fit
to the decay curve gave the final result
\begin{equation*}
\tau_{\mu} = 2~197~083 \pm 32 ({\rm stat}) \pm 15 ({\rm syst})~{\rm ps}
\end{equation*}
with an overall uncertainty of 35~ps (16~ppm).

\subsection{Negative muon lifetime}

The above experiments involved measurements
of the lifetime $\tau_{\mu}$
of the positively charged muon.
As discussed elsewhere,
for stopped negative muons,
in addition to muon decay
the process of muon capture
is possible.
Measurement of the negative muon lifetime in a particular stopping material
therefore determines a total disappearance rate, {\it i.e.}\
the sum of the decay rate and the capture rate.
The contribution of the capture rate to the disappearance rate
varies from roughly 0.1\% in hydrogen isotopes
to greater than 90\% in heavy elements.

In principle the lifetime of free negative muons
can be determined by the combined measurements of the
disappearance rate and the capture rate for a specific stopping target.
For muon stops in hydrogen and deuterium
there exist both measurements of the muon disappearance
rates ($\Lambda_D$) via the decay electron time spectra and the total
muon capture rates ($\Lambda_C$) via the capture neutron yields.
Thus the combination of these measurements allows the determination
of the free negative muon lifetime using
\begin{equation}
1 / \tau_{\mu^-} = \Lambda_D - \Lambda_C .
\label{eq:negativemuonlifetime}
\end{equation}

Unfortunately, the procedure is complicated by muon atomic and molecular
processes that occur in isotopes of hydrogen.
Consequently, to directly extract the lifetime $\tau_{\mu^-}$
the values of $\Lambda_D$ and $\Lambda_C$ must be determined
for the same muonic atomic and molecular state populations.
We note that together---the recent measurement of the muon disappearance rate \cite{Andreev:2012fj}
and the earlier measurements of the muon capture rate
\cite{Bystritskii:1974gd,AlberigiQuaranta:1969ip}
from the singlet $\mu^- p$ atom---yield using Eqn.\ \ref{eq:negativemuonlifetime}
a negative lifetime that is entirely consistent
with the positive muon lifetime.\footnote{Small corrections for
$\mu^- p$ bound state effects and $pp \mu$ molecular formation are required to
extract of the negative muon decay rate from the measured
values of the disappearance rate and the capture rate. See Ref.\ \cite{Andreev:2012fj} for details.}

\subsection{Results for the muon lifetime}

The uncertainty of the MuLan measurement of the muon lifetime
is about 30-40 times smaller than the previous generation
of lifetime experiments and about ten times smaller
than the published result from the FAST experiment.
While the MuLan result is in reasonable agreement
with the earlier experiments there is some tension
between the MuLan experiment and the FAST experiment.
The weighted average of all results
gives a lifetime $\tau_{\mu} = 2\, 196\, 981.1 \pm 2.2$~ps
with a chi-squared value 
that is dominated by the 2~$\sigma$ difference
between MuLan and FAST.

As discussed in Sec.\ \ref{sc:muz},
the accurate knowledge of $\tau_{\mu}$ is important
to precision measurements in muon capture.
The MuCap experiment \cite{Andreev:2007wg}
has recently measured the $\mu^-$p singlet capture rate $\Lambda_s$
and the MuSun experiment \cite{Andreev:2010wd} is currently  measuring
the $\mu^-$d  doublet capture rate $\Lambda_d$.
Both experiments derive these capture rates
from the tiny difference $( \Lambda_o - \Lambda )$ between the positive muon decay rate ($\Lambda_o = 1/\tau_{\mu^+}$) and
the muonic atom disappearance rates ($\Lambda = 1/\tau_{\mu Z}$).
Because the capture rates are very small,
very precise determinations of
both the muonic atom disappearance rates
and the free muon decay rate are necessary.

Prior to this work the muon decay rate $\Lambda_o$
was known to 9~ppm or 5~s$^{-1}$, an uncertainty
that limited the capture rate determinations
from disappearance rate experiments.
The recent work on the muon lifetime---yielding
a precision of 1~ppm or 0.5~$s^{-1}$ in
$\Lambda_o$---has eliminated this source of uncertainty.

\subsection{Results for the Fermi constant}

As mentioned, the value of the Fermi constant $G_F$
can be extracted from the muon lifetime $\tau_{\mu}$.
Until recently the uncertainty
in the theoretical relationship between the Fermi constant and the
muon lifetime was about 15~ppm.

Recent work by van Ritbergen and Stuart
\cite{vanRitbergen:1999fi,vanRitbergen:1998yd,vanRitbergen:1998hn}
and Pak and Czarnecki \cite{Pak:2008qt} have reduced this
theoretical uncertainty to about 0.14~ppm.
Van Ritbergen and Stuart
were first to calculate
the muon lifetime with 2-loop order QED corrections.
Pak and Czarnecki were subsequently responsible
for treating corrections from the mass of the electron.

The van Ritbergen and Stuart relation between the Fermi constant $G_F$
and the muon lifetime $\tau_{\mu}$---derived using the
$V$-$A$ current-current Fermi interaction with
QED corrections evaluated to 2-loop order---yields

\begin{equation}
G_F = \sqrt{\frac{192\pi^3}{\tau_\mu m_\mu^5} \frac{1}{1 + \Delta q^{(0)} + \Delta q^{(1)} +
\Delta q^{(2)}}}\
\label{eq:gf_inversion}
\end{equation}
where $\tau_{\mu}$ is the measured muon lifetime,
$m_{\mu}$ is the measured muon mass, and $\Delta q^{(0)}$,
$\Delta q^{(1)}$ and $\Delta q^{(2)}$ are theoretical corrections
with  $\Delta q^{(0)}$ accounting for the effects of the non-zero
electron mass on the muon-decay phase space and $\Delta q^{(1/2)}$ accounting
for the contributions of the 1-/2-loop radiative corrections
to the decay amplitude.

Using Eqn.\ \ref{eq:gf_inversion}, 
with the MuLan value for the muon lifetime $\tau_{\mu}$ \cite{Tishchenko:2012ie},
the 2010 CODATA recommended value of the muon mass $m_{\mu}$ \cite{Mohr:2012tt},
and the theoretical corrections $\Delta q^{(0)}$,
$\Delta q^{(1)}$ and $\Delta q^{(2)}$ gives,\footnote{This value for $G_F$ is also
listed in the 2014 compilation of the physical constants 
by the Particle Data Group
\cite{Agashe:2014kda}.}
\begin{equation*}
G_F =
1.166\, 378\, 7(6)\times 10^{-5}~{{\rm GeV}^{-2}}\ (0.5~\mathrm{ppm}) .
\label{eq:gf_mulan_final}
\end{equation*}
The result is a thirty-fold improvement over the
Particle Data Group value prior to the recent theoretical work
and the MuLan / FAST experiments.

The electroweak sector of the standard model involves three parameters,
the two gauge coupling constants $g$, $g\prime$
and the Higgs vacuum expectation value $v$.
Their values are fixed by measurements
of the fine structure constant, $\alpha$,
Fermi coupling constant, $G_F$,
and Z boson mass, $M_Z$. Consequently, the thirty-fold improvement
in the determination of $G_F$, together with other improvements
in the determinations of $\alpha$ and $M_Z$,
now permit improved tests of the electroweak sector
of the standard model.

Within the standard model the quantities $\alpha$, $M_Z$ and $G_F$
are related to other fundamental quantities that include
the charged weak boson mass $M_W$ and weak mixing angle $\theta_W$.
Such SM relationships have radiative corrections
that impart sensitivities to the top quark mass $m_t$
and the Higgs boson mass $m_h$. Historically, precision electroweak
data was important in constraining the then-unknown masses
of the top quark and the Higgs boson.

The top quark was discovered in 1995
at Fermilab by the CDF and D0 experiments \cite{Abe:1995hr,Abachi:1995iq}
and the Higgs boson was discovered in 2012
at CERN by the ATLAS and CMS experiments \cite{Aad:2012tfa,Chatrchyan:2012ufa}.
The spectacular agreement between the forecast and the
measurement of top mass $m_t = 173$~GeV/c$^2$
was a huge triumph for the predictive power
of the standard model.
Through the SM relations for the Higgs sector \cite{Leader:1996hk}
\begin{equation*}
G_F = {1  \over  \sqrt{2} v^2 }
\end{equation*}
\begin{equation*}
m_h = 2 \nu^2 \lambda
\end{equation*}
our knowledge of $G_F$ and $m_h$
are sufficient to determine the two parameters---the vacuum expectation value $v$
and self-interaction parameter $\lambda$---of the SM Higgs potential.





The Fermi constant obtained from the muon lifetime
is the anchor of the universality tests of the weak force.
Comparison between the purely leptonic decays
of the tau and the muon---{\em i.e.},  $\tau \rightarrow \mu \nu \bar{\nu}$,
$\tau \rightarrow e \nu \bar{\nu}$ and $\mu \rightarrow e \nu \bar{\nu}$---are
natural opportunities for testing the universality of
leptonic weak interactions across the three generations.
Using the available data on the $\tau$ lifetime
and its purely leptonic $\mu \nu \bar{\nu}$ and $e \nu \bar{\nu}$
branching ratios the leptonic universality of weak interactions
has been demonstrated to the levels of several parts-per-thousand \cite{Agashe:2014kda}.

The Fermi constant is also the metric
for extracting the Cabibbo-Kobayashi-Maskawa (CKM)
matrix elements which specify the flavor mixing
in weak interactions of quarks.
The concept of flavor mixing of quark states
was originally introduced by Cabibbo
to explain the different strengths
of strangeness-changing and strangeness-conserving
weak interactions and thereby save weak universality.
Modern tests of weak universality for
three quark generations are therefore based on testing whether
the sum of all couplings of an up-type quark (up, charm or top) to
all down-type quarks (down, strange and bottom) is equal to unity.
For the case of the up-quark, one obtains \cite{Hardy:2014qxa}
\begin{equation*}
| V_{ud} |^2  + | V_{us} | ^2 + | V_{ub} |^2 = 0.99978(55)
\end{equation*}
which demonstrates weak universality between the quark-lepton
sectors at the level of about 0.6 parts-per-thousand.

%% file: MuonMichel_final.tex
\section{Muon decay}
\label{sc:decayparameters}

\subsection{$V$-$A$ structure of the weak interaction}

Herein we discuss the decay $\mu^+ \rightarrow e^+ \nu_e \bar{\nu_{\mu}}$
and the investigation of the Lorentz structure of the
leptonic weak charged current.
Through combination  of its purely-leptonic nature---permitting
rigorous theoretical analysis---and its not-too-short, not-too-long
lifetime---permitting rigorous experimental analysis---the
decay $\mu^+ \rightarrow e^+ \nu_e \bar{\nu_{\mu}}$
is the quintessential test
of the $V$-$A$ structure
of the weak interaction.



\begin{figure}
\begin{centering}
  \includegraphics[width=\columnwidth]{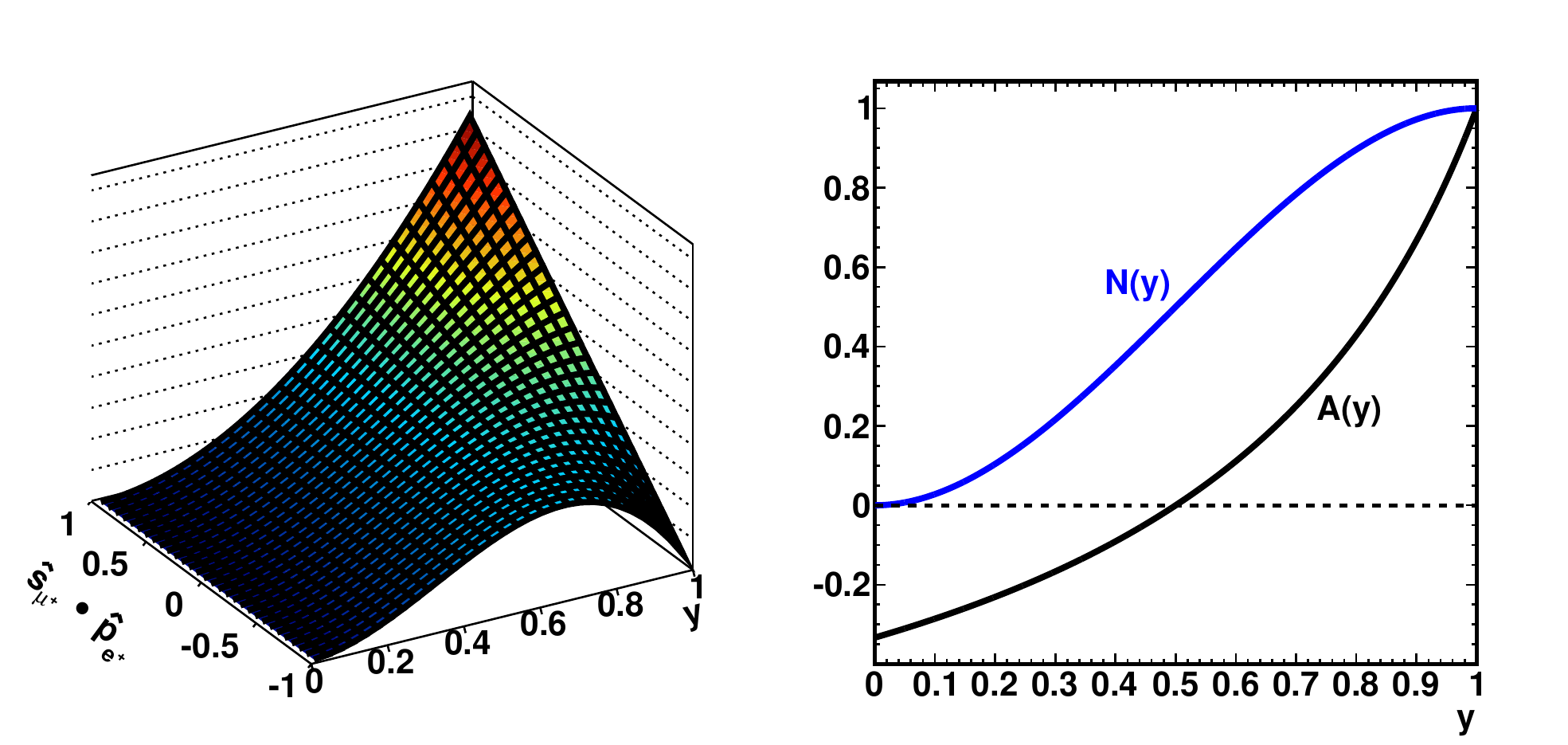}
  \caption{\label{fg:mudecay}
    Left: Relative probability in arbitrary units
of the emitted positron in ordinary muon decay as a function
of the muon spin direction $\hat{s_{\mu}}$ and the emitted positron
reduced energy $y$ for the $V$-$A$ weak interaction.
Right:  Projections of the relative number,
$N(y)$, and asymmetry, $A(y)$, vs. positron reduced energy.}
\end{centering}
\end{figure}

The measurable observables of emitted positrons
in $\mu^+ \rightarrow e^+ \nu \bar{\nu}$ decay
are the positron energy and directional
distributions, shown
in Fig.\ \ref{fg:mudecay}, and the positron polarization.
The $V$-$A$ interaction dictates
the precise form of the energy-angle
distribution and the polarization observables
which include the parity non-conserving
angular asymmetry and longitudinal polarization.
In general the presence of non-$V$-$A$ currents
would change these energy, angle and polarization
distributions.

Based upon a general current-current interaction with
possible scalar, pseudoscalar, tensor, vector and axial terms
the positron energy, angle and polarization distributions
may be parameterized by nine parameters
(see Refs.\ \cite{Michel:1950,PhysRev.93.354,PhysRev.107.593}).
The differential probability for emitting a
positron with a reduced energy $y$
at an angle $\theta$ from the
muon  spin axis
is given by
\begin{equation}
{d^2 \Gamma \over dy~d\cos{\theta}}
= {1 \over 4 \pi^3} m_{\mu} W^4_{e\mu} G_F^2 \sqrt{y^2-y_0^2} ~
\left( F_{IS}(y) + P_{\mu} \cos{\theta} F_{AS}(y) \right)
\end{equation}
where $P_{\mu} = | \vec{P_{\mu}} |$ is the muon polarization,
$W_{e\mu} = (m_{\mu}^2+m_e^2)/2m_{\mu} = 52.8$~MeV
is the maximum positron energy, $y = E_e/W_{e\mu}$
is the reduced energy, and $y_o = m_e/W_{e\mu} = 9.7 \times 10^{-3}$.
The y-dependent functions
\begin{equation}
F_{IS}(y) = y(1-y) + {2 \over 9} \rho (4y^2-3y-y_o^2) + \eta y_o (1-y)
\end{equation}
%
%
\begin{equation}
F_{AS}(y) = \frac{1}{3} \xi \sqrt{y^2-y_o^2} \left \{ 1-y+\frac{2}{3} \delta \left[ 4y -3 + (\sqrt{1-y_o^2}-1) \right] \right \}
\end{equation}
describe the isotropic and anisotropic parts of
the energy spectrum.  They are governed by the
decay parameters $\rho$, $\eta$, $\xi$, and $\delta$---conventionally
termed the Michel parameters.
For unpolarized muons, the two
parameters $\rho$ and $\eta$ completely determine
the energy distribution with $\rho$ governing the high-energy part
and $\eta$ governing the low-energy part of the spectrum.\footnote{The
contribution of $\eta$ is order $y_o$ and consequently very small.}
For polarized muons, the two parameters $\xi$ and $\delta$
additionally determine the angular distribution
with $\xi$ governing the energy-integrated positron asymmetry
and $\delta$ governing the energy dependence of this angular asymmetry.
The standard model values of the Michel parameters are $\rho = 0.75$, $\eta = 0.0$,
$\xi = 1.0$ and $\delta = 0.75$.

The energy-angle dependent positron polarization vector $\vec{P_e} (y, \theta)$
may be decomposed into a longitudinal component $P_L$ and two transverse
components $P_{T1}$, $P_{T2}$ according to
\begin{equation}
\vec{P_e} (y, \theta)  = P_L \cdot \hat{z}
+ P_{T1} \cdot ( \hat{z} \times \hat{P_{\mu}} ) \times \hat{z}
+ P_{T2} \cdot ( \hat{z} \times \hat{P_{\mu}} )
\end{equation}
where $\hat{z}$ and $\hat{P_{\mu}}$ are unit vectors
along the positron momentum and the muon polarization, respectively.
The transverse polarization $P_{T1}$ lies
in the decay plane of the positron momentum and the muon spin
while the transverse polarization $P_{T2}$
lies along the perpendicular to this decay plane.
Note that the transverse polarization $P_{T1}$
is a (time-reversal) T-conserving observable
while the transverse polarization $P_{T2}$
is a T-violating observable.

The energy-angle dependence of the longitudinal polarization $P_L$
is given by
\begin{equation}
P_L ( y, \theta ) = { F_{IP}(y) + P_{\mu} \cos{\theta} F_{AP}(y) \over
F_{IS}(y) + P_{\mu} \cos{\theta} F_{AS}(y) }.
\end{equation}
The two additional y-dependent functions are
\begin{equation}
F_{IP}(y) = {1 \over 54} \sqrt{y^2-y_o^2}  \left\{ 9 \xi^{\prime} \left( -2y+2+\sqrt{1-y_o^2} \right )
+ 4 \xi \left( \delta - {3 \over 4} \right) \left( 4y -4 + \sqrt{1-y_o^2} \right) \right\}
\end{equation}
\begin{equation}
F_{AP}(y) = {1 \over 6} \left \{ \xi^{\prime \prime} \left(2y^2 - y - y_o^2 \right) + 4 \left( \rho -{3 \over 4} \right)
( 4y^2 -3y -y_o^2) + 2 \eta^{\prime \prime} (1-y) y_o \right \}
\end{equation}
and involve two further decay
parameters---$\xi^{\prime}$
and $\xi^{\prime \prime}$.\footnote{An additional
parameter $\eta^{\prime \prime}$ also enters
$F_{AP}(y)$ but its role
is highly suppressed
by the factor $y_o$.}
Neglecting the positron mass and radiative
corrections, the standard model values
of the decay parameters $\xi^{\prime} = \xi^{\prime \prime} = 1.0$
yield an energy-angle independent longitudinal
polarization $P_L = 1$.

The T-conserving polarization $P_{T1}$ and its energy-angle dependence
is given by
\begin{equation}
P_{T1} ( y, \theta ) = {P_{\mu} \sin{\theta} F_{T1}(y) \over
F_{IS}(y) + P_{\mu} \cos{\theta} F_{AS}(y)}
\end{equation}
and the T-violating polarization $P_{T2}$ and its energy-angle dependence
is given by
\begin{equation}
P_{T2} ( y, \theta ) = {P_{\mu} \sin{\theta} F_{T2}(y) \over
F_{IS}(y) + P_{\mu} \cos{\theta} F_{AS}(y)}.
\end{equation}
The two additional y-dependent functions are
\begin{equation}
F_{T1}(y) = {1 \over 12} \left\{ -2 \left[ \xi^{\prime\prime} + 12 \left( \rho-{3 \over 4} \right) \right ]
(1-y) y_o -3 \eta (y^2 - y_o^2) + \eta^{\prime\prime} (-3y^2+4y-y_o^2)  \right\}
\end{equation}
\begin{equation}
F_{T2}(y) = {1 \over 3} \sqrt{y^2-y_o^2} ~ \left[ 3 {\alpha^{\prime} \over A} (1-y)
+ 2 {\beta^{\prime} \over A} \sqrt{1-y_o^2} \right].
\end{equation}
The T-conserving polarization $P_{T1}$ and its energy-angle dependence
are determined by the decay parameters $\eta$ and $\eta^{\prime \prime}$ and
the T-violating polarization $P_{T2}$ and its energy-angle dependence
are determined by the decay parameters $\alpha^{\prime}/A$ and $\beta^{\prime}/A$.
The standard model values of these parameters are all zero;
$\eta = \eta^{\prime \prime} = \alpha^{\prime}/A = \beta^{\prime}/A = 0$.
They yield---when neglecting
the positron mass and radiative corrections---an
energy-angle independent transverse polarization
$P_{T1}  = P_{T2} =  0$.
On accounting for positron mass effects
the average T-conserving transverse polarization component is
$\langle P_{T1} \rangle  = -3 \times 10^{-3}$
\cite{Danneberg:2005xv}.






The standard model and measured values of
the decay parameters are summarized in Table
\ref{tab:decayparameters}.
Until recently our knowledge
of the muon decay parameters was
based on experiments from twenty, thirty
and forty years ago.
For example, the best determination
of the energy spectrum parameter $\rho$
was due to Derenzo {\it et al.}\cite{PhysRev.181.1854} and
the best determinations of the
angular distribution parameters $\xi$ and $\delta$
were due to
Balke {\it et al.}\cite{PhysRevD.37.587},
Beltrami {\it et al.}\cite{Beltrami:1987ne}
and Jodidio {\it et al.}\cite{PhysRevD.34.1967}.

The past decade has seen a substantial improvement
in the  experimental knowledge of the decay parameters
of both the $\mu^+ \rightarrow e^+ \nu \bar{\nu}$ energy-angle
distribution and the various polarization observables.
A precision measurement of the positron energy-angle distribution
in polarized muon decay was conducted by the TWIST collaboration \cite{TWIST:2011aa,Bueno:2011fq}
at the TRIUMF cyclotron.
It led to roughly a factor ten improvement
in the determination of the decay parameters $\rho$, $\xi$ and $\delta$.
Other precision measurements of both
the transverse polarization \cite{Danneberg:2005xv}
and the longitudinal polarization \cite{Prieels:2014paa}
of the decay positrons were recently conducted
at PSI.
They have improved our knowledge
of the longitudinal polarization parameter $\xi^{\prime \prime}$
by roughly a factor of ten
and the transverse polarization parameters
by roughly a factor of three.

\subsection{TWIST experiment}

\begin{figure}
\begin{centering}
  \includegraphics[width=0.8\columnwidth]{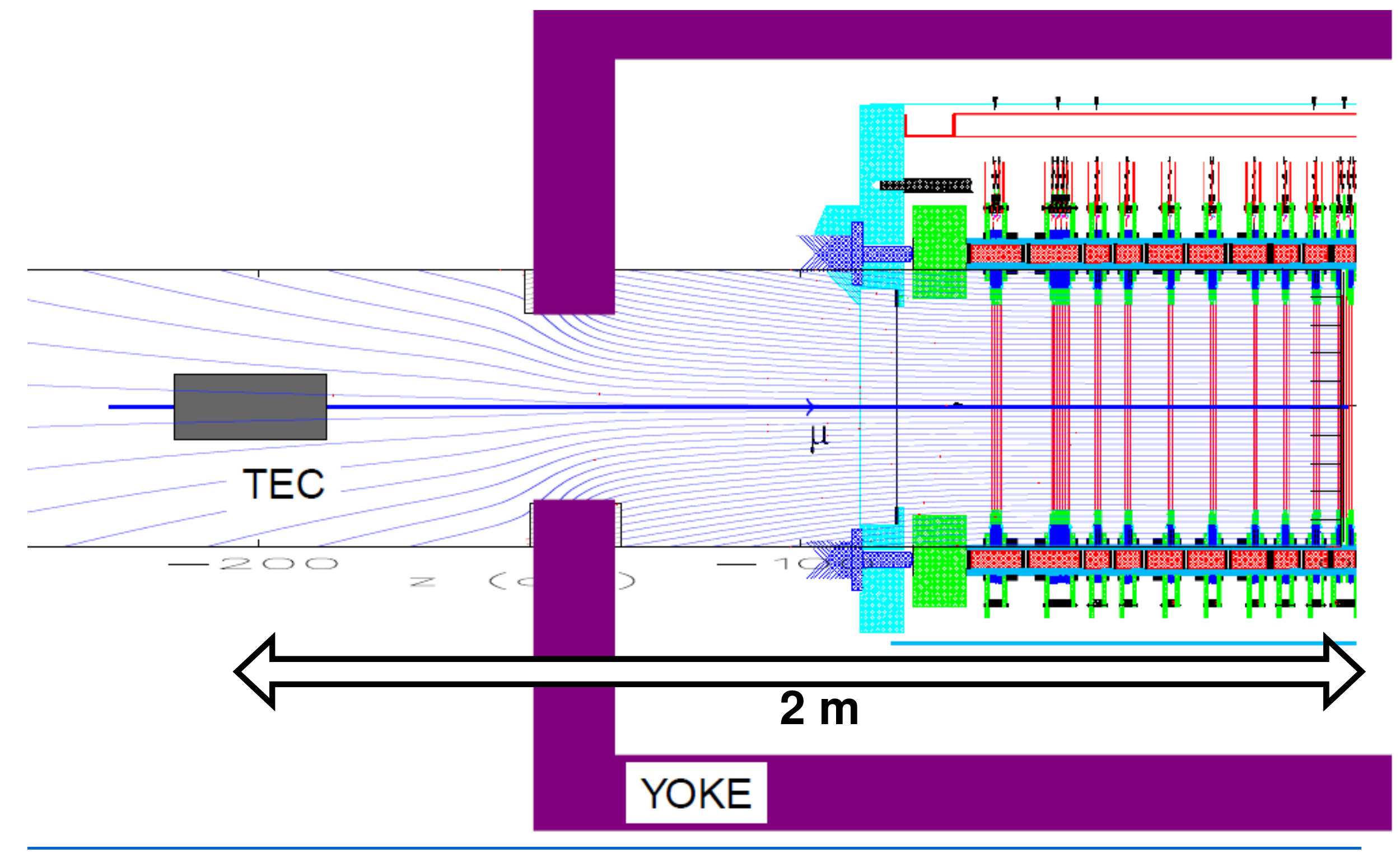}
  \caption{ \label{fg:TWISTSetup} Cross section view of the TWIST experiment
indicating the muon beam, time expansion chambers (TECs), beam scintillator,
planar drift and proportional chambers, stopping target,
and the iron yoke and the magnetic field of the cylindrical
superconducting magnet. Figure courtesy TWIST collaboration.}
\end{centering}
\end{figure}

TWIST---the TRIUMF Weak Interaction Symmetry
Test experiment \cite{TWIST:2011aa,Bueno:2011fq}---was
designed to measure with unprecedented accuracy the
energy-angle distribution of decay positrons from polarized muons.
The setup---including the muon beam, stopping target and
positron spectrometer---is shown in Fig.\ \ref{fg:TWISTSetup}.
The experiment involved stopping a highly polarized muon beam
in a very thin target foil and detecting the resulting decay positrons
in a low mass, large acceptance, high resolution magnetic spectrometer.

%

The experiment utilized a 0.7\% momentum bite, surface muon beam
with a near-100\% polarization at typical rates of several kHz.
To fully characterize and periodically monitor
the beam properties the beamline incorporated
two custom time expansion chambers (TECs).
Each TEC contained a separate horizontal and
vertical drift region to measure
the horizontal and vertical trajectories
of individual beam particles.
The TECs  were constructed with
ultra-thin entrance and exit windows and filled
with low pressure (80~mbar) gas.

The beam entered the spectrometer through a thin scintillator
and a gas degrader. A feedback system adjusted the degrader's
helium / CO$_2$ gas mixture to account for environmental changes
and maintain centering of the stopping distribution in the target foils.

The magnetic spectrometer permitted the measurement of both the
beam muons and the decay positron. The setup comprised
a symmetric array of planar drift chambers and planar proportional
chambers along the common beam and spectrometer axis.
The drift chambers enabled the precision determination
of the momentum-angle coordinates from the helical trajectories
of the decay positrons.
The proportional chambers rendered the $\mu$ / $e$ particle discrimination
via the above-threshold time interval of the resulting chamber signals.
The low-mass chambers were constructed with ultra-thin windows  and filled
with  slow-drift gas to optimize position resolution.
The support structure was constructed from low-thermal expansion
ceramics  to reduce the effects of temperature fluctuations
on spectrometer resolution.

The experiment utilized two different stopping targets;
a 31~$\mu$m thickness, high-purity silver foil 
and a 72~$\mu$m thickness, high-purity aluminum target. 
To optimize the detector resolution the target foils were employed
as the window material separating the two central drift chambers
of the magnetic spectrometer.

A cylindrical superconducting magnet with symmetric upstream and downstream holes
provided a uniform magnetic field along the beam axis.
The field uniformity in the tracking region was a few parts-per-thousand
and the field stability was continuously monitored by NMR probes.
The field was both mapped with a movable Hall probe
and calculated using a finite element analysis.
The track reconstruction used the calculated field distribution
anchored by the Hall probe measurements.

The experiment accumulated large datasets
of decay positrons from  each target
as well as dedicated measurements
of beam properties and other systematics uncertainties.
Both experimental data and simulated data
were stored in a common data format
and processed with a common analysis code.

The analysis involved the reconstruction of the helical particle tracks from the raw TDC data
and the construction of the positron momentum-angle (p-$\theta$)
distribution from the track parameters.
A sophisticated simulation
of particle interactions,
detector geometry, and
data readout incorporated:
the TEC measurements of beam particle trajectories,
the discontinuous nature of chamber ionization,
a Garfield calculation of the position-dependent drift velocities,
an OPERA calculation of the magnetic field distribution,
and rate-dependent effects of overlapping tracks.
The simulated p-$\theta$ coordinates of decay positrons
included full radiative corrections to order $\mathcal{O}(\alpha^2)$
and leading-log radiative corrections to order $\mathcal{O}(\alpha^3)$.

Importantly the measured and simulated p-$\theta$ distributions
are functions of both the muon decay parameters and the p-$\theta$ dependence
of the spectrometer response function and the track reconstruction efficiency.
Therefore the fitting of p-$\theta$ distributions
is used to extract only the differences
between the simulated values and the measured values of
the decay parameters, {\it i.e.}\ $\Delta \rho$, $\Delta \xi$ and $\Delta \delta$.
This ``relative''  method facilitated
a blind analysis
where the exact values
of the decay parameters
were not known
to the collaboration
until the completion
of the data analysis.

The final results for the Michel parameters extracted from the
combined analysis of the experimental data and the simulated data
were
\begin{eqnarray*}
\rho &=& 0.74977 \pm 0.00012~({\rm stat}) \pm 0.00023~({\rm syst})\, \\
\delta &=& 0.75049 \pm 0.00021~({\rm stat}) \pm 0.00073~({\rm syst})\, \\
P^{\pi}_{\mu} \xi  &=& 1.00084 \pm 0.00029~({\rm stat}) \pm~^{0.00165}_{0.00063}~({\rm syst}).
\end{eqnarray*}
Note that the
decay parameter $\xi$ and the muon polarization in pion decay $P^{\pi}_{\mu}$
are experimentally inseparable. Contrary to $\rho$ and $\delta$
the uncertainty in $P^{\pi}_{\mu} \xi$ is dominated by
muon depolarization effects including the spectrometer-beam alignment
and spectrometer fringe fields.

\subsection{Measurement of the positron longitudinal polarization}


A measurement of the positron longitudinal polarization $P_L$ in 
muon decay was recently
reported by Prieels {\it et al.}\ \cite{Prieels:2014paa}.
The experimental setup was configured to
measure the longitudinal polarization
near the positron energy endpoint 
and opposite to the muon spin direction. 
This kinematical region is uniquely sensitive to $\xi^{\prime \prime}$,
the decay parameter that characterizes the angle-energy dependence of $P_L$.
A polarization of $P_L \neq 1$ for high-energy, backward-angle positrons
would signal a departure of this decay parameter from the SM value $\xi^{\prime \prime} = 1$.

The experiment utilized a high intensity, longitudinally polarized,
surface muon beam at PSI. The incident muons
were stopped in either an aluminum target or a sulfur target that
were located in a  0.1~T longitudinal holding field.
In aluminum the initial muon polarization is only weakly depolarized
thus yielding a large time-averaged muon polarization along the polarimeter axis.
In sulfur the initial polarization is more strongly depolarized
thus yielding a small time-averaged muon polarization along the polarimeter axis.

The detection system consisted of a small acceptance, high precision,
cylindrical spectrometer followed by a positron polarimeter.
The spectrometer was orientated to accept only decay positrons
that were emitted directly opposite to the muon polarization direction.
It consisted of three parts: a filter, tracker and lens.
The filter comprised an initial cylindrical magnet with collimators arranged
to transmit only high energy positrons.
The tracker comprised a superconducting cylindrical magnet with three planes
of position-sensitive silicon detectors to measure the
trajectories and determine the energies of decay positrons.
The lens comprised a final cylindrical magnet
that focused the decay positrons into parallel trajectories
at the polarimeter entrance.

The polarimeter utilized the Bhabha scattering $e^+ e^- \rightarrow e^+ e^-$
and in-flight annihilation $e^+ e^- \rightarrow \gamma \gamma$ of
decay positrons on polarized electrons in magnetized foils.
The cross section for Bhabha scattering and in-flight annihilation
is different for the two orientations
with the $e^+ e^-$-spins parallel
and the $e^+ e^-$-spins anti-parallel.
These spin-dependent cross sections thus induce
a signal rate from scattering and annihilation
that depends on $P_L$.

\begin{figure}
\begin{centering}
  \includegraphics[width=0.8\columnwidth]{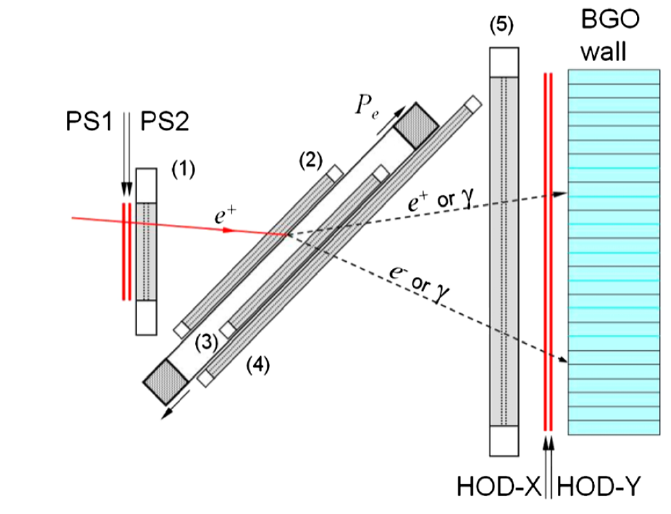}
  \caption{ \label{fg:PrieelsSetup} Schematic diagram of the
double foil positron polarimeter
in the Prieels {\it et al.}\ experiment \cite{Prieels:2014paa}.
It shows the two oppositely-magnetized foils,
the arrangement of wire chambers and plastic scintillators that discriminate between $\gamma$-ray
and $e^+ e^-$ events and determine their vertex coordinates, and the bismuth germanate array. Figure courtesy Prieels {\it et al.}}
\end{centering}
\end{figure}

The polarimeter employed (see Fig.\ \ref{fg:PrieelsSetup})
a double layer of oppositely magnetized foils
that were orientated with their magnetization axes
at $\pm$45$^{\circ}$ angles to the spectrometer axis.
The $\pm$45$^{\circ}$  configuration yielded an electron polarization
along the spectrometer axis of roughly 5\%.
A series of wire chambers and plastic scintillators
at the location of the foils was used to
discriminate the struck foil
and determine the scattering / annihilation vertex.
A bismuth germanate detector array
was used to measure the energies and position coordinates
of the annihilation $\gamma$-rays or the scattered $e^+$$e^-$.
Asymmetries in rates of annihilation
and scattering were investigated
by reversing the foil magnetization
and inverting the  $\pm$45$^{\circ}$ foil orientation.
The measured energies and opening angles of $\gamma$-rays or $e^+$$e^-$
pairs---along with the measured energy of the decay positron in the magnetic
spectrometer---over determined the reaction kinematics and allowed for backgrounds suppression.

The analysis involved constructing the super-ratio
\begin{equation*}
s_{\alpha} = { r^+_{\alpha} - r^-_{\alpha} \over r^+_{\alpha} + r^-_{\alpha} }
\end{equation*}
where $r^{\pm}_{\alpha}$ is the event ratio from the two foils
and the superscript $\pm$ denotes the two magnetization orientations. The subscript
$\alpha$ distinguishes the eight measurement configurations
consisting of (i) $\pm$45-degree foil orientations, (ii) aluminum / sulfur targets,
and (iii) annihilation / scattering events.
The super-ratios are proportional
to the longitudinal polarization $P_L$
of the decay positrons.
Moreover the use of $s_{\alpha}$ cancels the effects
of differing detection efficiencies for scattering / annihilation events
from the two foils.

The key characteristic of $\xi^{\prime \prime}$ differing from unity
is the introduction of an energy-dependent
longitudinal polarization $P_L$ of the backward-emitted decay positrons.
Consequently, a value of $\xi^{\prime \prime} \neq 1$ can be identified
through a comparatively strong energy dependence of $s_{\alpha}$
for the polarization-preserving Al target
with a comparatively weak energy dependence of $s_{\alpha}$
for the polarization-destroying S target.
No evidence of such an effect was observed
in the comparison of the measured super-ratios for the two target materials.

A combined fit to the super-ratio data for all experimental configurations
yielded the value\footnote{The
fitting procedure incorporated energy-independent attenuation factors
that empirically account for globally lower than predicted values
for the analyzing powers for the double foil arrangement.
Although such effects as a smaller than expected magnetization
and a larger than expected background were investigated,
the cause of the attenuator factors was not convincingly established.}
\begin{equation*}
\xi^{\prime \prime} = 0.981 \pm 0.045 ({\rm stat}) \pm 0.003 ({\rm syst}).
\end{equation*}
This result represents a  ten-fold improvement over earlier experimental
work and is consistent with the SM prediction $\xi^{\prime \prime} = 1.0$.

\subsection{Measurement of the positron transverse polarization}

Danneberg {\it et al.}\ \cite{Danneberg:2005xv} have recently reported a measurement
of the transverse polarization of the positrons emitted
in the $\mu \rightarrow e \nu \bar{\nu}$ decay.
In the limit of vanishing positron mass
the $V$-$A$ theory implies
neither a T-conserving transverse polarization ({\it i.e.}\ $P_{T1} = 0$)
nor a T-violating transverse polarization ({\it i.e.}\ $P_{T2} = 0$)
for the emitted positrons.
After accounting for positron mass effects
the average T-conserving transverse
polarization in $V$-$A$ theory is about $\langle P_{T1} \rangle = -3 \times 10^{-3}$.

\begin{figure}
\begin{centering}
  \includegraphics[width=0.8\columnwidth]{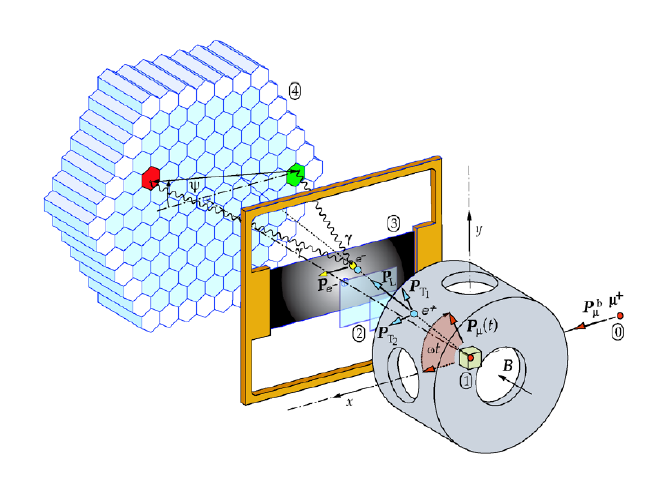}
  \caption{ \label{fg:DannebergSetup} Schematic diagram
of the Danneberg {\it et al.}\ experiment \cite{Danneberg:2005xv}
indicating the longitudinally polarized muon beam, beryllium stopping target,
transverse precession field, magnetized iron-cobalt foil and the bismuth germanate detector array. Figure courtesy Danneberg {\it et al.}}
\end{centering}
\end{figure}

The experiment
was conducted at PSI.
The experimental setup---shown in Fig.\ \ref{fg:DannebergSetup}---incorporated a polarized muon beam,
beryllium stopping target and
positron polarimeter.
To optimize the sensitivity
to $P_T$ the polarimeter was
orientated at a 90-degree angle
to the muon polarization axis.

%

The high-rate muon beam possessed
the cyclotron radio frequency time structure
that consisted of pulses
at 20~ns intervals
with 4~ns durations.
A 0.37~T transverse magnetic field
was used to maintain
the ensemble polarization
of multiple muon stops via the matching
of the Larmor precession period
to the $20$~ns pulse period.
The 4~ns pulse duration of beam particles limited
the ensemble polarization of stopped muons to roughly 80\%.

The polarimeter employed the detection of $\gamma$-ray pairs
from  the in-flight $e^+ e^-$ annihilation
of the decay positrons on the polarized electrons
in a magnetized iron-cobalt foil.
The resulting angular distribution of gamma-ray pairs
is correlated with the relative orientation of the $e^+$-$e^-$ polarization axes.
The time and position coordinates of the positron annihilation on the magnetized foil
were measured by combination of plastic scintillators and planar drift chambers.
The energy and position coordinates of the coincident $\gamma$-rays
were measured using a bismuth germanate (BGO) detector array.

The Larmor precession of stopped muons is
important in understanding the experimental
signature of transverse polarization.
This precession would cause
the corresponding rotation
of any non-zero components
of the positron transverse polarization.
Consequently, the angular distribution
of the annihilation $\gamma$-rays
would likewise rotate at the Larmor frequency,
with an amplitude and a phase
that is governed by the transverse
polarization ($P_{T1}$,$P_{T2}$).
By flipping the magnetization of the polarimeter foils
or reversing the direction of muon precession,
the associated signal of transverse polarization
could be distinguished from other systematic effects
that might induce periodic signals
in the $\gamma$-ray time
distributions.\footnote{Specifically, ``off-axis'' positron annihilation
can also induce sinusoidal variations
in the $\gamma$-ray time distribution.
This off-axis signal was used
for polarimeter calibration.}

The data analysis involved constructing time distributions $N_{ij} (t)$
of the $\gamma$-ray coincidences in the BGO detector elements $(i, j)$
as a function of the angle $\psi$
between the $\gamma$-ray pair plane and the
horizontal magnetization plane. A non-zero transverse
polarization would impart on $N_{ij} (t)$ a
$\psi$-dependent sinusoidal signal with the Larmor
frequency and an overall amplitude and phase determined
by ($P_{T1}$,$P_{T2}$).

The experiment recorded about  10$^6$ annihilation $\gamma$-ray pairs
from decay positrons and permitted measurement
of $P_{T1}$ and $P_{T2}$ for positron energies 10-50~MeV.
A Monte Carlo simulation was performed to determine
the dependence of the transverse polarizations on each decay parameter.
A fit was then performed of the Monte Carlo polarizations
to the measured polarizations to determine the ``best fit''
values for $\eta$, $\eta^{\prime \prime}$, $\alpha^{\prime} / A$ and $\beta^{\prime} / A$.
The procedure yielded values of
\begin{eqnarray*}
\eta &=& ( 71 \pm 37 \pm 5 ) \times 10^{-3} \\
\eta^{\prime \prime} &=& ( 100 \pm 52 \pm 6 ) \times 10^{-3}
\end{eqnarray*}
for the T-conserving polarization component $P_{T1}$
and
\begin{eqnarray*}
\alpha^{\prime}/A &=& ( -3.4 \pm 21.3 \pm 4.9 ) \times 10^{-3} \\
\beta^{\prime}/A &=& ( -0.5 \pm 7.8 \pm 1.8 ) \times 10^{-3}
\end{eqnarray*}
for the T-violating polarization component $P_{T2}$.
Using these best-fit values for the decay parameters
the authors then derived corresponding values for energy-averaged polarizations
of
\begin{eqnarray*}
\langle P_{T1} \rangle ~=~ ( 6.3 \pm 7.7 \pm 3.4 ) \times 10^{-3} \\
\langle P_{T2} \rangle ~=~ ( -3.7 \pm 7.7 \pm 3.4 ) \times 10^{-3}
\end{eqnarray*}

The determinations of the transverse polarizations
and their decay parameters represent a three-fold improvement
over previous experimental work. The results
are consistent with zero transverse polarization
and the SM values for the decay parameters,
$\eta = \eta^{\prime \prime} =  \alpha^{\prime}/A = \beta^{\prime}/A = 0$.

\subsection{Global analysis of decay parameters and theoretical implications}

A common representation for the matrix element
of the most general, Lorentz invariant, derivative-free,
current-current interaction
is based on definite electron and muon chiralities
\begin{equation}
M \propto\sum\limits _{\gamma = S, V, T}^{\epsilon , \mu = L, R} g^{\gamma}_{\epsilon \mu}
\langle \bar{e_{\epsilon}} | \Gamma_{\gamma} | \nu_e \rangle
\langle \bar{\nu_{\mu}} | \Gamma^{\gamma} | \mu_{\mu} \rangle
\end{equation}
where $\Gamma^S$, $\Gamma^V$, $\Gamma^T$ represent the possible
(scalar-pseudoscalar, vector-axial, tensor) interactions
between a left / right handed muon and a left / right handed electron.
The parameters $g^{\gamma}_{\epsilon \mu}$
are the ten associated complex coupling constants
of the current-current interaction terms
($g_{RR}^T$ and $g_{LL}^T$ are identically zero).
In the case of the $V$-$A$ weak interaction,
the  coupling $g^V_{LL}$ is unity and all the remaining
couplings are exactly zero.

The nine decay parameters from ordinary decay $\mu^+ \rightarrow e^+ \nu \bar{\nu}$,
and one additional parameter from radiative decay $\mu^+ \rightarrow e^+ \nu \bar{\nu} \gamma$,
appear insufficient at first glance to unambiguously determine the
ten independent {\em complex} couplings
$g^{\gamma}_{\epsilon \mu}$.\footnote{One coupling is contrained
by normalization thus leaving the Fermi constant
$G_F$ and eighteen additional independent parameters to be experimentally
determined.}
Therefore the global analyses
of measured decay parameters
involve determining values for smaller sets
of intermediate bilinear combinations
of complex couplings.
The possible limits on non-$V$-$A$ interactions---{\it i.e.}\ all
couplings $g^{\gamma}_{\epsilon \mu}$ excepting $g^V_{LL}$---are
then derived from the values determined
for the bilinear combinations.

Sets of intermediate combinations of complex couplings
for parameterizing the constraints obtained from decay parameters
have been introduced by Kinoshita and Sirlin \cite{PhysRev.107.593}
and Fetscher, Gerber and Johnson \cite{Fetscher:1986uj}.
For example,
the scheme of Fetscher {\it et al.}\
incorporates the maximum number of
positive semidefinite bilinear combinations
of complex couplings $g^{\gamma}_{\epsilon \mu}$.
Of ten parameters, the four bilinears
\begin{equation*}
Q_{LL} = {1 \over 4} |g^S_{LL}|^2 + |g^V_{LL}|^2 ,
\end{equation*}
\begin{equation*}
Q_{LR} = {1 \over 4} |g^S_{LR}|^2 + |g^V_{LR}|^2 + 3 |g^T_{LR}|^2 ,
\end{equation*}
\begin{equation*}
Q_{RL} = {1 \over 4} |g^S_{RL}|^2 + |g^V_{RL}|^2 + 3 |g^T_{RL}|^2 ,
\end{equation*}
\begin{equation*}
Q_{RR} = {1 \over 4} |g^S_{RR}|^2 + |g^V_{RR}|^2
\end{equation*}
have special significance
from their correspondance
to the transition probabilities $Q_{\epsilon\mu}$
from a muon with handedness $\mu = L, R$
to an electron with handedness $\epsilon = L, R$.
In the standard model the quantity $Q_{LL} = 1$
and the other bilinears are exactly zero.

In Table \ref{tab:decayparameters} we compile the
current determinations of the decay parameters
including the recent results from TWIST,
Danneberg {\it et al.}\
and Prieels {\it et al.}
Also listed are the SM values for the decay parameters,
{\it i.e.}\  for $g^V_{LL} = 1$ and all other couplings zero.
The world data on decay parameters are
consistent with the $V$-$A$ charged current
weak interaction to levels of
$10^{-3}$-$10^{-4}$ in many cases.

\begin{table}[htb]
\centering \caption {Compilation of the most recent determinations
of the nine decay parameters in ordinary muon decay
and the additional $\bar{\eta}$ decay parameter in radiative muon decay.
With the exception of the ordinary decay parameter $\xi^{\prime}$
and the radiative decay parameter $\bar{\eta}$, we give the
results of the most recent precision measurements. For $\xi^{\prime}$
and $\bar{\eta}$ we give the world averages compiled in Ref.\
\cite{Agashe:2014kda} (the average for $\xi^{\prime}$ is dominated by
the measurement of Burkard {\it et al.}\ \cite{Burkard:1985kf} and the average
for $\bar{\eta}$ is dominated by the measurement of Eichenberger
{\it et al.}\ \cite{Eichenberger:1984gi}).}
\vspace{0.3cm}
\begin{tabular}{cccc}
\hline
  Parameter   & Measured value & SM value & Ref.\ \\
\hline
$\rho$ & $0.749~77 \pm 0.000~12 \pm 0.000~23$ & 0.75 & \cite{TWIST:2011aa} \\
$\delta$ & $0.750~49 \pm 0.000~21 \pm 0.000~27$ & 0.75 &  \cite{TWIST:2011aa} \\
$P^{\pi}_{\mu} \xi$  & $1.00084 \pm 0.00029 \pm ^{0.00165}_{0.00063}$ & 1.0 &  \cite{Bueno:2011fq} \\
$\xi^{\prime}$ & $1.00 \pm 0.04$ & 1.0 &  \cite{Agashe:2014kda} \\
$\xi^{\prime \prime}$ & $0.981 \pm 0.045 \pm 0.003$ & 1.0 &  \cite{Prieels:2014paa} \\
$\eta$ & $( 71 \pm 37 \pm 5 ) \times 10^{-3}$ & 0.0 &  \cite{Danneberg:2005xv} \\
$\eta^{\prime \prime}$ & $( 100 \pm 52 \pm 6 ) \times 10^{-3}$ & 0.0 &  \cite{Danneberg:2005xv} \\
$\alpha^{\prime}/A$ & $( -3.4 \pm 21.3 \pm 4.9 ) \times 10^{-3}$ & 0.0 & \cite{Danneberg:2005xv} \\
$\beta^{\prime}/A$ & $( -0.5 \pm 7.8 \pm 1.8 ) \times 10^{-3}$ & 0.0 & \cite{Danneberg:2005xv} \\
$\bar{\eta}$ & $0.02 \pm 0.08$ & 0.0 &  \cite{Agashe:2014kda} \\
\hline
\end{tabular}
\label{tab:decayparameters}
\end{table}

A global analysis of decay parameters has
recently been performed by the TWIST collaboration \cite{TWIST:2011aa}.
It followed the procedure of
Gagliardi, Tribble and Williams \cite{Gagliardi:2005fg}
but updated their input values
for decay parameters with the
final results of the TWIST experiment.\footnote{This global analysis
precedes the recent measurement of $\xi^{\prime \prime}$
by Prieels {\it et al}.}
The procedure involved determining the
joint probability distributions
for a hybrid set
of nine bilinears
of coupling $g^{\gamma}_{\epsilon \mu}$
from the muon decay parameters
via a Monte Carlo procedure (see Ref.\ \cite{Burkard:1985wn} for details).
From these results,
they then derived limits
on the various couplings $g^{\gamma}_{\epsilon \mu}$.
Their results for the bilinears $Q_{e\mu}$
and the couplings $g^{\gamma}_{\epsilon \mu}$
are reproduced in Table \ref{tab:couplings}


\begin{table}[htb]
\centering \caption {Results for the experimental limits
on the bilinear quantities $Q_{\epsilon\mu}$ and the
coupling constants $g^{\gamma}_{\epsilon \mu}$
of the general current-current interaction
derived from the global analysis of the
muon decay parameters \cite{TWIST:2011aa}.
In $V$-$A$ theory the bilinear $Q_{LL}$ and
coupling $g^V_{LL}$ are unity and
all other bilinears and coupling constants are exactly zero.
Note the limits on $|g^S_{LL}|$ and $|g^V_{LL}|$
are derived from inverse muon decay,
$\nu_{\mu} e^- \rightarrow  \nu_e \mu^-$.}
\vspace{0.3cm}
\begin{tabular}{cccc}
\hline
\\
$ Q_{RR}  < 3.0\times10^{-4}$ & $ Q_{LR}  < 6.3\times10^{-4}$ & $ Q_{RL}  < 0.044$ & $ Q_{LL}  > 0.955$ \\
\\
$| g^S_{RR} | < 0.035$ & $| g^S_{LR} | < 0.050$ &
$| g^S_{Rl} | < 0.420$ & $| g^S_{LL} | < 0.550$ \\
$| g^V_{RR} | < 0.017$ & $| g^V_{LR} | < 0.023$ &
$| g^V_{RL} | < 0.105$ & $| g^V_{LL} | > 0.960$ \\
$| g^T_{RR} | \equiv 0$ & $| g^T_{LR} | < 0.015$ &
$| g^T_{RL} | < 0.105 0$ & $| g^T_{LL} | \equiv 0$ \\
\\
\hline
\end{tabular}
\label{tab:couplings}
\end{table}

Of special interest are
the upper limits derived on the
bilinear sums $Q^{\mu}_R \equiv Q_{RR} + Q_{LR}$
and $Q^{e}_R \equiv Q_{RR} + Q_{RL}$
A non-zero value of $Q^{\mu}_R$ would
signal a weak interaction contribution
from a right-handed muon current and
a non-zero value of $Q^{e}_R$ would
signal a weak interaction contribution
from a right-handed electron current.
The decay parameter $\xi$ of the
positron angular distribution is
particularily sensitive to $Q^{\mu}_R$
and the decay parameter $\xi^{\prime}$
of the positron longitudinal polarization
is particularily sensitive to $Q^{e}_R$.
The global analysis yielded limits
on  contributions from right-handed muon interactions
of $Q^{\mu}_R < 8.2 \times 10^{-4}$
and right-handed electron interactions
of $Q^{e}_R < 0.044$.

Danneberg {\it et al.}\
and Prieels {\it et al.}\
also discuss limits
on right-handed currents involving specific exotic scalar, vector
and tensor interactions.
For example, the
decay parameters $\alpha^{\prime} / A$
and $\beta^{\prime} / A$ derived
from the transverse polarization $P_{T2}$
are uniquely sensitive  to T-violating contributions
in purely leptonic interactions.
Danneberg {\it et al.}\ obtained a limit on
possible T-violating scalar interactions
between right-handed charged leptons currents of
Im$(g^s_{RR}) = (5.2 \pm 14.0 \pm 2.4) \times 10^{-3}$.

Muon decay is particularly valuable
in imposing constraints on various left-right symmetric (LRS) extensions
of the standard model electroweak interaction \cite{Herczeg:1985cx}.
Such models introduce a new $V$+$A$ interaction coupling to
right-handed currents that partners the known $V$-$A$ interaction
coupling to left-handed currents in order to restore parity
conservation at high energies.
In LRS models the $V$+$A$ and  $V$-$A$ interactions are mediated by
$W_R$ and $W_L$ gauge bosons with couplings constants
$g_R$ and $g_L$. A mass mixing angle $\zeta$ and CP-violating phase $\omega$
together determine the relation between the weak eigen-states $W_{L/R}$
and the mass eigen-states $W_{1/2}$
\begin{equation*}
W_L = W_1 \cos{\zeta} + W_2 \sin{\zeta}
\end{equation*}
\begin{equation*}
W_R = e^{i \omega} ( - W_1 \sin{\zeta} + W_2 \cos{\zeta} )
\end{equation*}
with  low-energy parity violation emerging when the $W_R$-boson mass
exceeds the $W_L$-boson mass. In {\it manifest} LRS models
the two couplings are equal and in {\it generalized}
LRS models the two coupling are distinct.

\begin{figure}
\begin{centering}
  \includegraphics[width=0.8\columnwidth]{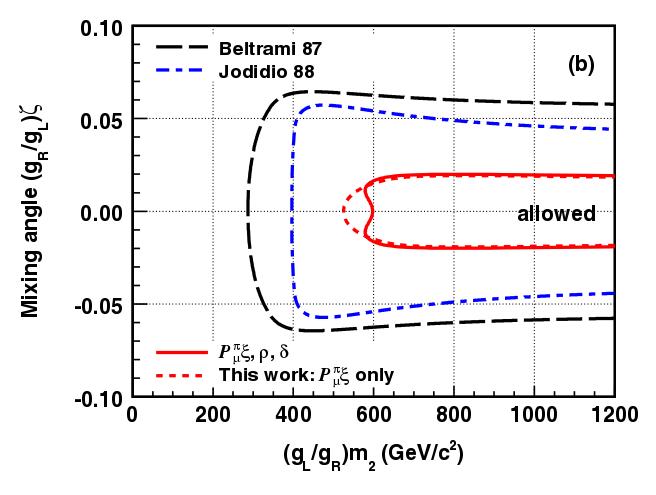}
  \caption{ \label{fg:LRSmodel} Allowed region of $W_2$ boson mass
versus mass mixing angle $\zeta$ in the generalized left-right symmetric
model. Shown are the limits derived from the recent TWIST measurement
of $P_{\mu} \xi$ and the earlier results from
Beltrami {\it et al.}\ \cite{Beltrami:1987ne}
and Jodidio {\it et al.}\ \cite{PhysRevD.34.1967}.}
\end{centering}
\end{figure}

A variety of approaches---from direct searches for $W_R$-boson production
at pp colliders to setting limits on $W_R$-boson virtual contributions
in $K^0$-$\bar{K^0}$ mixing, $\beta$-decay and muon decay---have been used
in the investigation of the various LRS extensions of the standard model
(for further details see the reviews
\cite{Agashe:2014kda,Severijns:2006dr}).
The results from pp colliders and
$K^0$-$\bar{K^0}$ mixing set impressive
lower bounds on $W_R$-boson masses
while results from unitarity tests
that utilize $0^+$$\rightarrow$$0^+$ nuclear
$\beta$-decay set impressive bounds
on $W_L$-$W_R$ mixing.
However, unlike these processes
the decay of muons
is purely leptonic and
essentially free
of any assumptions concerning
the CKM matrix elements
of the hypothetical $W_R$ boson
(many analyses assume
a $W_R$ boson with
standard model-like
couplings and CKM matrix elements).
Consequently, the TWIST result
for $P_{\mu} \xi$ \cite{Bueno:2011fq}
yields complementary and restrictive limits
on the parameter space of the generalized LRS model.
Their limits on $ g_L / g_R ~m_2 $ versus
$ g_L / g_R ~ \zeta $ are reproduced in Fig.\ \ref{fg:LRSmodel}.




%% file: MuonCLFV_final.tex
\section{Charged Lepton Flavor Violating decays}
\label{sc:cLFV}

\subsection{Lepton flavor and physics beyond the standard model}
Once a muon, always a muon---or at least the {\em flavor} of a muon seems to be preserved. The conservation of a ``muon'' flavor was not expected when this particle was discovered, nor does it associate a conserved quantity with a fundamental symmetry, as required by Noether's theorem.   But, following non-observations of its violation over decades of trials in reactions such as the simple decay mode $\mu \rightarrow e \gamma$, it emerged as a standard model foundational statement; a muon cannot become an electron.

In a two-neutrino (later, three) world of massless neutrinos, the transformation of muon flavor to electron flavor (or tau flavor) is strictly forbidden. The conservation of separate $e-, \mu-,$ and $\tau$-type lepton numbers holds; that is  $\Delta L_e = \Delta L_\mu = \Delta L_\tau = 0$, where the $L_i$ are assigned lepton flavor numbers.
The additive sums are unchanged in any given reaction or decay.  The rules worked perfectly prior to the period from 1998 - 2001 when the atmospheric, solar, and reactor-based neutrino experiments demonstrated that neutrino flavors mix~\cite{Fukuda:1998mi,Ahmad:2002jz,Eguchi:2002dm}. These discoveries changed the rules.  Neutrinos must have finite mass, and they can transform within the three generations. In neutral processes at least, lepton flavor is not a good quantum number\footnote{A current hot topic in neutrino physics is the search for neutrinoless double beta decay, $0\nu\beta\beta$.  If observed, this  process violates lepton {\em number} conservation and serves to prove that neutrinos are their own antiparticles---Majorana fermions.  It further supports the leptogenesis explanation of why we have a baryon asymmetry in the universe.}.

\begin{figure}
\begin{centering}
  \includegraphics[width=.4\columnwidth]{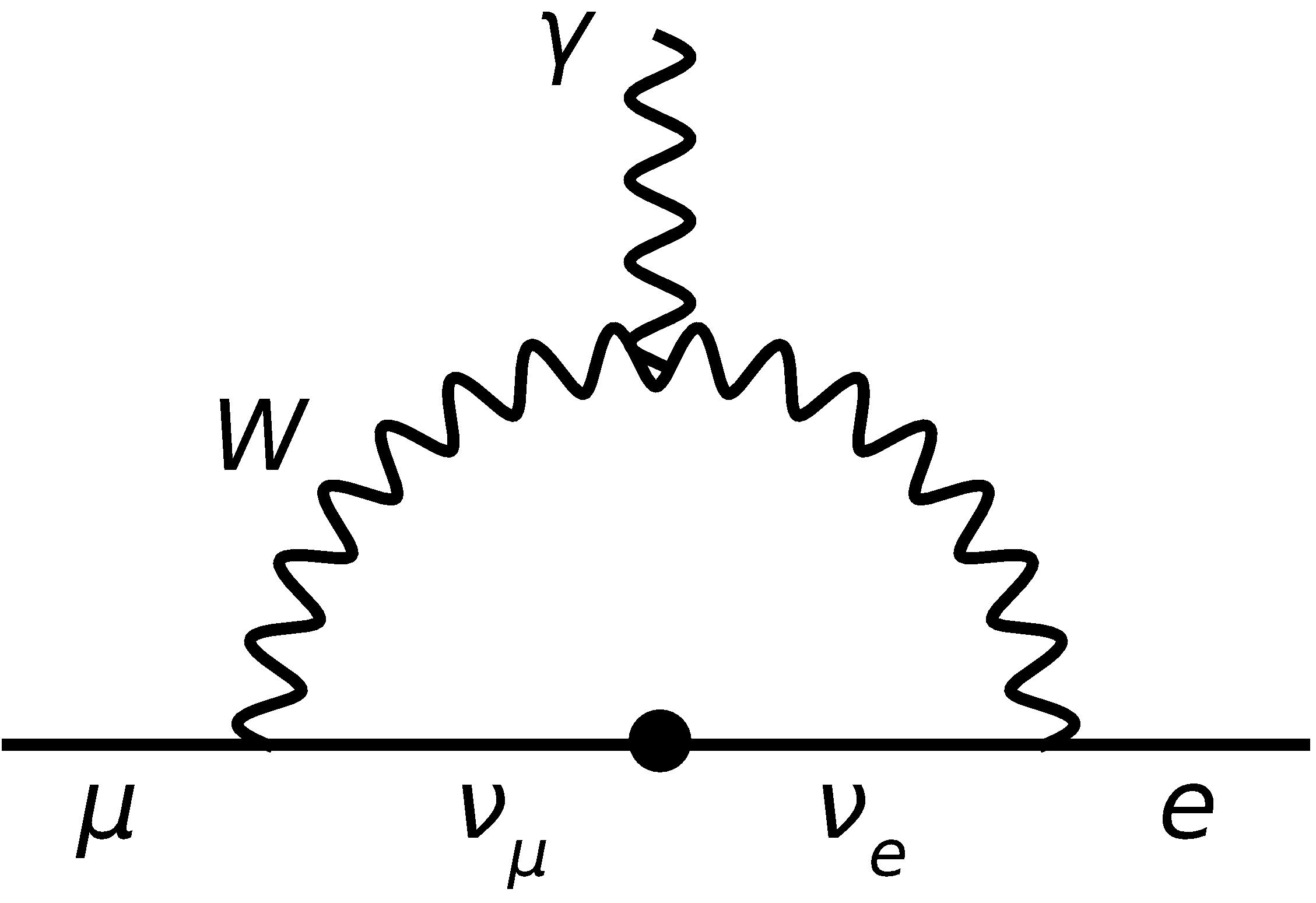}
  \includegraphics[width=.4\columnwidth]{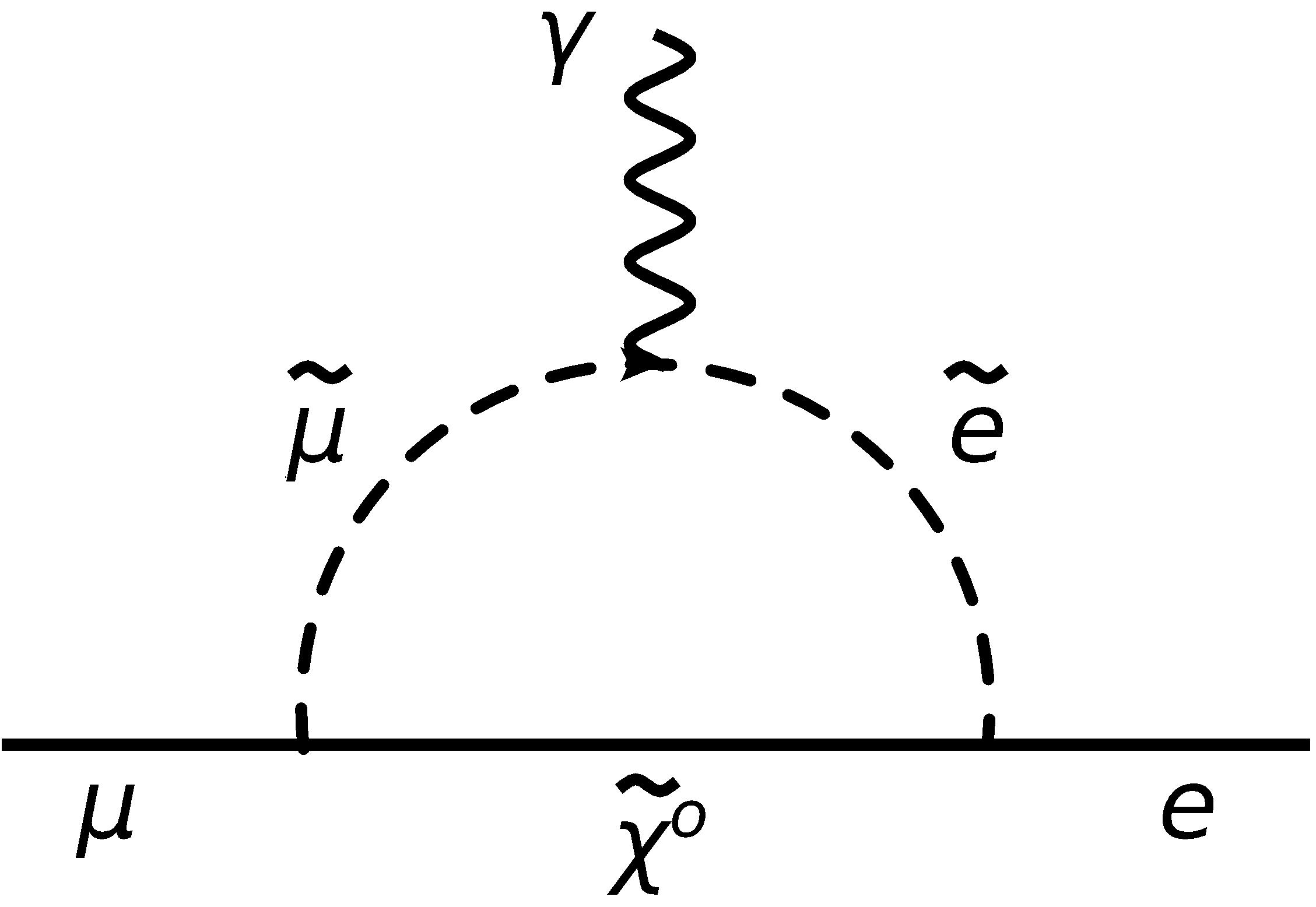}
  \caption{\label{fg:megfeynman}
    a) The standard model allowed decay $\mu \rightarrow e \gamma$, which proceeds through a loop process involving neutrino mixing at the unmeasureably low branching ratio below $10^{-54}$.  b) A SUSY based diagram depicting smuon-selectron mixing inducing the same process.}
\end{centering}
\end{figure}

Violation of lepton flavor in the neutrino sector, in turn, implies that the decay $\mu \rightarrow e \gamma$ must occur through loop processes.  However, using upper limits on the neutrino masses and the measured mixing angles, a branching ratio below $10^{-54}$ is deduced for the diagram depicted in Fig.~\ref{fg:megfeynman}a.  Needless to say, this is unmeasurably small.  The bright side, is that this standard model allowed rate---which is representative of the other {\em charged} lepton flavor violation (cLFV) processes---is so tiny that any non-null measurement must indicate new physics.  Indeed, calculations based on many popular SM extensions suggest large cLFV effects should appear near to current experimental limits.  That is the subject of the current chapter. The richness and importance of this topic is documented in numerous scholarly reviews~\cite{Kuno:1999jp,Marciano:2008zz,deGouvea:2013zba,Bernstein:2013hba,Mori:2014aqa}. Each generally takes on a different emphasis between theoretical expectations, experimental techniques and history, or an overall picture.  Here, we highlight our view of the key topics being discussed and the three most sensitive and promising experimental programs going forward.

Comprehensive and high-sensitivity studies of tau lepton cLFV decay modes are being carried out now and further plans exist at the upgraded $B$ factory in Japan.~\cite{Abe:2010gxa}. Tagged $\tau$ decays to muons and electrons, such as $\tau \rightarrow eee$, $\tau \rightarrow \mu\mu\mu$, $\tau \rightarrow e\gamma$, and $\tau \rightarrow \mu\gamma$ as well as many other combinations are being studied.  Impressive sensitivity has been achieved with branching ratios typically close to the $10^{-8}$ level and the future promises order-of-magnitude improvements with Belle-II.  Additionally, the LHCb experiment has already set a competitive limit on the  $BR(\tau \rightarrow \mu\mu\mu)$ of  $< 4.6 \times 10^{-8}$, which was based on the 3\,fb$^{-1}$ of data acquired in the first LHC run~\cite{Aaij:2014azz}.  Presumably this limit will be strengthened in the future as the systematics do not dominant the uncertainty.

\begin{figure}
  \includegraphics[width=\columnwidth]{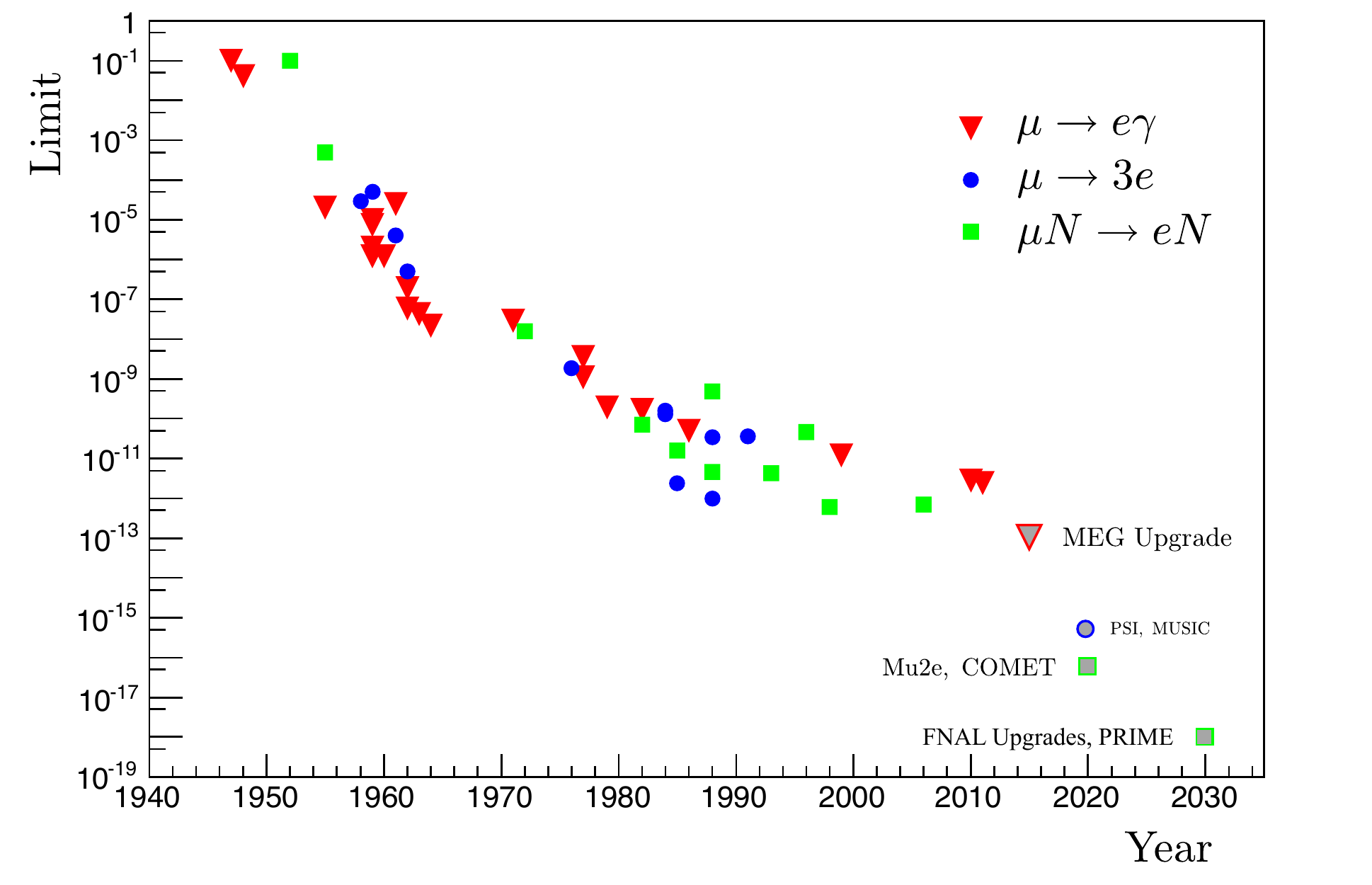}
  \caption{\label{fg:cLFV-history}
    The solid shapes indicate the achieved 90\% sensitivity limits for $\mu^+ \rightarrow e^+\gamma$ (triangle), $\mu^+ \rightarrow e^+ e^- e^+$ (circle), and the coherent conversion process $\mu^-N \rightarrow e^-N$ (square). Expected improvements from approved experimental programs have light grey interiors and the same shapes.  Figure updated from~\cite{Bernstein:2013hba}, courtesy R.\,Bernstein.}
\end{figure}

However, it is the three low-energy muon reactions
\begin{enumerate}
  \item \tab{$\mu^+ \rightarrow e^+ \gamma$}
        \tab{($BR: < 5.7 \times 10^{-13}$)}
        \tab{Ref.~\cite{Adam:2013mnn}}
  \item \tab{$\mu^+ \rightarrow e^+ e^- e^+$}
        \tab{($BR: < 1.0 \times 10^{-12}$)}
        \tab{Ref.~\cite{Bellgardt:1987du}}
  \item \tab{$\mu^- + N \rightarrow e^- N$}
        \tab{($BR: < 7   \times 10^{-13}$)}
        \tab{Ref.~\cite{Bertl:2006up}}
\end{enumerate}
that have the more impressive limits and greater sensitivity to new physics. Processes 1 and 2 represent ultra-rare decay modes.  Process 3 involves the formation of a muonic atom with a nucleus---the limit quoted here is from the $\mu^-$Au atom---followed by coherent conversion of the muon to an electron, which is ejected with an energy close to the muon rest mass.  The history of these measurement sensitivities is compiled in Fig.~\ref{fg:cLFV-history}, which includes projections for experiments now being upgraded or constructed.

Note that other processes also exist.  For example, the spontaneous conversion of the $M \equiv \mu^{+}e^-$ muonium atom to anti-muonium, $\mu^{-}e^+$, has been searched for and a limit of $P_{M\bar{M}} \leq 8.2\times 10^{-11}$ (90\% C.L.) in a 0.1\,T magnetic field has been set~\cite{Willmann:1998gd}.  Here, electron and muon number are violated in the same exotic process.  The physics reach is not as competitive if directly compared to current cLFV efforts and there are no current plans to improve it on the horizon~\cite{Mori:2014aqa}.  But, it should be noted that this is a rather complementary process that might be induced by different physics; for example, see~\cite{PhysRevD.79.015001}.

A relative comparison remark is in order here.  Generically, for a loop-induced process, similar to those that might cause the deviation from the SM for the muon anomaly, the expected rates for processes $1\,:\,2\,:\,3$ scale as $389\,:\,2.3\,:\,1$; (here, $\mu$\,Al is assumed for  3)~\cite{Marciano:2008zz}. The ratio implies that the present $\mu^+ \rightarrow e^+\gamma$ limits greatly exceed current limits from the $3e$ decay or the $\mu \rightarrow e$ conversion.  However, the ambitious goals of modern $\mu \rightarrow e$ conversion experiments aim for 4 orders of magnitude improvements, which will close the gap.

\begin{figure}
  \includegraphics[width=.5\columnwidth]{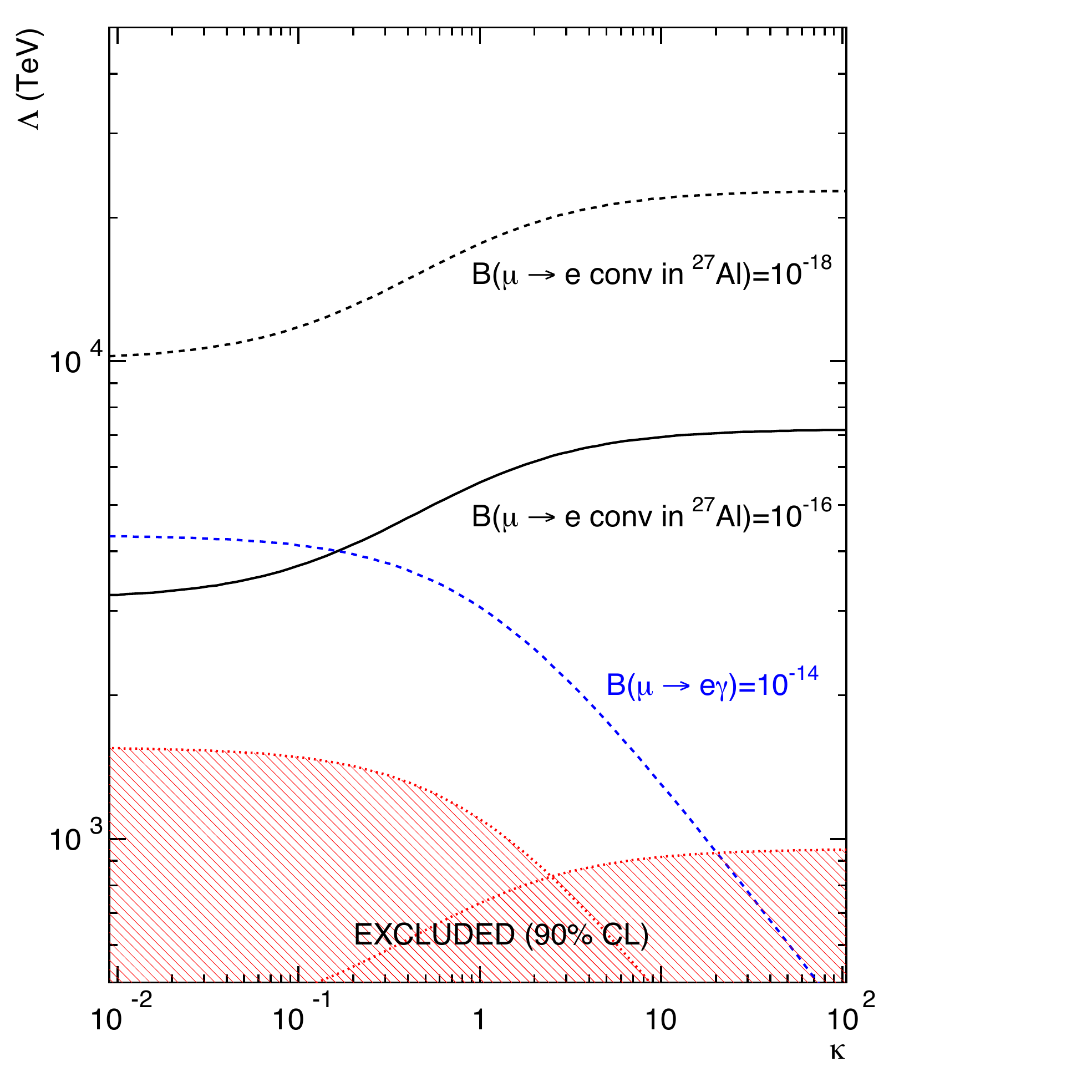}
  \includegraphics[width=.5\columnwidth]{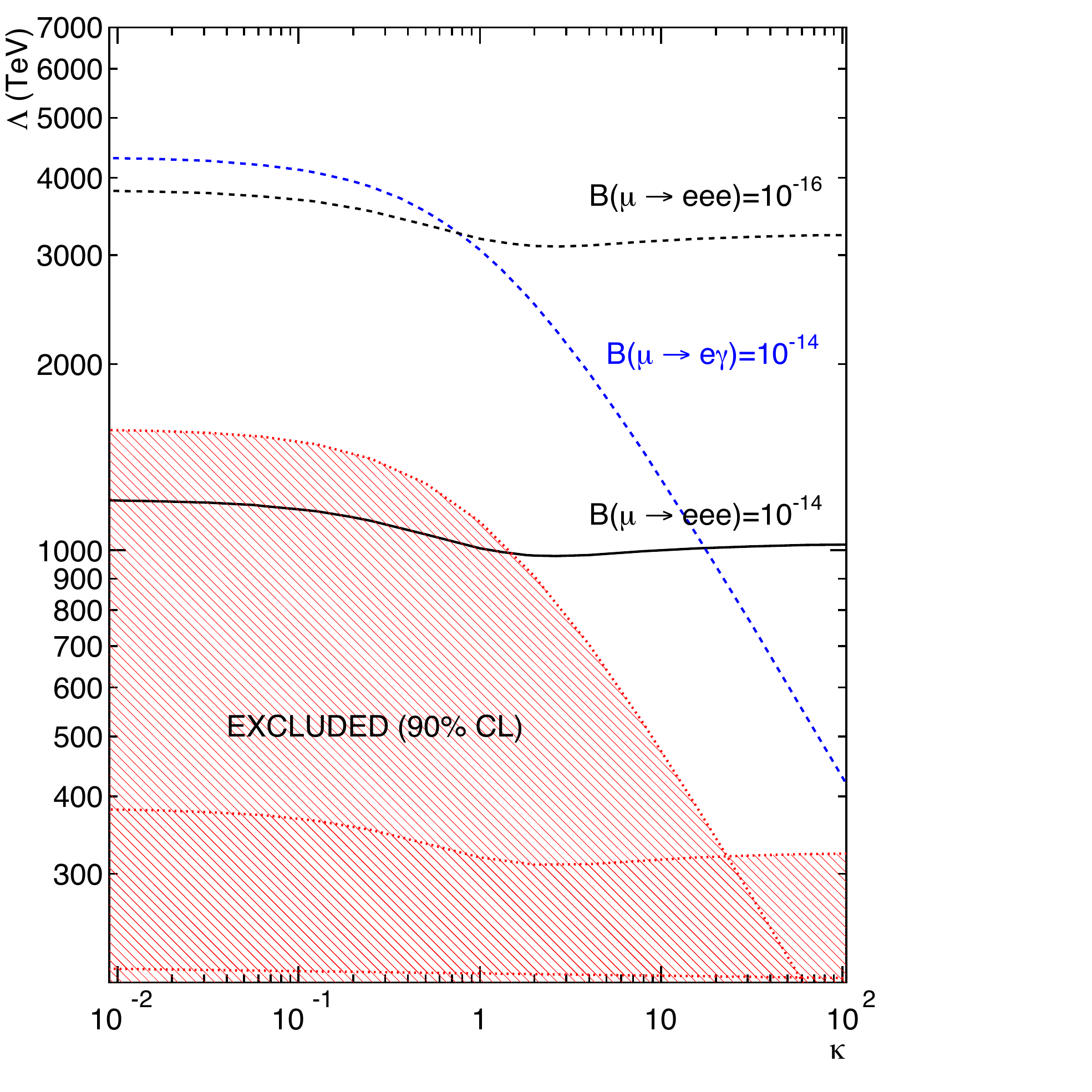}
  \caption{\label{fg:cLFV}
    Comparison of the new physics mass scale reach for the cLFV experiments using current achieved sensitivities and goals of planned programs.  In both panels, $\kappa \ll 1$ corresponds to loop-like processes, transitioning to 4-fermion contact interactions for large values of $\kappa$.  Left panel: $\mu^+ \rightarrow e^+\gamma$ versus $\mu^-N \rightarrow e^-N$ coherent conversion experiments.  Right panel: $\mu^+ \rightarrow e^+\gamma$ versus $\mu^+ \rightarrow e^+ e^- e^+$.  Figure updated from~\cite{deGouvea:2013zba}, courtesy A.\,de Gouv\^ea.}
\end{figure}

The comparison using loop-like scaling is not the whole story, since the sensitivities vary depending on what kind of new physics leads to cLFV.  Ultimately, one would like to have measured deviations in all three channels to sort out the underlying mechanism.  Although there are many compelling standard model extensions, such as the SUSY process involving smuon-selectron mixing as depicted in Fig.~\ref{fg:megfeynman}b, we will follow the more generic approach developed by de\,Gouv\^ea~\cite{deGouvea:2013zba}. In this model-independent analysis, the cLFV processes can be compared for their respective new-physics sensitivities in terms of an effective energy scale versus a parameter that slides from loop-like exchanges to contact interactions.  de\,Gouv\^ea considers an effective Lagrangian to describe processes (1) and (3) of the form
\begin{eqnarray*}
\nonumber
  L_{\rm{cLFV}} &=& \frac{M_\mu}{(\kappa + 1)\Lambda^2}\bar{\mu}_R \sigma_{\mu\nu}e_{L}F^{\mu\nu} \\
    &+& \frac{\kappa}{(\kappa + 1)\Lambda^2}\bar{\mu}_L \gamma_\mu e_L (\bar{u}_L \gamma^\mu u_L  + \bar{d}_L \gamma^\mu d_L ).
\end{eqnarray*}
Here $L$ and $R$ are fermion field chiralities and $F^{\mu\nu}$ is the photon field. If the dimensionless parameter $\kappa \ll 1$ the process is dominated by a magnetic moment type operator (as above). In contrast, if  $\kappa \gg 1$, the four-fermion operators (2nd line) dominate, representing a point-like contact interaction.  The parameter $\Lambda$ is an effective energy scale that extends for current experimental sensitivity goals to 1000's of TeV---for {\em optimized} coupling---well in excess of any collider reach.  This comparison is plotted in the left panel of Fig.~\ref{fg:cLFV} for both current experimental limits and the future projects described below.   The  sensitivity from $\mu^+ \rightarrow e^+\gamma$ searches falls off quickly for contact-like interactions, unlike $\mu - e$ conversion, which rises.

A similar Lagrangian (the second term is different) can be used to describe reactions (1) and (2):
\begin{eqnarray*}
\nonumber
  L_{\rm{cLFV}} &=& \frac{M_\mu}{(\kappa + 1)\Lambda^2}\bar{\mu}_R \sigma_{\mu\nu}e_{L}F^{\mu\nu} \\
    &+& \frac{\kappa}{(\kappa + 1)\Lambda^2}\bar{\mu}_L \gamma_\mu e_L (\bar{e}_L \gamma^\mu e),
\end{eqnarray*}
where $\kappa$ and $\Lambda$ correspond to similar definitions.  The right panel of Fig.~\ref{fg:cLFV} plots these sensitivities.  The $\mu \rightarrow e e e$ process will be several orders of magnitude less sensitive  compared to $\mu^+ \rightarrow e^+\gamma$ for loop-like processes; however, for contact-type interactions there is a part of the new-physics search space that favors the $3e$ decay mode.

An important word of caution is in order here so as not to over-interpret the vertical axis in Fig.~\ref{fg:cLFV}.  The new physics ``mass scale'' $\Lambda_{{\rm cLFV}}$ is not to be interpreted strictly as mass. It is  product of mass and flavor-violation mixing, where the latter is similar to the mixing angles in a CKM or PMNS matrix for quarks and neutrinos.  The mixing can be large---neutrino like---or very small---quark like, or it can be anything at all.  With a cLFV experiment there is but one-measurement and two parameters, which emphasizes the need for multiple measurements (channels) and comparisons to other precision measurements.  For example, consider the situation where the new physics impacts both $g-2$ (flavor conserving) and cLFV.  Then the new physics scale can be related by  $\Theta_{e\mu}\Lambda_{{\rm cLFV}}^2 = \Lambda_{g-2}^2$, where $\Theta_{e\mu}$ represents the level of flavor-mixing in the new-physics process; $\Theta_{e\mu} \rightarrow 1$ is a flavor indifferent, maximally mixed scenario, and $\Theta_{e\mu} \ll 1$ implies small mixing~\cite{deGouvea:2013zba}.

\subsection{Experimental challenge of muon cLFV}
The goals for modern charged lepton flavor violation experiments all aim at single event branching ratios of a few times $10^{-14}$ to $10^{-17}$ range. To be sensitive to such stunningly rare processes requires an event signature that is unique with respect to backgrounds, and exceptional design and execution by the experimental team.  Because all processes involve muons at rest, the energy scale of the emitted particles is low---below $m_\mu$ for the decay studies and equal to $\sim m_\mu$ for the conversion process. From a detector design perspective, this is a challenge, requiring ultra-thin tracking detectors and especially high-resolution calorimetry.  The sub-dominant, but ordinary, muon decay modes are problematic. Radiative muon decay, $\mu^+ \rightarrow e^+ \nu_e \bar{\nu}_\mu  \gamma$, occurs with a branching ratio of $1.4(4)\%$ and radiative decay with internal conversion,
$\mu^+ \rightarrow e^+ e^- e^+ \nu_e \bar{\nu}_\mu$, occurs at the $(3.4 \pm 0.4) \times 10^{-5}$ level. For the corners of phase space where the neutrinos escape with very little energy, these decay modes can mimic the cLFV processes (1) and (2), at a many orders of magnitude greater rate.

To achieve sensitivities below $10^{-14}$ a considerable number of stopped muons is required (well above $10^{14}$ after factoring in efficiencies).
In context, the MuLan lifetime experiment described in Sec.~\ref{ssc:mulan} measured $\tau_\mu$ to 1\,ppm using $2 \times 10^{12}$ decays, and utilizing a high-intensity beamline at PSI.  The cLFV program requires, in contrast, muon samples larger by factors of $10^3 - 10^5$.  The highest-intensity $\pi E5$ beamline at PSI can deliver a dc rate of $10^8$\ muons/sec, which is 10 times higher than MuLan used.  The dc nature of the beam structure is ideal for the decay programs that observe one event at a time.  However, at high rates, multiple events are in the detector at once, requiring excellent coincident timing and position resolution to distinguish individual decays.  To reach the ultimate goal of the $3e$ experiment of $10^{-16}$ will require a new ultra-high intensity beamline having $10\times$ greater flux than presently exists.  For the next-generation $\mu \rightarrow e$ conversion experiments---the goals there approach the $10^{-17}$ level---an entirely different concept is required, which we describe below.

\subsubsection{Measuring  $\mu^+ \rightarrow e^+ \gamma$}
\begin{figure}
  \includegraphics[width=\columnwidth]{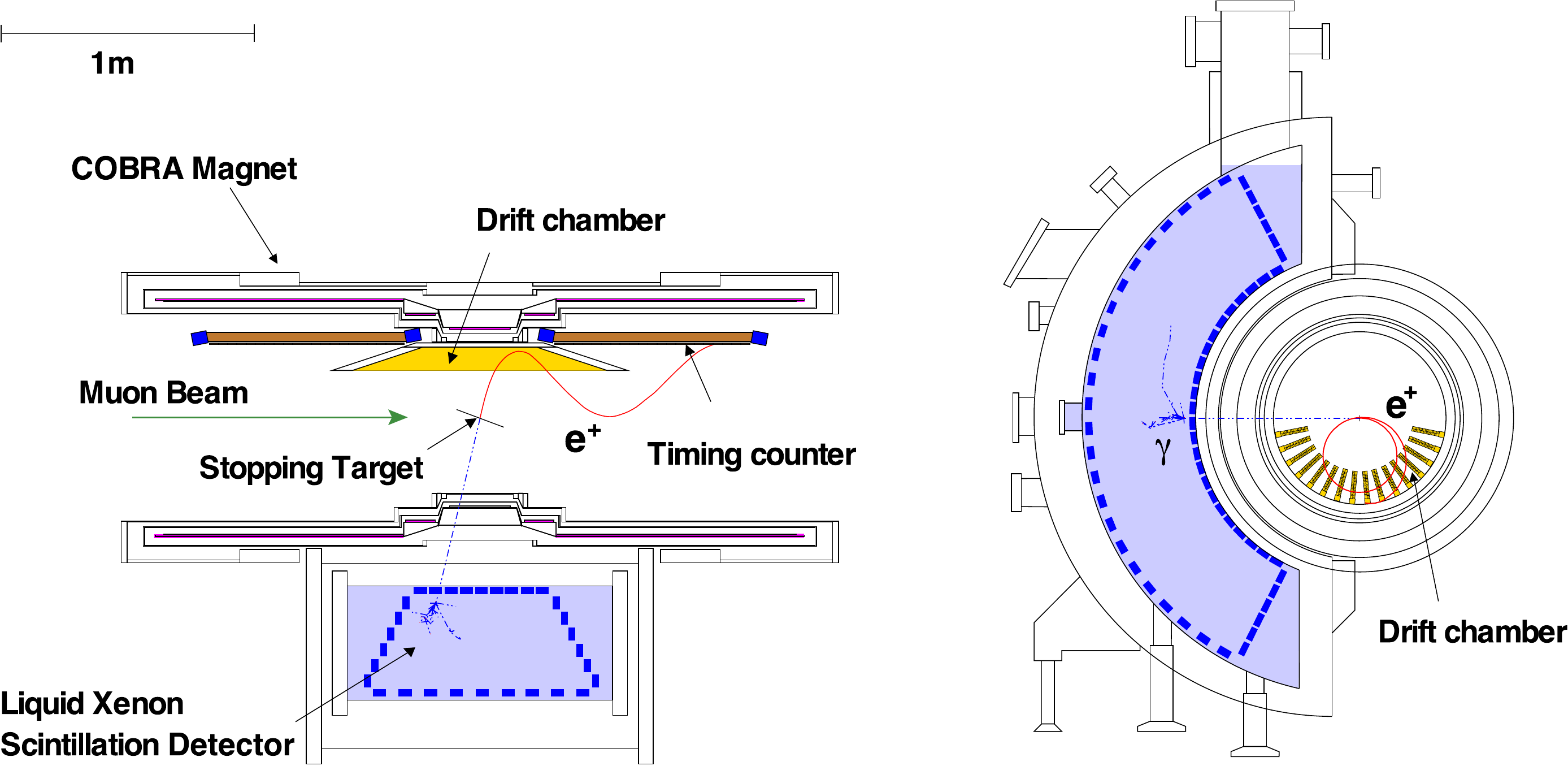}
  \caption{\label{fg:MEG-detector}
    The side and front schematic of the MEG detector with a hypothetical $\mu \rightarrow e \gamma$ event superimposed.  Figure courtesy MEG collaboration.}
\end{figure}

The decay mode $\mu^+ \rightarrow e^+ \gamma$ is being pursued by the MEG collaboration at PSI. Their results to date set the limit $BR_{e\gamma} < 5.7 \times 10^{-13}$; 90\% C.L.~\cite{Adam:2013mnn}. The decay mode features the back-to-back emission of a positron and a gamma ray, each having energies equal to $m_\mu / 2 \approx 53$\,MeV.  Positive muons at a dc rate of up to 30\,MHz are stopped in a thin polyethylene disk centered in a cylindrical geometry.  A tracking drift chamber system and a high-resolution liquid xenon (LXe) calorimeter are key components, see Fig.~\ref{fg:MEG-detector}.

A unique feature of the experiment is the COnstant Bending RAdius (COBRA) magnet. Its gradient field shape is designed to maintain a near-constant bending radius of decay positrons, independent of their initial pitch angle, and further, to sweep away the decay trajectories to the upstream or downstream side of the stopping target.  The unique calorimeter uses 862 PMTs, submerged in LXe and mounted on all surfaces to sum the light and to therefore provide energy, time-of-arrival, and location of the interaction of the gamma in the xenon bath. The ultra-low-mass drift chamber sits in a half circle below the target. A scintillator array is used to precisely establish the timing needed.  At signal energies $E_\gamma = E_{e^+} = m_\mu/2$, the tracker resolution is 1.5\% and the calorimeter resolution is 4.5\%.  The overall acceptance for decays of interest is $\sim18\%$.~\cite{Adam:2013vqa}

\begin{figure}
  \includegraphics[width=0.49\columnwidth]{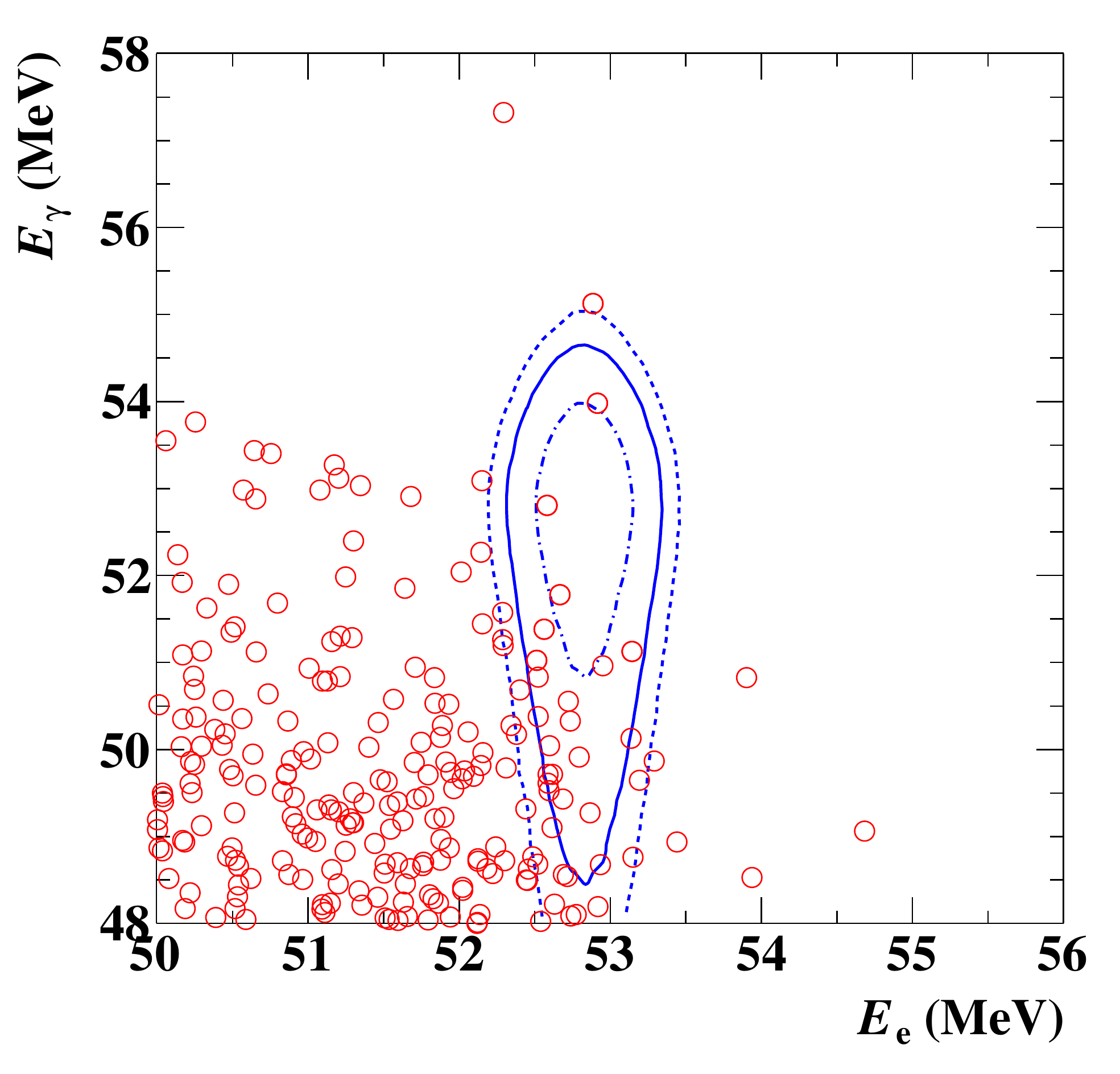}
  \includegraphics[width=0.49\columnwidth]{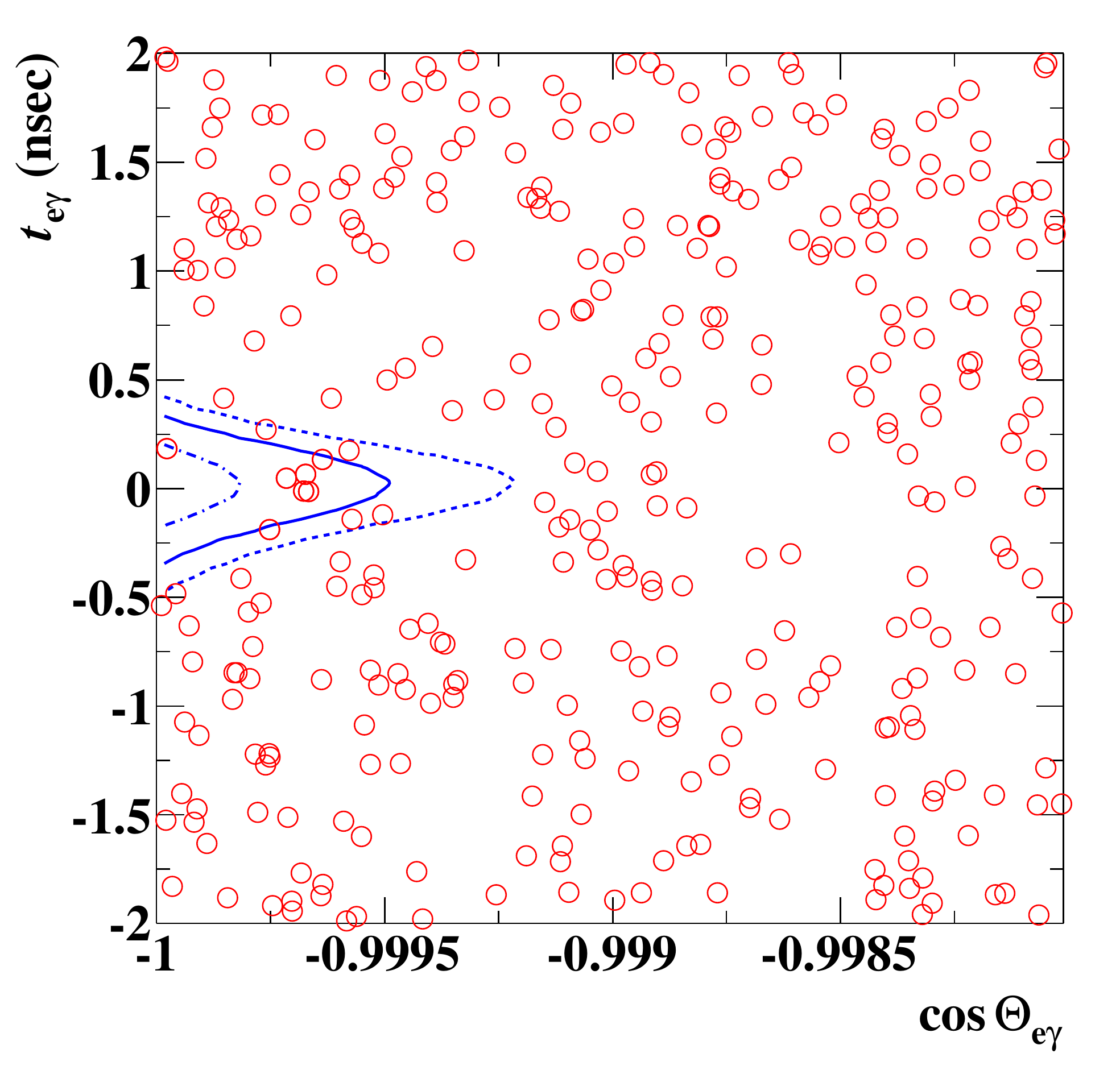}
  \caption{\label{fg:MEG-data}
    The event search regions from the MEG publication~\cite{Adam:2013mnn}, which illustrates the importance of a comprehensive approach to improving detector resolutions and timing in order to narrow the allowed good event region.  The left panel shows the $E_\gamma - E_e$ plane, where the allowed region is centered at $m_\mu / 2$ in both projections.  The right panel is the $\Delta t - \cos\Theta_{e\gamma}$ plane where the region of interest picks out simultaneous, back-to-back decay products. The blue enclosed regions are signal probability distribution function contours of $1-, 1.64-,$ and $2\,\sigma$. Figures courtesy MEG collaboration, Ref.~\cite{Adam:2013mnn}}
\end{figure}

Two types of backgrounds enter, those from radiative decay, and those from pileup events.  The key to distinguish these from signal are the performance parameters of the detectors.  We reproduce Fig.~\ref{fg:MEG-data} from \cite{Adam:2013mnn} because it perfectly illustrates the case by displaying the allowed signal event regions, which are dictated by detector resolutions.  If the resolutions can be improved, the signal region can be reduced, increasing the overall sensitivity. The right panel shows the $cos\Theta_{e\gamma}$ vs. $\Delta t$ plane, where $\Theta_{e\gamma}$ is the difference between the emitted angles of the electron and gamma, and $t_{e\gamma}$ is the difference in their timing.  The left panel shows the $e^+ - \gamma$ energy plane with the enclosed ellipses defining the signal region.  Improvements in energy-, angle-, and timing-resolution of the involved detectors allows one to shrink the good-event windows proportionally.

The MEG\,II approved upgrade program~\cite{Baldini:2013ke} is designed to handle muon stopping rates on a thinner target at rates up to 70\,MHz; the final statistics will require three years of running, beginning in 2016.  Resolutions will be improved by about a factor of 2 on all detectors.  New UV-sensitive SiPMs will replace the PMTs and provide more uniform light collection.  A single-volume drift chamber system will replace the current vane structures. The positron scintillator counters provide 30\,ps timing resolution.  The overall goal for the branching ratio is $< 4 \times 10^{-14}$.

\subsubsection{Measuring $\mu^+ \rightarrow e^+ e^- e^+$}
The unique signature of the $\mu^+ \rightarrow e^+ e^- e^+$ reaction is the three coincident $e^\pm$s emanating from a common vertex and carrying, in sum, a total energy corresponding to the full muon mass, and a net momentum of zero.  This measurement is a particular challenge, as the three tracks must be accurately measured with very high resolution for each event, and the triple coincidence implies that the tracker efficiency be very high.  Candidate three $e^\pm$ events having a total energy near to the muon mass must be considered.  Background processes such as the internal-conversion decay, generally yield lower event energy sums and distort the momentum balance such that the event should be outside the search window.  Multiple scattering and energy loss in the tracking chambers could distort the interpreted energy if the detector design does not factor that in properly.  Of the three processes discussed, this one has been idle for a long time---see~\cite{Bellgardt:1987du} with $BR_{3e} < 10^{-12}$---, but the new Mu3e collaboration at PSI is seeking to study it in stages. They are approved with a goal of $10^{-15}$ using existing beamlines.  A final phase is imagined but requires a muon beam with a rate of $10^9$\,Hz, which is a non-trivial development and investment~\cite{Berger:2014vba}.

The key to the ultimate precision is an extremely high muon stopping rate distributed over a double cone thin aluminum target.  The decay trajectories are then imaged precisely using state-of-the-art thin silicon pixel detectors assembled into cylindrical geometries.  Timing scintillators between tracking layers mark the event times. To achieve the ultimate sensitivity of $10^{-16}$, the resolution on the reconstructed muon mass must be roughly 0.5\,MeV (the mass of a single electron!), just to suppress the internal conversion look-a-like decay mode $\mu^+ \rightarrow e^+ e^- e^+ \nu_e \bar{\nu}_\mu$.  This further assumes excellent sub-ns coincident timing of the tracks, and does not factor in accidental backgrounds.

\begin{figure}
  \includegraphics[width=\columnwidth]{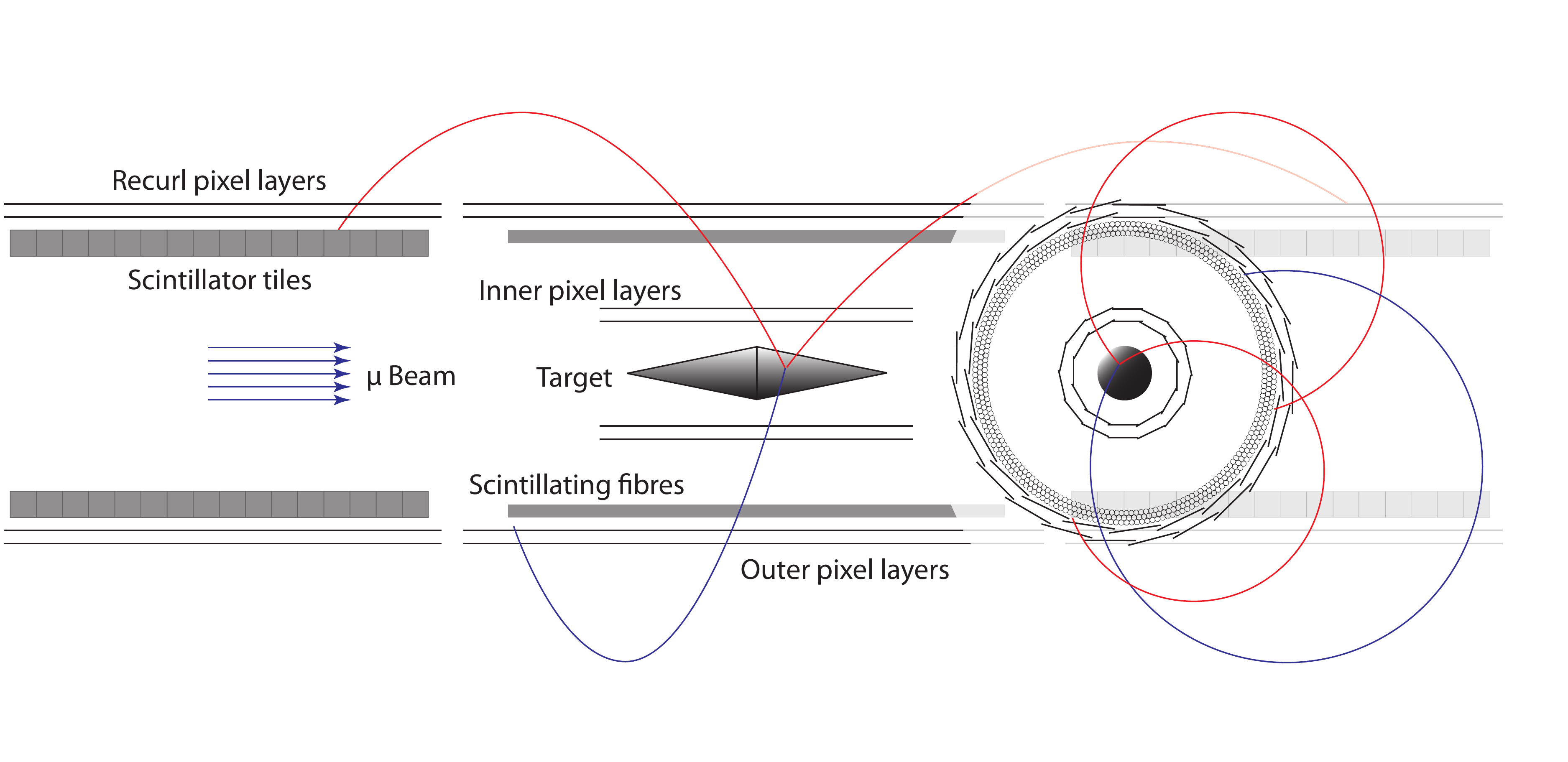}
  \caption{\label{fg:Mueee-det}
Side and end views of the proposed Mu3e detector with a single simulated $\mu^+ \rightarrow e^+ e^- e^+$ event displayed.  The conical target distributes the high rate of muon stops in different locations forming distinct vertices in the reconstruction of tracks.  The ultra-thin pixelated silicon tracker extends up- and down-stream. The scintillating fiber hodoscope provides precision timing.  The experiment will be assembled and run in phases; the approved Phase IB configuration is shown. Figure courtesy Mu3e collaboration.}
\end{figure}

A schematic diagram of the approved Mu3e experiment is shown in Fig.~\ref{fg:Mueee-det}.  In the first phase, only the central pixel detector surrounding the target will be used and the beam rate in the $\pi E5$ line will be limited to $10^7 \mu^+$/s.  The physics goal is a BR sensitivity of roughly $10^{-14}$. The next step increases the tracking volume, adds scintillating-fiber trackers and other components, and increases the muon stopping rate to $10^8$\,Hz.  The aim is a sensitivity of $10^{-15}$.

\subsubsection{Measuring $\mu^-N \rightarrow e^-N$ coherent conversion}
\label{ssc:mu2e}
The quantity of interest in a $\mu \rightarrow e$ experiment is the ratio
\begin{equation}
\label{eq:mue}
R_{\mu e} = \frac{\Gamma[\mu^- + A(Z,N) \rightarrow e^- + A(Z,N)]}{\Gamma[\mu^- + A(Z,N) \rightarrow \nu_\mu + A(Z-1,N)]}
\end{equation}
where the rate of ordinary capture in the denominator is already quite well known.  For example, for typical stopping targets such as aluminum, which we will continue to assume below, one can observe that the muon lifetime of a $\mu$-Al atom is reduced to $\approx 864$\,ns.  From
\begin{equation}
\frac{1}{\tau_{\mu Al}} = \Gamma_{{\rm tot}} = \Gamma_{{\rm decay}} + \Gamma_{{\rm capture}}
\end{equation}
and knowing the free lifetime, one obtains the capture rate
$\Gamma_{\mu Al} \approx 7 \times 10^5$\,s$^{-1}$.  The numerator in Eq.~\ref{eq:mue}, on the other hand, takes on a new level of challenge.

While the history of $\mu \rightarrow e$ experiments is well covered in Ref.~\cite{Bernstein:2013hba}, here we concentrate on the two major efforts\footnote{We note that the less ambitious  DeeMe~\cite{Natori:2014yba} experiment at the J-PARC Material Life Science Facility aims to turn on sooner than these efforts, with the single event sensitivity of $2 \times 10^{-14}$. If achieved, that level will be more than an order of magnitude improvement compared to current limits.} that ambitiously aim at single-event sensitivities approaching $\approx 3 \times 10^{-17}$, a 4 order of magnitude improvement compared to present limits.
The two projects---Mu2e\,\cite{Bartoszek:2014mya}  at Fermilab and COMET\,\cite{Cui:2009zz} at J-PARC---share many similar conceptual design features, while individual technical solutions differ.  We discuss the generic method here using some examples for clarity.

A principle requirement is a sample of $\sim 6 \times 10^{17}$ muonic atoms.  No existing secondary decay beamline can provide such a flux to accumulate that many events in a few years of running.  Furthermore, the muons should be delivered with a beam-on / beam-off time structure commensurate with the muon lifetime in the target material. Ideally, bursts of muons should arrive and stop in a suitable nuclear target during an accumulation period lasting some 100's of ns at most, followed by a beam-off and background-free quiet measuring period.  During that time, the detector is sensitive to the unique $\mu \rightarrow e$ signature event having a mono-energetic electron emitted with $E_e = 105$\,MeV; the value represents the muon mass, less the atomic binding energy.  For maximally efficient data taking, the combined accumulation-measuring cycle should be $\mathcal{O}(2\tau_{(\mu Al)}) \approx 1.7\,\mu$s and repeat as efficiently and continuously as possible, subject to the main accelerator macro cycle timing.

To provide some numbers as an example, the Mu2e statistical budget assumes an average data-taking rate of $\sim 10^{10}$ formed muonic atoms per second and $6 \times 10^7$ s of running time.  Using the reduced fraction of the Fermilab 1.33\,s accelerator cycle available, they must collect nearly 60,000 atoms in a $1.7\,\mu$s cycle.  How is this possible?

\begin{figure}
  \includegraphics[width=\columnwidth]{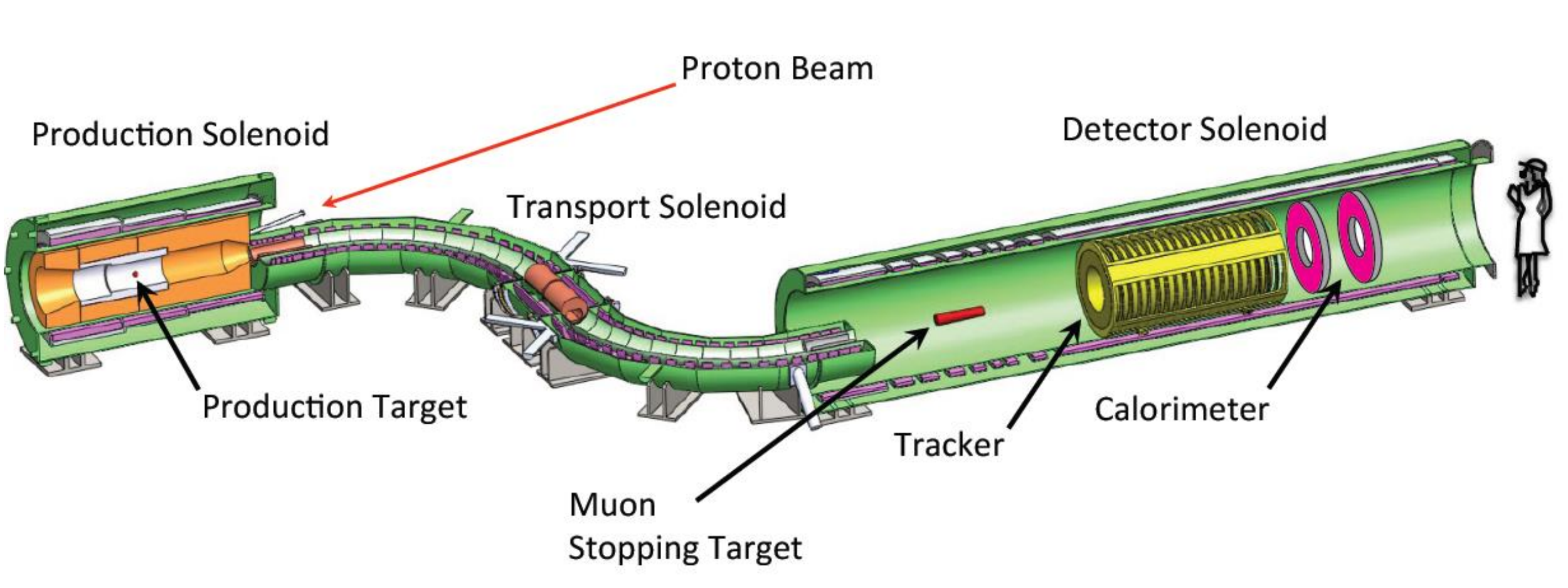}
  \caption{\label{fg:Mu2e}
The Mu2e Experimental concept.  The pulsed proton beam enters the production solenoid and passes through a tungsten rod. Pions produced there, decay to muons which are directed via magnetic gradients into the $S-$shaped transport solenoid, arriving in the detector solenoid region. The momentum-selected negative muons stop in an Al target.  The signature decay 105\,MeV electron is measured with a tracking chamber followed by an electromagnetic calorimeter. Figure courtesy Mu2e collaboration.}
\end{figure}

Both Mu2e and COMET follow a recipe credited to Lobashev and proposed for MELC in 1989~\cite{Dzhilkibaev:1989zb}.  A pulsed proton beam is slammed through a target that is enveloped in a superconducting ``production'' solenoid.  The produced pions are contained by the magnetic field, as are their decay muons below a momentum of $\approx 40$\,MeV/$c$. The solenoid field has a strong gradient to direct the spiraling secondaries out along the upstream direction into a curved ``transport'' solenoid. This either $S-$ (Mu2e) or $U-$ (COMET) shaped device provides momentum and sign selection of the captured muons and rejects by line-of-sight the transport of neutral particles from the production target.  It is long enough to allow most of the pions to decay before emerging at the exit. The final stage is yet another solenoid---the ``detector'' region---where the muons will first stop in thin Al blades.  In the same volume, and downstream further, a specialized spectrometer system is positioned to intercept the highest energy decay electrons (well above the Michel endpoint).
Figure~\ref{fg:Mu2e} illustrates the geometry and scale of the Mu2e interpretation of this concept. As noted, the transport section for COMET is $U$ shaped,and the detector section is also $U$ shaped, the bend between target and tracker.

\begin{figure}
\begin{centering}
  \includegraphics[width=0.49\columnwidth]{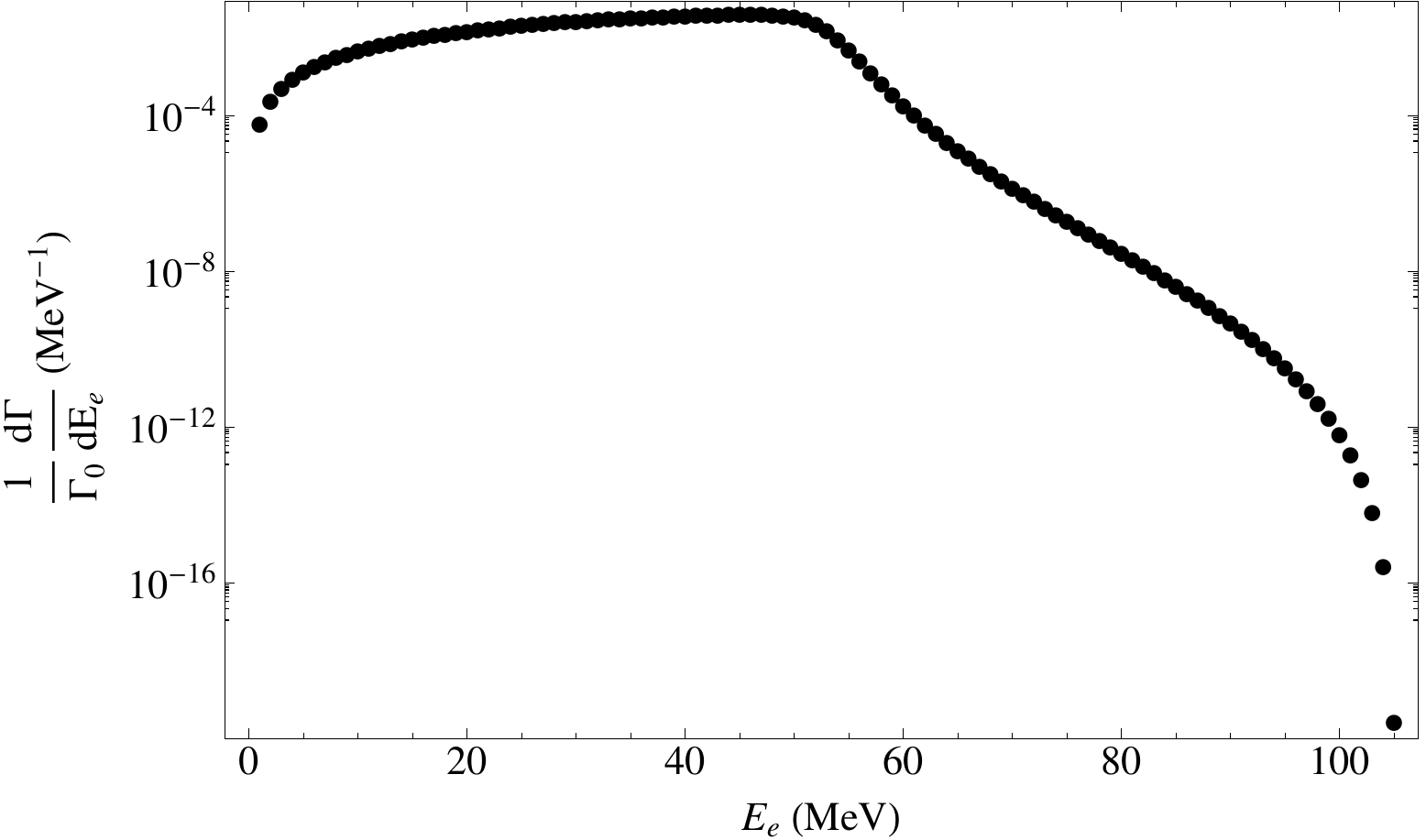}
  \includegraphics[width=0.49\columnwidth]{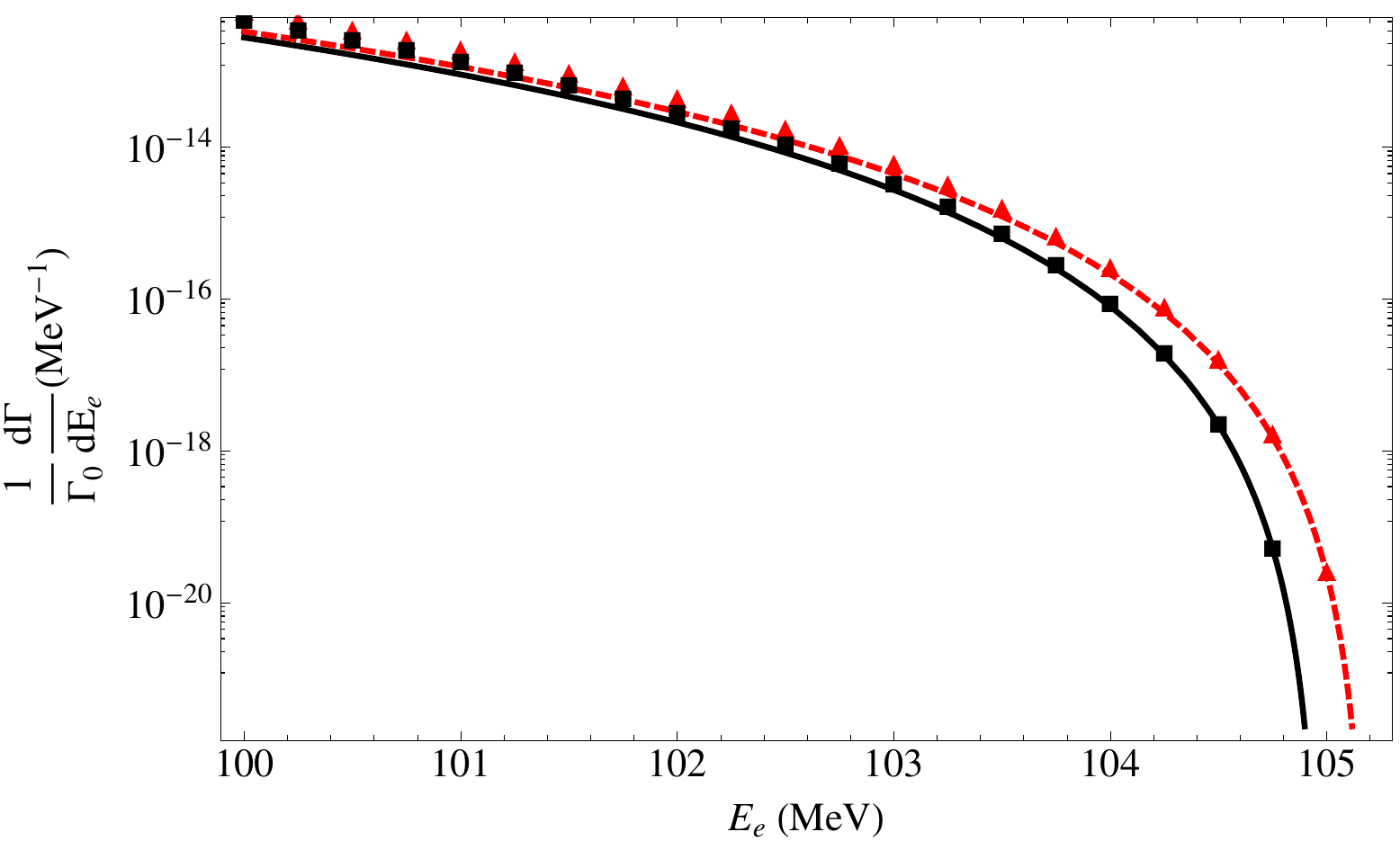}
  \caption{\label{fg:DIO}
Left: Decay in orbit electron energy spectrum assuming an aluminum target for the $\mu$Al atom. Right:  Blowup of the critical signal region from 100 to 105\,MeV where the black squares include recoil effects; red triangles do not.  Figures courtesy A.\,Czarnecki, Ref.~\cite{Czarnecki:2011mx}}.
\end{centering}
\end{figure}

Assuming now that one has a sample of formed $\mu$Al atoms, with the muon having cascaded to the 1S ground state, the next critical issue is associating a high-energy emerging electron with $\mu \rightarrow e$ conversion.  One might naively assume this is quite far away from the Michel endpoint $E_e = m_{\mu/2}$, which it is; however, that only applies for an unbound muon.  The muon in a $\mu$Al atom can ``decay in orbit" (DIO) about 39\% of the time. The corresponding kinematics of the in-motion system allows---in principle---the emitted electron to carry away the entire muon mass in energy, with the nucleus participating to conserve momentum.  This is a well-known effect and the probability of an electron close to the 105\,MeV signal region is indeed quite rare (or else, conversion experiments would have long ago ceased).  With the approaching efforts of Mu2e and COMET, the exactness of the DIO spectrum had to be revisited to determine the real sensitivity of these major new efforts.  In a recent calculation, Czarnecki, Tormo and Marciano included proper nuclear recoil effects and generated the expected spectra of electron rate vs. energy shown in Fig.~\ref{fg:DIO}.  The left figure provides the full range of energies; the right panel is a blowup of the critical region from 100 to 105\,MeV. The scale is in 1\,MeV bins of the relative sensitivity units projected by the experimentalist. It is clear that very high resolution is thus required above 100\,MeV in order to keep the DIO fraction below the single event sensitivity.
At the time of this writing, both Mu2e and COMET are in construction phases.

%% file: Muon-g2_final.tex
\section{Muon Dipole Moments}
\label{sc:muondipole}

\subsection{Terminology for muon magnetic and electric dipole moments}

The muon's magnetic dipole moment (MDM) $\vec{\mu}_{\mu}$ is related to its intrinsic spin $\vec{s}$ with a proportionality that includes the $g$-factor, which is embedded in the relation
\begin{equation}
\vec{\mu}_{\mu} = g_{\mu}\left(\frac{q}{2m}\right) \vec{s}.
\end{equation}
The provision for a possible nonzero electric dipole moment (EDM) $\vec{d}_\mu$, with a magnitude parameterized by $\eta$, is given by
\begin{equation}
\vec{d}_\mu = \eta_\mu \left( \frac{q\hbar}{2mc} \right) \vec{s}.
\end{equation}
Notice that any EDM must be aligned along the angular momentum axis, the only vector in the system.
We first discuss the important topic of the MDM, where the measured value is in disagreement with the standard model (SM) prediction.  The measured EDM is compatible with zero, but it is not yet determined to high enough precision to challenge SM completeness.  The EDM prospects are discussed in Sec.~\ref{ssc:electricdipole}.

\subsection{Magnetic dipole moment}
\label{ssc:magneticmoment}
\begin{figure}
\begin{centering}
  \includegraphics[width=\columnwidth]{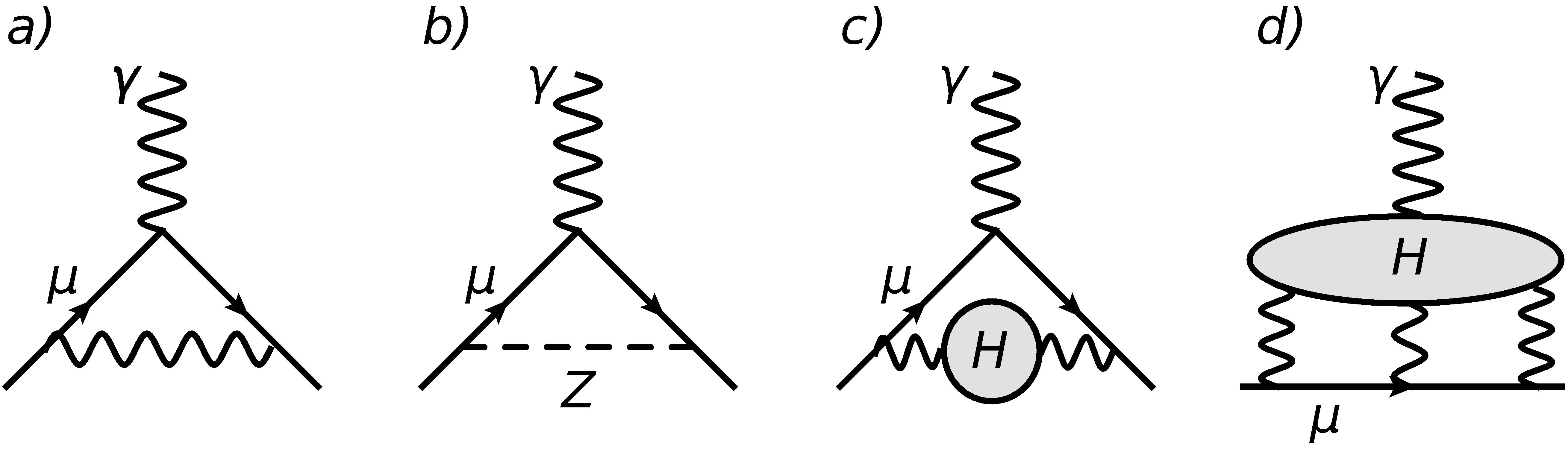}
  \caption{\label{fg:diagrams}
    Example diagrams that contribute to the muon anomalous magnetic moment.  a) Leading-order QED ``Schwinger'' term; b)  Electroweak $Z$ exchange diagram; c) Lowest-order hadronic vacuum polarization; d) Hadronic light-by-light scattering.}
\end{centering}
\end{figure}

\begin{figure}
\begin{centering}
  \includegraphics[width=\columnwidth]{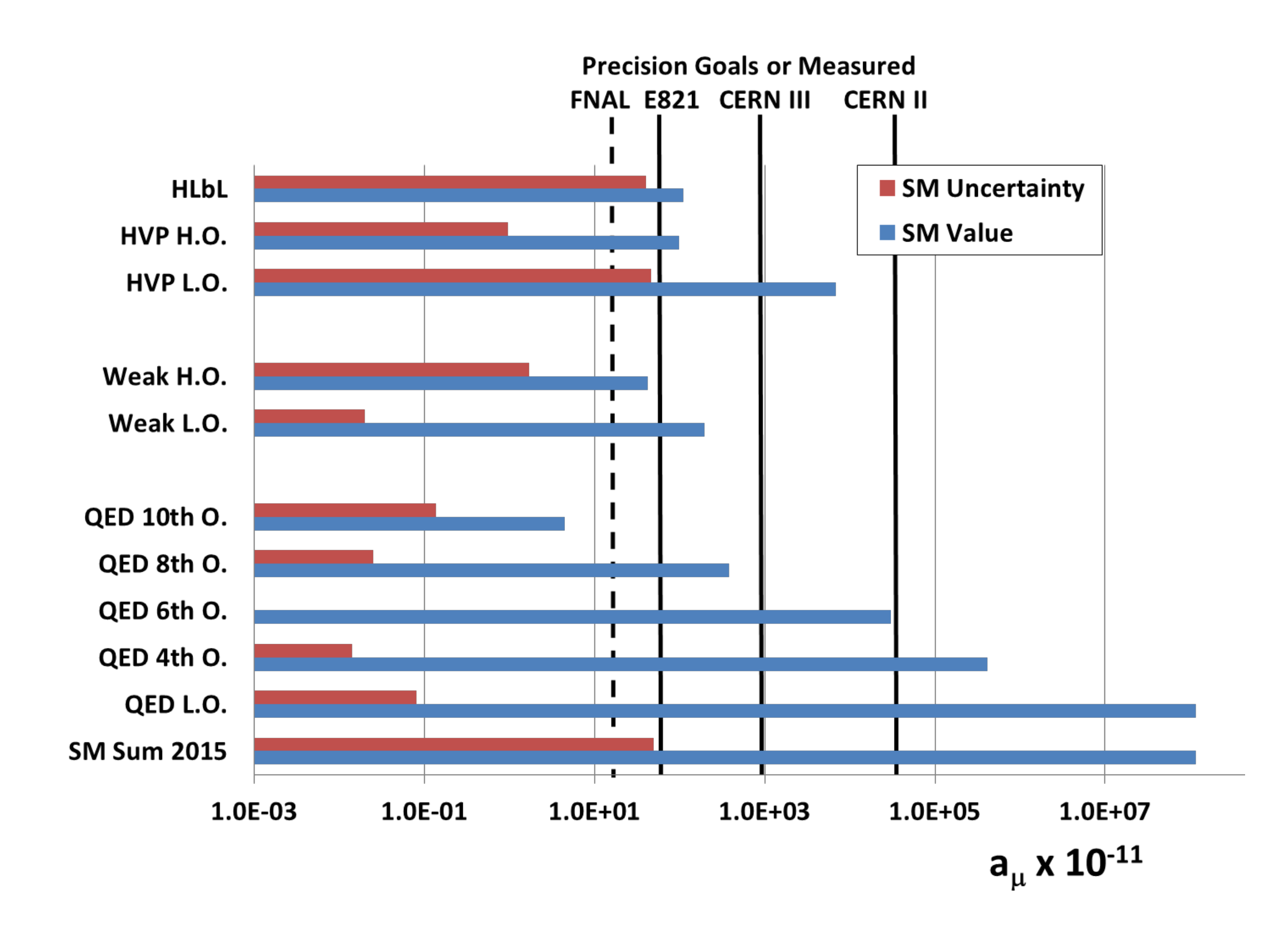}
  \caption{\label{fg:SM-contributions}
    The horizontal bars display the magnitudes of the standard model contributions (blue) and their present uncertainties (red).  Terms include QED through 10$^{th}$ order, leading- and higher-order weak, leading- and higher-order hadronic vacuum polarization, and hadronic light-by-light.  The 2014 summary is given along with its uncertainty of $49 \times 10^{-11}$.  The solid vertical black lines are placed at the achieved final precisions of the {\tt CERN\,II} (27-\,ppm), \cernexp\ (7.3\,ppm), and \bnlexp\ experiments (0.54\,ppm).  The dashed black line illustrates the precision goal of \fnalexp\ (0.14\,ppm)}
\end{centering}
\end{figure}

The Dirac equation predicts $g_\mu \equiv 2$ for the structureless, spin-1/2 muon.  Radiative corrections from electromagnetic (QED), weak and hadronic loops give rise to a so-called anomalous magnetic moment \amu; that is, $g_\mu = 2(1 + \amu)$, or more commonly stated: $a_\mu \equiv (g-2)/2$.  The anomaly is a small correction, with $\amu \approx 1/850$. Figure~\ref{fg:diagrams} illustrates example Feynman diagrams that must be evaluated to arrive at the SM value for \amu. In practice, more than 10,000 of these topologies have been calculated.  These include QED through 5 loops (10$^{th}$ order), weak exchanges through 3 loops, leading-order hadronic vacuum polarization (Had-LO)---which is determined from experimental data---and higher-order hadronic light-by-light scattering (HLbL), which is difficult to evaluate owing to the non-perturbative nature of QCD at low energies.  An excellent and detailed summary can be found in the textbook and review by Jegerlehner~\cite{Jegerlehner:2008zza,Jegerlehner:2009ry}.
Figure~\ref{fg:SM-contributions} displays the magnitudes and uncertainties of the SM contributions along with the current summary.

The experimental history of \gm\ measurements is quite rich and well reviewed; see, for example Ref.~\cite{Farley:2004hp}.  A series of storage ring experiments was conducted at CERN in the 1960's and 70's and the highest precision effort was completed at Brookhaven National Laboratory in 2001.  As this review will describe, next-generation efforts are being prepared~\cite{Grange:2015bea,Iinuma:2011zz}.  The vertical black bars in Figure~\ref{fg:SM-contributions} indicate the achieved uncertainties of the three most recent experiments (solid lines) and the precision goal (dashed line) of the new Fermilab experiment.

Both experiment and theory are now known to similar sub-ppm uncertainty and the comparison provides a sensitive test of the completeness of the standard model. If the SM accounting is accurate, and the experimental result is correct, the present comparison already begins to suggest the existence of some new physics process that affects \amu.

Currently accepted SM predictions for $a_{\mu}$(SM) are
\begin{eqnarray*}
a_{\mu}({\rm SM_a}) &=& 1\,165\,918\,02\, (49) \times 10^{-11} ~~~ (0.42\,{\rm ppm})  ~~~ {\rm and}\\
a_{\mu}({\rm SM_b}) &=& 1\,165\,918\,28\, (50) \times 10^{-11} ~~~ (0.43\,{\rm ppm}).
\end{eqnarray*}
The subscripts a (\cite{Davier:2010nc}) and b (\cite{Hagiwara:2011af}) represent slightly different evaluations of the leading-order hadronic vacuum polarization contributions.  In both cases a theoretically sound dispersion relationship is used, which evaluates \amu\ from an appropriately weighted integral of $e^+e^- \rightarrow hadrons$ absolute cross section data, summed over all energies.  The uncertainty here is largely experimental. The HLbL contribution, on the other hand, is model based and its quoted theoretical uncertainty is only estimated at present.

The experimental \amu\ is based solely on the BNL E821 measurement~\cite{Bennett:2006fi},
\begin{equation*}
a_{\mu}({\rm Exp}) = 1\,165\,920\,91\, (63) \times 10^{-11} ~~~ (0.54\,{\rm{ppm}}).
\end{equation*}

The difference between experiment and theory---the ``\gm\ test''---is
\begin{eqnarray*}
\Delta a_{\mu}({\rm Exp - SM_a}) &=& 289(80) \times 10^{-11}  ~~~ (3.6\,\sigma) ~~~ {\rm or} \\
\Delta a_{\mu}({\rm Exp - SM_b}) &=& 263(80) \times 10^{-11}  ~~~ (3.3\,\sigma).
\label{eq:deltaamu}
\end{eqnarray*}
In both cases, the result is quite provocative.  The magnitude of the difference is large---it exceeds the electroweak contributions---and the statistical significance is large. This persistent discrepancy has led to many speculations of new physics scenarios and to challenges to the accounting procedures, data selection, and hadronic models used to determine the SM expectation.  The resolution---new physics or some type of theory or experimental error---has led to the launching of two major new experimental thrusts, with one of them that aims to reduce $\delta \amu({\rm exp})$ fourfold to an absolute uncertainty of  $16 \times 10^{-11}$.  In parallel,  a vigorous theory campaign is taking place~\cite{Blum:2013xva,Benayoun:2014tra}.

As Fig.~\ref{fg:SM-contributions} illustrates, the QED and weak uncertainties are already well below any experimental reach.  Improvements by a factor of 2 in the dominant Had-LO uncertainty are expected from new data sets at VEPP-2000 (Novosibirsk) and BESIII (Beijing), along with continued analyses of BaBar and Belle existing data sets~\cite{Blum:2013xva}.

While the magnitude of the complete HLbL contribution is $\sim 110 \times 10^{-11}$, the quoted uncertainties can be as large as 35\%.  A recent review of the status of the HLbL contributions is given in \cite{Masjuan:2014rea}. Initiatives going forward include those based on lattice QCD~\cite{Blum:2013qu}, and new $\gamma^*$ physics measurement programs at BESIII and KLOE (Frascati) that aim to build a data-driven approach to leading HLbL terms~\cite{Blum:2013xva}.

\subsubsection{New physics possibilities from \gm}
\begin{figure}
  \includegraphics[width=0.68\columnwidth]{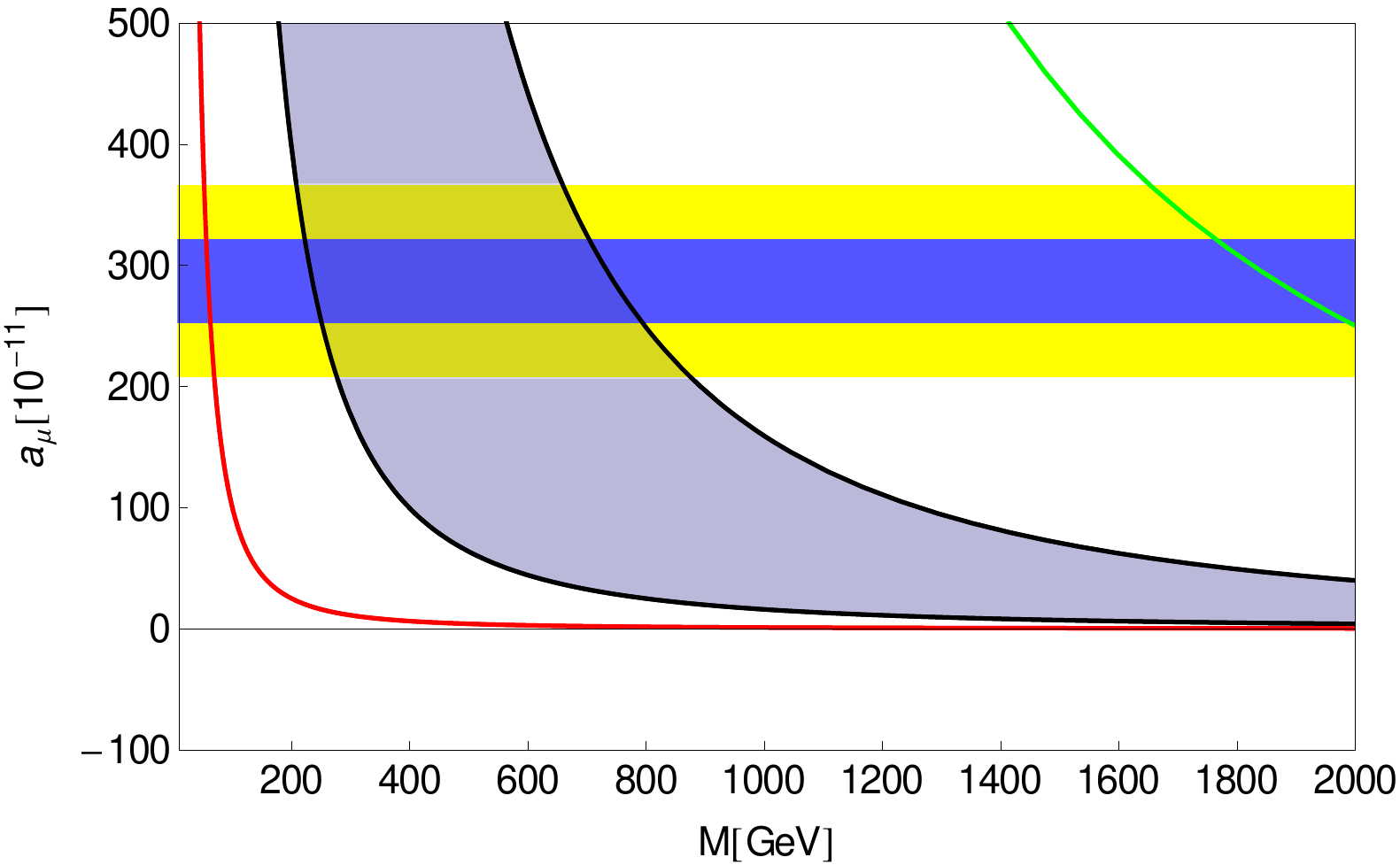}
  \includegraphics[width=0.3\columnwidth]{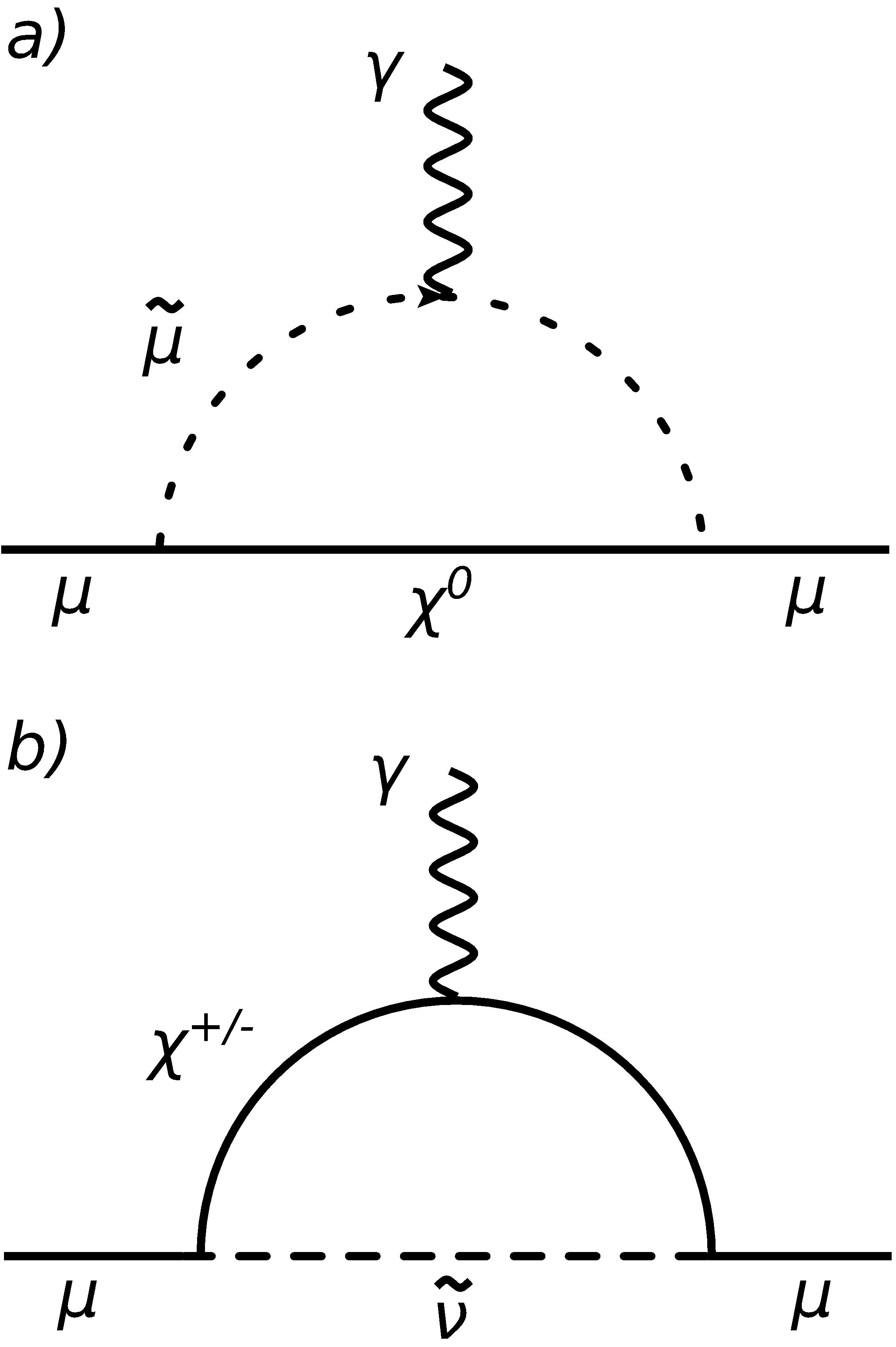}
  \caption{\label{fg:stockinger}
    Left: Generic classification of mass scales vs. $a_\mu$ contributions from new physics sources.  Green line: radiative muon mass generation;  Red line: $Z', W'$, universal extra dimensions, or Littlest Higgs models with typical weak-interaction scale coupling; Purple band: unparticles, various extra dimension models, or SUSY models where the coupling is enhanced.  The width illustrates a $\tan\beta$ range of $5 - 50$ for SUSY models.  The yellow horizontal band corresponds to the current difference between experiment and theory and the blue band is an improvement with a combined theory and experimental error of $34 \times 10^{-11}$. Figure courtesy D. Stockinger; see also \cite{Hertzog2015}.  Right: Neutralino-smuon (a) and Chargino-sneutrino (b) one-loop SUSY contributions to \amu~\cite{Stockinger:2006zn}}
\end{figure}

What is nature trying to tell us if the current discrepancy $\Delta a_\mu$ remains as large as it is and the significance eventually exceeds $5\,\sigma$?  While numerous explanations exist in the literature, it is instructive to take a more generic approach here as first outlined by Czarnecki and Marciano~\cite{Czarnecki:2001pv}, and elaborated on by Stockinger~\cite{Stockinger}.
The magnetic moment is a flavor- and CP-conserving, chirality-flipping, and loop-induced quantity. Any new physics (N.P.) contributions will typically contribute to \amu\ as
\begin{equation}
\delta a_\mu (N.P.) = \mathcal{O}[C(N.P.)]\times \frac{m_{\mu}^2}{M^2},
\end{equation}
where $M$ is a new physics mass scale and $C(N.P.)$ is the model's coupling strength, which is then also common to the same new physics contributions to the muon mass.  That means
\begin{equation}
C(N.P.) \equiv \frac{\delta m_\mu(N.P.)}{m_\mu}.
\end{equation}
Different predictions for $a_\mu(N.P.)$ are illustrated in Fig.~\ref{fg:stockinger} for various coupling strengths, $C(N.P.)$ versus the present and future $\delta a_\mu$ limits, the latter being the combined uncertainty from SM theory and experimental sensitivity. For radiative muon mass generation,  $C(N.P.) = \mathcal{O}(1)$, which implies $a_\mu$ probes the multi-TeV scale. For models with typical weak interaction coupling, $C(N.P.) = \mathcal{O}(\alpha/4\pi)$, the implied mass scale is very light, arguably ruled out by direct measurements.  In contrast, models with enhanced coupling such as supersymmetry, unparticles and extra dimensions are represented by the central band, where the overlap with $a_\mu$ corresponds to the TeV-scale physics regime.  Focussing on SUSY, the expected contribution to $a_\mu$  has the following behavior:
\begin{equation}
\amu^{\rm SUSY} \approx 130 \times 10^{-11}\left(\frac{100\,{\rm GeV}}{M_{\rm SUSY}}\right)^2\,\tan\beta\,{\rm sign}(\mu).
\end{equation}
Here, $\tan\beta$ is the ratio of vacuum expectation values for the two Higgs doublets and sign$(\mu)$ is the sign of the higgsino mass parameter, taken to be +1 from current \gm\ constraints.  A wide, yet natural, range for $\tan\beta$ of $5 - 50$ provides the width of the central band. It is clear from this figure that the LHC reach overlaps well with the simplest SUSY expectations, which have been highly motivated given the compatibility with the current \gm\ result~\cite{Stockinger:2006zn}.  However, the lack of any signal in the 7-8\,TeV data taking at the LHC has pushed the mass scale near to and even above 1\,TeV, which has also spawned many variants on the simplest SUSY models that must remain compatible with a wide suite of experimental limits and the \gm\ ``signal;'' see reviews:~\cite{ellis2014,ross2014}.

Hypothetical dark photons are very weakly interacting and very light particles that could produce a large enough contribution to \amu\ to explain the discrepancy between experiment and theory~\cite{Davoudiasl:2012ig}.  In Fig.~\ref{fg:stockinger}, a dark photon mechanism would correspond to a new band to the left of the red band that crosses the $\Delta a_\mu$ region in the 10 - 100\,MeV mass range.  The coupling strength would be appropriately tuned and small.  While initially suggestive of a neat explanation for \gm\, recent experiments~\cite{Batley:2015lha} have all but ruled out the simplest versions of the theory and its implied parameter space; however, more complex scenarios---ones that imply a dark $Z'$ for instance---remain viable~\cite{Davoudiasl:2014kua}.

\subsection{Experimental considerations}
Given the high impact of the \gm\ experiments, numerous general and detailed reviews have been written~\cite{Farley:1990gm,Hertzog:2004gc,Miller:2007kk}, along with a comprehensive publication by the E821 collaboration~\cite{Bennett:2006fi}.  Here we will describe the essential features that enable such a precision measurement.  The persistent $> 3\,\sigma$ discrepancy between experimental and theoretical \amu\ over the past 10 years has led to the development of two new experiments that are being designed to resolve the situation.  \fnalexp~\cite{Grange:2015bea} and \jparcexp~\cite{Iinuma:2011zz} are following different approaches. In both cases, intense bunches of polarized muons are injected into a uniform-field magnet, where the muon spin precession frequency is proportional to \amu.  The decay positrons are measured as a function of time, the rate of which features a sinusoidal modulation imprinted on the exponential decay.
Beyond this, the difference in implementation is quite significant. We first outline common considerations for a \gm\ experiment and follow with brief descriptions of the two new experimental campaigns.

The spin of a muon at rest will precess in a magnetic field $\vec{B}$ at the Larmor frequency,
$\vec{\omega}_L = -gq\vec{B}/2m_\mu$.  A precision measurement of $\omega_L$ together with an equally precise determination of $\vec{B}$ gives $g$.   Since \gm\ is known to sub-ppm already, a direct measurement of $g$ would need to be at the sub-ppb level to be competitive.  However, the muon mass
$m_\mu$ is ``only'' known to 34\,ppb, so that quantity too would need a major improvement.

In contrast, a measurement using in-flight muons in a magnet can directly determine \gm.  It is based on the difference between the cyclotron and spin-precession frequencies for a polarized ensemble of muons that circulates in the horizontal plane of a uniform vertical magnetic field.   The cyclotron frequency when $\vec{B} \cdot \vec{P}_\mu = 0$ is
\begin{equation}
{\vec \omega_c} = -{q {\vec B} \over m \gamma}
\end{equation}
and the spin turns at frequency
\begin{equation}
{\vec\omega_s} = -{gq{\vec B} \over 2 m} - (1-\gamma) {q {\vec B} \over
\gamma m}
\end{equation}
owing to the torque on the magnetic moment and including the Thomas precession effect for the rotating reference frame~\cite{Bargmann:1959gz}.

If $g$ were exactly equal to 2, ${\vec \omega_s} = {\vec \omega_c}$; however, for $g \neq 2$,
\begin{equation}
{\vec \omega_a} \equiv {\vec \omega_s} - {\vec \omega_c} = -\left( {g-2
\over 2} \right) {q{\vec B} \over m} = -\amu {q{\vec B} \over m},
\label{eq:simpleomega}
\end{equation}
where we have defined \wa\ as the anomalous precession frequency.  It is this quantity that must be measured to determine \gm.

The recent storage ring experiments used electric quadrupoles to provide vertical containment---effectively creating a large Penning trap.  The motional magnetic field seen by a relativistic muon in an electric field $\vec{E}$ will contribute an important term to the spin precession rate. Additionally, a nonzero muon EDM will also require modification to Eqn.~\ref{eq:simpleomega}.  The full expression is then
\begin{equation}
   \vec \omega_{net} =  -{q \over m }\left[ a_{\mu} \vec B -
   \left( a_{\mu}- {1 \over \gamma^2 - 1}\right)
   \frac{\vec \beta  \times \vec E}{c} +
   \frac{\eta}{2} \left( \vec \beta  \times \vec B + \frac{\vec E}{c} \right) \right],
   \label{eq:omega}
\end{equation}
where $\vec \omega_{net} = \vec \omega_a + \vec \omega_{EDM}$.

The proposed \jparcexp\ experiment will not use electric focusing, which simplifies Eqn.~\ref{eq:omega} and separates the contributions of \wa\ and $\omega_{EDM}$. In contrast, the modern storage ring experiments operate with a relativistic gamma of 29.3, ($P_\mu = 3.094$~GeV/$c$), which makes the first expression in parentheses vanish. In practice, a combined correction to \wa\ in \bnlexp\ owing to the electric-field and related pitch correction was $+0.77 \pm 0.06$\,ppm~\footnote{The finite momentum spread, $\delta P_\mu/P_\mu \approx 0.15\%$, means the term in parenthesis does not vanish completely. The vertical betatron oscillations cause $\vec{B} \cdot \vec{P}_\mu$ to not always equal zero.}.

Parity violation in the muon decay chain $\mu^- \rightarrow e^-{\bar{\nu}}_{e}\nu_\mu$ provides the necessary polarimetry that is required to access the average muon spin direction vs. time; that is, the link to \wa. The CM correlation between the emitted angle and energy of the decay electron with respect to the muon spin direction is illustrated in Fig.~\ref{fg:mudecay}.  To a good approximation, the electron {\em energy} in the boosted laboratory frame is related to the energy and {\em angular} distribution in the CM as
\begin{equation}
E_{e,lab} \approx \gamma E_{e,CM}(1+\cos\theta_{CM}).
\end{equation}
This is the key relation. The highest energy electrons are preferentially emitted when the muon spin is opposite to its momentum (alternatively, along its momentum for $e^+$ from $\mu^+$ decay).  Every decay electron has a momentum smaller than its parent muon and consequently curls to the inside of the storage radius where a detector is positioned to intercept it and measure its energy and arrival time.

\begin{figure}
  \includegraphics[width=0.6\columnwidth]{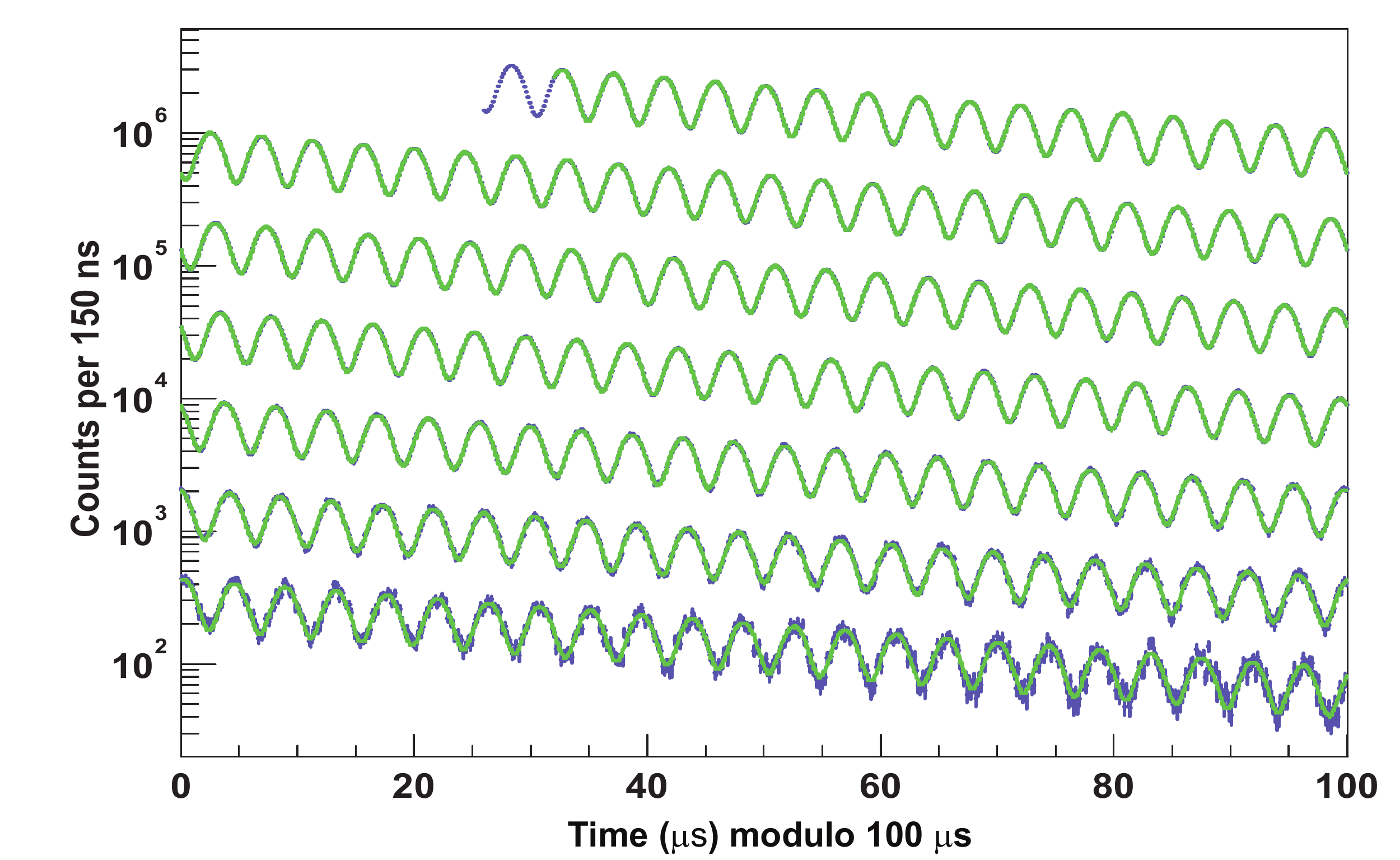}
  \includegraphics[width=0.38\columnwidth]{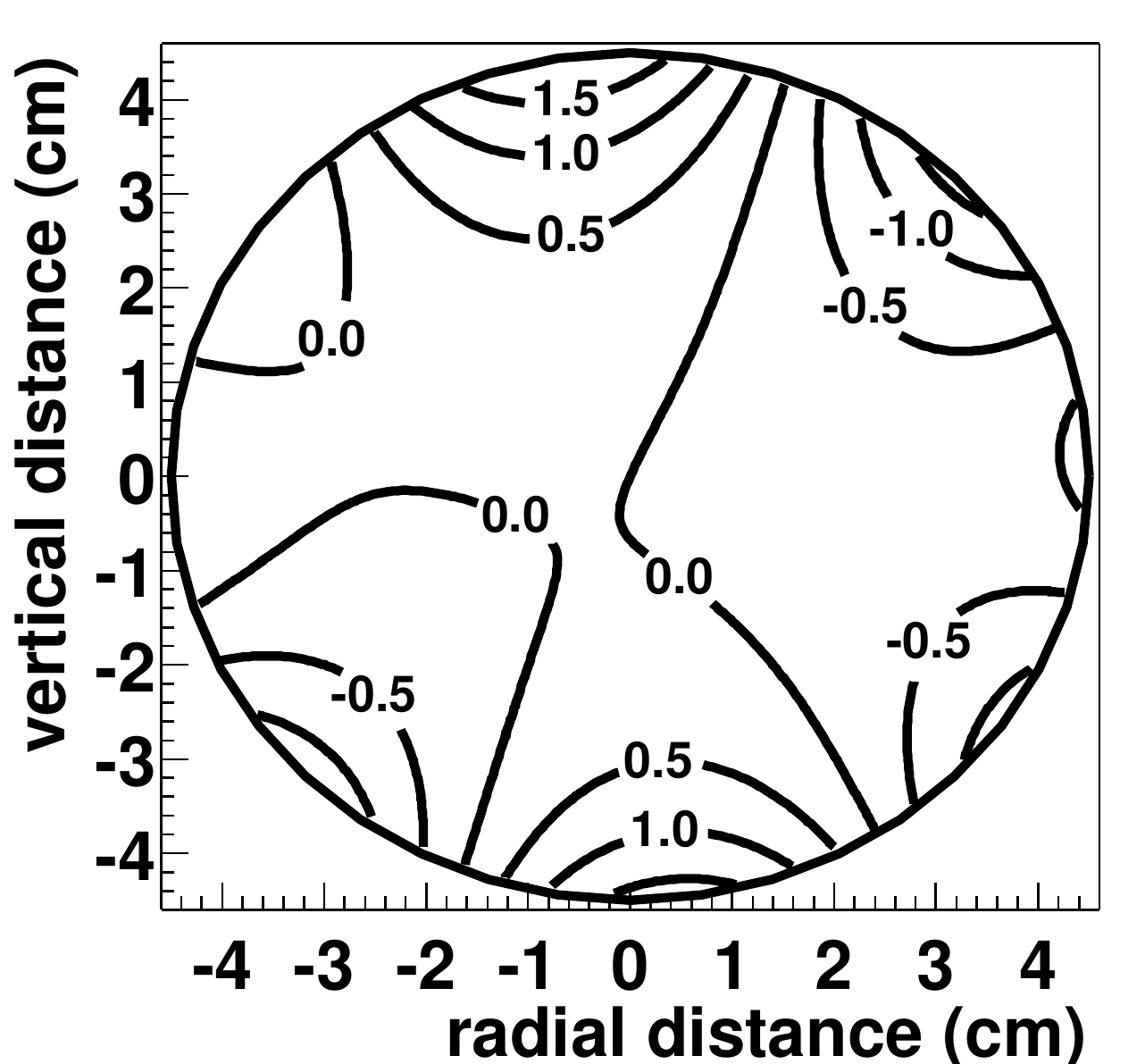}
  \caption{\label{fg:E821data}
    Left: E821 anomalous precession data, including fit.  The data is wrapped around every $100\,\mu$s. Right: The final, azimuthally averaged magnetic field contours; the scale is in ppm with respect to the average. Figures courtesy E821 collaboration.}
\end{figure}

The accumulated number of electrons having an energy greater than $E_{th}$ forms a distribution vs. time that has the structure
\begin{equation}
 N(t;E_{th}) = N_{0} e^{-t/\gamma \tau_{\mu} } \left[1 +
 A\cos(\wa t + \phi\right)] \label{eq:fivepar}
\end{equation}
for a 100\% polarized beam.  The normalization $N_0$ and asymmetry $A$ depend on $E_{th}$. The ensemble-averaged spin direction at $t=0$ is represented by $\phi$, which can have a subtle energy dependence because the time of decay measured at the detector might have an energy-dependent time-of-flight component from the time of the muon decay. A representative data set from \bnlexp\ is shown in Fig.\,\ref{fg:E821data}a. The time-dilated lifetime of $\approx 64.4~\mu s$ is evident, upon which is the modulation from \wa.  The actual $N(t;E_{th})$ distribution can be more complicated compared to~Eqn.~\ref{eq:fivepar} because of coherent betatron oscillations that give rise to further modulations of $N$, $A$ and $\phi$.\footnote{True for the storage ring experiments that use quadrupole focussing; absent for the \jparcexp\ configuration.}  The statistical uncertainty on $\wa$ has been described in detail in Ref.~\cite{Bennett:2007zzb}. Different fitting methods and weighting schemes can produce reduced statistical uncertainties from the same data; however, sensitivity to leading systematic errors typically increases when one is using the more aggressive analysis procedures.  The simple threshold method that is robust and tested has a relative uncertainty on $\delta\wa /\wa$ that behaves as
\begin{equation}
\delta\wa /\wa = \frac{1}{\wa\gamma\tau_{\mu}}\sqrt{\frac{2}{NA^{2}\langle P\rangle^{2}}},
\label{eq:statistical}
\end{equation}
where we include the ensemble averaged polarization $\langle P \rangle$ for completeness. To minimize $\delta\wa /\wa$, it is advantageous to:  1) Use a high magnetic field ($\wa \propto B$); 2) Run at high energy (increases $\gamma\tau_{\mu}$); 3) Employ a highly polarized muon source; 4) Optimize the figure-of-merit (FOM), $NA^2$.  The latter occurs when $E_{th}/E_{max} \approx 0.6$.

To obtain $\amu$ from the measurement of \wa\ requires an equally precise measurement of the magnetic field.  This was accomplished in \bnlexp\ using a suite of pulsed NMR probes.  They were used to establish the absolute field magnitude, control the time stability of the field via feedback to the magnet power supply, and to periodically map the field {\em in situ} by use of a multi-probe NMR trolley that could traverse the circumference of the storage ring orbit without breaking the vacuum.  The water or petroleum jelly filled probes provided a field value in terms of the free proton precession frequency $\wp$.  Minimization of the systematic uncertainty in the field integral is intimately related to the intrinsic field uniformity.  Minimizing multipoles having higher order than the dipole depends on the care and precision of the shimming tasks carried out prior to physics data taking. The final azimuthally averaged magnetic field is represented by a contour map as shown in Fig.\,\ref{fg:E821data}b.

With \wa\ and \wp\ measured, the muon anomaly was obtained from
\begin{equation}
\amu =  \frac{\wa/\tilde{\wp}}{\omega_{L}/\tilde{\wp} - \wa/\tilde{\wp}} =  \frac{R}{\lambda - R}.
      \label{eq:amuR}
\end{equation}
In the ratio $R \equiv \wa/\tilde{\wp}$, $\tilde{\wp}$ is the free proton precession frequency in the average magnetic field experienced
by the muons.  The muon-to-proton magnetic moment ratio $\lambda = 3.183\,345\,107\,(84)$ is determined\footnote{Here we quote the updated value from the 2010 CODATA recommended values of the fundamental physical constants~\cite{Mohr:2012tt}.} from muonium hyperfine level structure measurements~\cite{Liu:1999iz} together with QED, see Sec.\,\ref{sc:muonium} and Ref.\cite{Mohr:2012tt}.

\subsubsection{Fermilab muon \gm\ experiment}

The precision goal of E989 is $\delta\amu = 16 \times 10^{-11}$ (140\,ppb), a four-fold improvement compared to \bnlexp.  This error includes a 100\,ppb statistical component in the measurement of \wa, where 21 times the data from BNL will be required. Equal systematic uncertainties of 70\,ppb are budgeted for both \wa\ and \wp, corresponding to reductions by factors of 3 and 2, respectively from what has been achieved.
While \bnlexp\ improved on the \cernexp\ experiment in a revolutionary manner---primarily by the invention of direct muon injection into the storage ring---the \fnalexp\ experiment, in contrast, will introduce a broad suite of  refinements focussed on optimizing the beam purity and rate, the muon storage efficiency, and modernizing the instrumentation used to measure both \wa\ and \wp.

A limiting factor at BNL was the 120\,m beamline between the pion production target and the storage ring. Because the decay length of a 3.11\,GeV/$c$ pion is $\approx 173$\,m, the beam injected into the storage ring contained both muons and a significant number of undecayed pions, the latter creating an enormous burst of neutrons when intercepting materials.  Their subsequent capture in scintillator-based detectors impacted detector performance adversely.
The Fermilab accelerator complex will deliver pure, high-intensity muon bunches to the storage ring at a fill-rate frequency increase of $\sim 3$ compared to BNL.  Proton batches at 8\,GeV from the Booster are divided into short bunches in the Recycler Ring.  Each bunch is then extracted and strikes a target station tuned to collect 3.1\,GeV/$c$ $\pi^+$ and transport them along a 270\,m FODO\footnote{alternating pairs of focusing and defocusing quadrupoles magnets tuned to transport the pions and the forward-decay muons} lattice.  The highest energy decay muons in the $\pi^+ \rightarrow \mu^+\nu_\mu$ decay chain are captured and transported along the same beamline with a longitudinal polarization of $\sim97\%$.  These muons (together with pions and protons) are injected into the repurposed $\bar{p}$ Delivery Ring (DR).  There, they make several revolutions to reduce the pion contamination by decay and to separate protons by their velocity difference. A kicker in the DR extracts the pure muon bunch into a short beamline that terminates at the storage ring entrance.

The centerpiece of the experiment is the 1.45\,T superconducting storage ring~\cite{Danby:2001eh} that was re-located to Fermilab from Brookhaven in 2013. It has recently been reassembled in a custom temperature-controlled building having
a firm foundation for the magnet support; both are critical for the magnetic field stability. Three storage ring subsystems---the superconducting inflector, four electric quadrupoles, and a fast kicker---determine the fraction of incoming muons $\epsilon_{store}$ that become stored, and their subsequent beam properties such as betatron oscillations. Improvements and replacements in these devices are aiming a factor of 2 or more increase in $\epsilon_{store}$ compared to BNL.


The storage ring magnetic field will be shimmed following procedures developed for E821, with small improvements owing to the need for a more highly uniform final field.  Retooling of the pulsed NMR probes, modern 3D \opera\ model guidance, upgrades to the in-vacuum shimming trolley, and NMR probe readout using waveform digitizers, represents just some of the work. The absolute NMR probe is the same one used in the muonium hyperfine experiment that established the muon-to-proton magnetic moment ratio $\lambda$ in Eqn.~\ref{eq:amuR} and it will be cross calibrated with the new J-PARC experiment (see Sec.~\ref{sc:muonium}) New absolute probes are also being developed as cross checks.

The entire suite of detectors, electronics, calibration and data acquisition systems will be new and modern.  These include highly segmented lead-fluoride Cherenkov calorimeters, where each crystal is read out by silicon photomultipliers~\cite{Fienberg:2014kka}.  The signals are recorded using custom 800\,MHz, 12-bit-depth waveform digitizers.  A distributed laser-based calibration system is designed to maintain gain stability at the 0.04\% level. Several stations of in-vacuum straw tracking detectors will provide beam dynamics measurements that are required to control various systematic uncertainties.  Finally, a high-speed, GPU-based DAQ is prepared to process Gbytes/s of data in a deadtime free operation.

The systematic uncertainty estimates are based on the considerable experience of E821 and the targeted experimental upgrades that have been made to address each one of them.
At the time of this review, the storage ring is being commissioned, with a 9-month shimming period planned for 2015. The beamlines are being built, and the detector and electronics systems have undergone various test-beam runs at Fermilab and SLAC.  Physics data taking is expected to begin in 2017.

\subsubsection{J-PARC muon ($g$-$2$) experiment}
\label{ssc:e34}

While the initial $\delta\amu$ statistical precision goal of the proposed J-PARC experiment was also at the 100\,ppb level, current estimates based on the muon source intensity and polarization now aim at $\sim 400$\,ppb, similar to the BNL experiment~\cite{Beer:2014ooa}.  The collaboration is developing a creative new method that will feature different systematic uncertainties compared to the storage ring experiment; thus, the systematic error budget is difficult to anticipate at this time.

The most striking difference will be the use of a 10-times lower momentum, but {\em ultra-cold}, muon beam. This choice is motivated by the desire to eliminate the electric field---and thus the spin-precession terms that are affected by it---used for vertical focusing in storage rings. To accomplish this goal, a muon beam must be made with a negligible transverse momentum component, $\Delta P_T/P_\mu \approx 10^{-5}$. This is accomplished by accelerating from rest a source of ultra-cold muons created by the re-ionization of muonium atoms in vacuum.  How the overall experiment is then designed, both upstream---creating the muon source---and downstream---where the positrons are measured---is quite unique.

A 3\,GeV proton beam strikes a graphite target, producing pions that can decay at rest near to the surface of the target. The 28.4\,MeV/$c$, 100\% longitudinally polarized  $\mu^+$ surface muon beam is directed to a thin target optimized to stop muons and form muonium atoms, $M \equiv \mu^+e^-$.  The target is designed to permit muonium to diffuse into the vacuum on the downstream surface.  Aggressive efforts over the past few years have realized promising results in raising the net yield of muonium that emerges into the vacuum.  A recent study at TRIUMF used a silica aerogel target where micro-channels in the target were created using a laser ablation technique~\cite{Beer:2014ooa}.  This target resulted in an 8-fold improvement compared to the previous measured yield using the same material. When the J-PARC beam rates are combined with the TRIUMF muonium yield measurements, an expected production of $0.2 \times 10^6$/s is found.  While a factor of 5 lower than originally planned, it still represents a major step forward and a viable rate for a \gm\ experiment.  Studies will continue aimed at further rate improvements.

The muonium atom~\cite{Hughes:1966yz} and its hyperfine structure measurements are described in Sec.~\ref{sc:muonium}; we also direct the reader to Fig.~\ref{fg:HFSexpt}, which illustrates the external field dependent quantities discussed herein.  Muonium formed in the {\em weak} magnetic field limit can be described in terms of its total angular momentum and associated magnetic quantum numbers $(F,M_F)_i$, where the triplet and singlet combinations are given as: $(1,1)_1, (1,0)_2, (1,-1)_3$ and $ (0,0)_4$, and the subscript $i$ is our  shorthand label for the four states. If muonium is formed in zero magnetic field, the relative population of the four states $i = 1$ to 4 is: $\frac{1}{2}, \frac{1}{4}, 0$ and $\frac{1}{4}$, respectively.  Allowing for a weak field, and choosing the axis of quantization along the incoming muon polarization direction, $\hat{z}$, the net muon polarization as a function of time is given by~\cite{KimSiang}
\begin{equation*}
P_{z} = \frac{1}{2}\left(\frac{1 + 2x^2_B + \cos 2\pi\nu_{24}t}{1+x^2_B}\right) \rightarrow \frac{1}{2}(1 + \cos 2\pi\nu_{24}t) {\rm ~~~as~~~} x_B \rightarrow 0.
\end{equation*}
Here, the field strength is traditionally expressed by $x_B$, which is a ratio of the sum of the electron and muon Zeeman interactions to the muonium ground-state hyperfine interval, all expressed in terms of frequencies. To obtain a sample of at-rest, but polarized muons, the muonium atoms must be re-ionized, which
is accomplished in the vacuum by two simultaneous laser bursts having wavelengths $\lambda_1 = 122$\,nm and $\lambda_2 = 355$\,nm.  The first excites the 1S to 2P transition and the second ionizes the atom, leaving the free $\mu^+$ essentially at rest in the vacuum.
Muons ionized from state 1, the (1,1) triplet, retain their initial polarization.  In contrast, those populating states 2 and 4 can make transitions at the rate $\nu_{24} \approx 4.5$\,GHz, which is so high that the emerging muon spin will be, for all practical purposes, unpolarized. Consequently, the maximum net polarization from the muonium-at-rest source is $P_z = 50\%$.

The ultra-cold, ``at rest,'' muons will be rapidly accelerated by a linac to a longitudinal momentum of ${\vec P}_\mu = 300\,$MeV/$c$. A novel feature of E34 is that the polarization direction can be flipped at production, prior to acceleration, by use of a low magnetic field.  This feature might become important for anticipated systematic uncertainties.

\begin{figure}
\begin{centering}
  \includegraphics[width=.8\columnwidth]{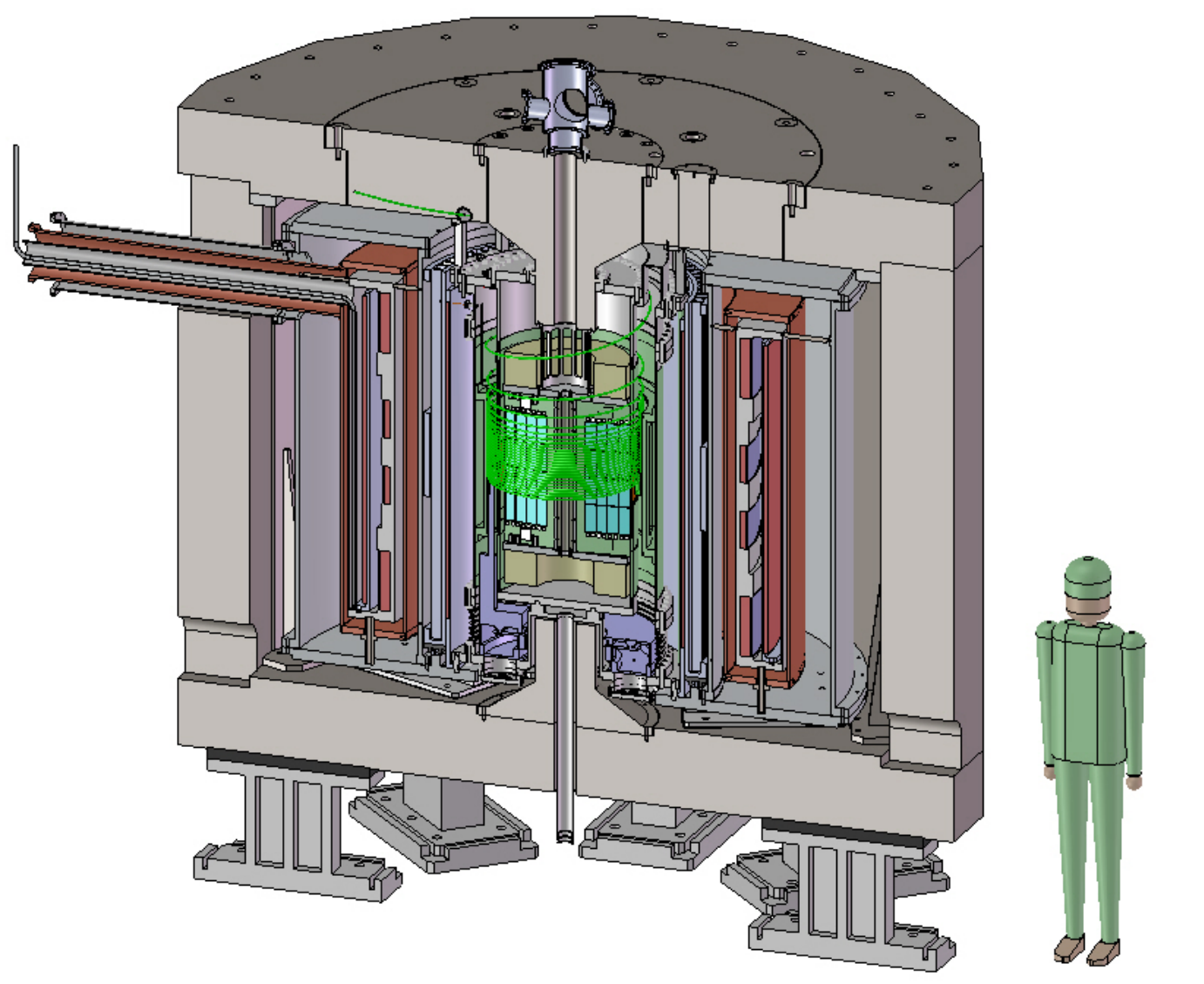}
  \caption{\label{fg:jparcg-2} The proposed setup for the J-PARC \gm\ Experiment.  Muons enter at the top left (green trajectory) and spiral into the highly uniform magnetic field region.  Their decay positrons curl inward to an array of silicon tracking detectors. Figure courtesy T.\,Mibe.}
\end{centering}
\end{figure}

The muon beam will be directed through the top of a highly uniform 3\,T MRI-type magnet.  A spiral injection path allows the muons to enter the field region at a steep angle, which softens such that the beam eventually orbits a plane perpendicular to the magnetic field, see Fig.~\ref{fg:jparcg-2}.   Custom fringe-fields and a vertical kicker are key elements in the design. Very weak magnetic focusing is required, but it causes a negligible perturbation to the \wa\ frequency.

Apart from the kinematic differences and uniqueness of the source, once the muons begin to circulate in the field, the experiment is much like the higher-energy storage ring designs.  The muon spin precesses proportionally to \gm\ and the anomalous precession frequency is encoded in the modulation of rate vs. time of the higher-energy positron decays.

The muon momentum and field strength values imply that the orbit radius is 33\,cm and the  cyclotron period is 7.4\,ns.  Decay positrons curl to the inside of the central orbit where vanes of silicon strip detectors are positioned.  The radius of the reconstructed positron tracks provides the momentum (energy) determination with good acceptance, and the location of the detectors is tuned for good performance for the higher-energy range that optimizes the FOM. The detector system must have a stable acceptance over the measuring period and withstand a total initial hit rate approaching $10^9$ hits/s~\cite{MIBE:2011kaa}. Sorting of hits into tracks presents a unique challenge here.  However, if solvable, the systematics of complete track reconstruction could be lower than those inherent using calorimeter techniques, and in any case, they will be different.

\subsubsection{Comparison of methods}
A comparison of many of the parameters between the two new \gm\ experiments is shown in Table~\ref{tb:comparison}.
Equation \ref{eq:statistical} can be used to evaluate the number of required events necessary to meet the desired statistical precision.  We used a polarization of 50\% for the J-PARC muonium beam and 97\% for the decay-in-flight beam at Fermilab.

\begin{table}
    \centering
\caption{Comparison of various parameters for the Fermilab and J-PARC \gm\ Experiments}\label{tb:comparison}
\begin{tabular}{lcc}
  \hline
  Parameter & Fermilab E989 & J-PARC E24 \\ \hline
  Statistical goal & 100\,ppb & 400\,ppb  \\
  Magnetic field & 1.45\,T & 3.0\,T \\
  Radius & 711\,cm & 33.3\,cm \\
  Cyclotron period & 149.1\,ns & 7.4\,ns \\
  Precession frequency, \wa\ & 1.43\,MHz & 2.96\,MHz \\
  Lifetime, $\gamma\tau_\mu$ & $64.4\,\mu$s & $6.6\,\mu$s \\
  Typical asymmetry, $A$ & 0.4 & 0.4 \\
  Beam polarization & 0.97 & 0.50 \\
  Events in final fit & $1.5 \times 10^{11}$ & $8.1 \times 10^{11}$ \\
  \hline
\end{tabular}
\end{table}

%% file: MuonEDM_final.tex
\subsection{Electric Dipole Moment}
\label{ssc:electricdipole}
A permanent electric dipole moment (EDM) of a particle or fundamental system violates the discrete symmetries of parity (P) and time-reversal (T).
Because quantum field theories are CPT invariant, a T-violation observation leads to CP violation (CPV). Finding a new source of CPV is a major quest in atomic, nuclear and particle physics because of its implications in any resolution of the baryon-antibaryon asymmetry problem: $(n_B - n_{\bar B})/n_\gamma = 6 \times 10^{-10}$ excess baryons per photon in the present universe, where $n_B + n_{\bar B} = 0$ is assumed at the Big Bang.  The three Sakharov conditions required to arrive at an excess of baryons are: 1) at least one $B$-number violating process; 2) a source of C- and CP-violation; and 3) interactions that occur outside of thermal equilibrium.  The challenging task of knitting together these ingredients into a quantitatively complete explanation is not yet complete and the reader is referred to discussions of this fascinating topic; see \cite{Dine:2003ax,Morrissey:2012db}.  One important fact, is that the CP violation that occurs in the lone phase in the CKM mixing matrix is insufficient by many orders of magnitude to explain a mechanism for baryogenesis.  Consequently, a new source of CPV is required.  Vigorous experimental efforts are underway at quark-flavor factories, in neutrino oscillation experiments, and in EDM searches using atoms, molecules and neutrons.

We begin with a caveat.  In a hadronic system, an EDM can be accommodated {\em within} the SM owing to the CP-violating $\Theta_{{\rm QCD}}$ term.  The non-observation of a neutron EDM with $d_n < 10^{-26} e \cdot$cm, implies $\Theta_{{\rm QCD}} < 10^{-10}$. The SM-allowed term appears to be very finely tuned---the strong CP problem---if it is finite at all.  The situation can be completely resolved if the physical axion exists; the Peccei-Quinn mechanism allows $\Theta_{{\rm QCD}} = 0$ at the price of a new particle~\cite{Agashe:2014kda}.  An axion of the right mass is also highly motivated to fulfil the role of the missing dark matter particle and current searches by the ADMX collaboration will soon largely explore that parameter space completely.\cite{Agashe:2014kda}

New sources of CPV are not unexpected in popular extensions of the standard model, such as supersymmetry, and two-Higgs doublet models (see Refs. and discussion in \cite{Morrissey:2012db}).  Current experimental limits are listed in Table~\ref{tb:EDMtable}.  The non-observations are beginning to have severe consequences on all of the models.  For example, they already severely constrain many supersymmetric CP-violating phases.  In the simple lepton sector, the electron imbedded in a polar molecule is probed to very impressive limits.  However, the fundamental source of any observed CPV in this kind of a system can have multiple interpretations as to its origin~\cite{Chupp:2014gka}. Is it singularly from the electron? The muon, on the other hand, is the {\em only} fundamental particle that can be directly tested and therefore easiest to interpret.  It is also the only 2nd-generation particle being probed, which can have important implications in certain BSM models.  In general, the sensitivity to new physics is expected to scale linearly owing to the mass term in the denominator of the dipole moment definition. Current $d_e$ limits are then nearly 7 orders of magnitude more sensitive to new physics than the muon.

However, there are BSM scenarios in which non-linear scaling occurs.  Babu et al.\cite{Babu:2000cz} presented an argument for $(m_\mu/m_e)^3$ scaling, a $\sim 10^7$ fold enhancement, meaning a next-generation muon EDM search would be competitive to the electron. This analysis deserves an update based on new information from lepton-flavor-changing tau decay limits, the improved electron EDM, and direct LHC bounds, but it remains an intriguing consideration.

More recently,  Hiller et al.~\cite{Hiller:2010ib} proposed a supersymmetric model with CP violation from lepton flavor violation that can achieve rather large values for $d_\mu$---as large as $10^{-22} e \cdot$\,cm---in the extremes of the model parameter space.  The practical range is constrained by the limits on the flavor-mixing decay $BR(\tau \rightarrow \mu\gamma) < 10^{-8}$, which will be improved with Belle\,II running in the future; lower BR's there imply smaller $d_\mu$.

The searches for SUSY at the colliders primarily focus on $R$-parity conserving models.  Considering $R$-parity-violating (RPV) models opens up the parameter space considerably~\cite{Yamanaka:2014nba}.  In such an analysis, it is observed that the muon is unique from the other systems being probed and limits as high as $d_\mu \sim 10^{-24} e \cdot$\,cm can be imagined.  While these very different examples do predict relatively ``large'' values for $d_\mu$, the relevant range sets a challenging experimental goal for muon enthusiasts as we discuss below.
\begin{table}
  \centering
  \caption{Selected EDM limits for the electron, Hg atom, neutron and muon.}\label{tb:EDMtable}
    \begin{tabular}{cclc}
    \hline
    Type & System & EDM Limit (e-cm) & Ref. \\
    \hline
    Paramagnetic & YbF & $d_e = (-2.4\pm5.9)\times10^{-28}$ & \cite{Hudson:2011zz}\\
    Paramagnetic & ThO & $d_e = (-2.1\pm4.5)\times10^{-29}$ & \cite{Baron:2013eja}\\
    Diamagnetic & $^{199}$Hg & $d_A = (0.5\pm1.5)\times10^{-29}$ & \cite{Griffith:2009zz}\\
    Nucleon &  Neutron & $d_n = (0.2\pm1.7)\times10^{-26}$ & \cite{Baker:2006ts}\\
    Lepton &  Muon & $d_\mu = (-0.1\pm0.9)\times10^{-19}$ & \cite{Bennett:2008dy}\\
    \hline
  \end{tabular}
\end{table}

\subsubsection{Experimental Considerations}
A typical atomic, molecular, or neutron EDM experiment involves a measurement of the difference in the spin precession frequency of a system subject to parallel and, alternatively,  antiparallel magnetic and electric fields. In practice, it is vital to work with the strongest possible electric fields, which are often found in the interior of atoms or molecules, where they greatly exceed laboratory capabilities.  For a relativistic muon circulating in a plane orthogonal to a pure dipole magnetic field, the situation is quite different.  Here, the muon will feel a transverse induced motional electric field $\vec E_m \propto \vec \beta  \times \vec B$.  For the \gm\ storage rings, where $\gamma = 29.3$, the electric field strength is nearly 13\,GV/m!

To understand the measurement concept, we rewrite Eqn.~\ref{eq:omega} in the absence of an external electric field.  While J-PARC will not use one at all, the focussing electric field at Fermilab has $(E/c \ll B)$ and the $\gamma$ is selected to eliminate the affect on the precession of the spin. Thus, we have simply
\begin{equation}
   \vec \omega_{net} = \vec \omega_a + \vec \omega_{EDM} = -{q \over m }\left[ a_{\mu} \vec B +
   \frac{\eta}{2} ( \vec \beta  \times \vec B) \right].
   \label{eq:omega-noE}
\end{equation}

\begin{figure}
\begin{centering}
  \includegraphics[width=.5\columnwidth]{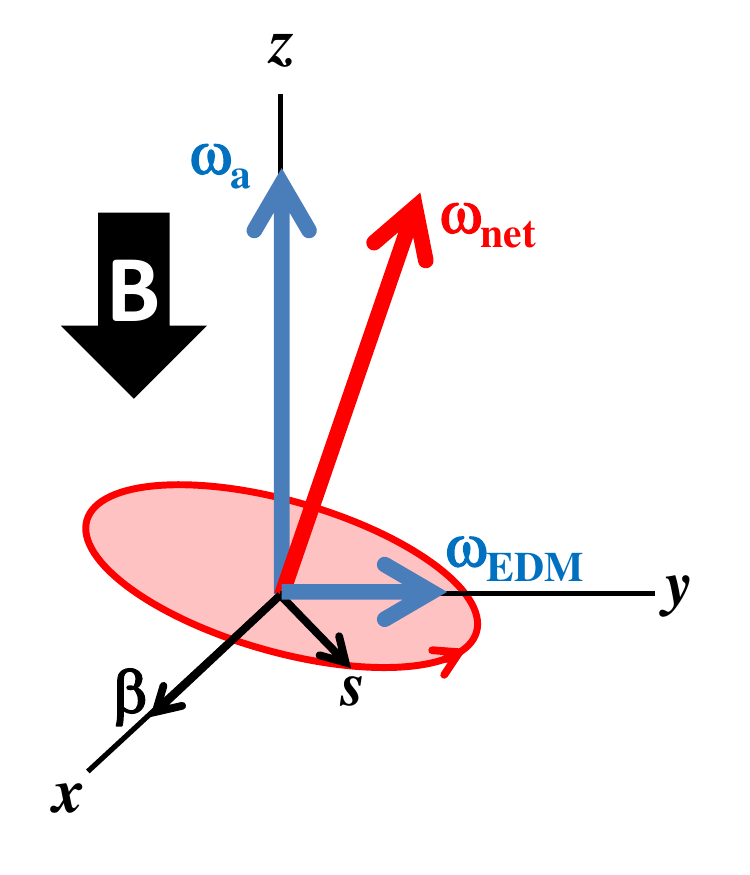}
  \caption{\label{fg:muonEDM} The net precession frequency is a vector sum of \wa\, caused by the anomalous magnet moment and $\omega_{EDM}$ caused by a possible permanent electric dipole moment. For a $\mu^+$ in the $\hat x$ direction, its spin would rotate counterclockwise in the $x-y$ plane in the absence of an EDM.  The $\beta \times \vec B$ term in Eqn.~\ref{eq:omega} points along the $\hat y$ axis, orthogonal to \wa.  This tips the precession plane as shown.  Note,  $|\omega_{EDM}| \ll |\wa|$ in practice.  It is exaggerated in the figure for clarity.
    }
\end{centering}
\end{figure}

The precession orientations for the magnetic and electric moments are orthogonal, as illustrated in Fig.~\ref{fg:muonEDM}.  For a non-zero EDM, the precession plane would be tilted inward toward the center of the cyclotron orbit by the very small angle
\begin{equation}
\delta = \tan^{-1}\left(\frac{\omega_{EDM}}{\wa}\right) = \tan^{-1}\left(\frac{\eta \beta}{2\amu}\right)
\end{equation}
and the observed precessional frequency would be $\omega_{tot} = \sqrt{\omega_{EDM}^2 + \wa^2}$. The key for the experimentalist is that the tilt of the plane is the signal.

Because the muon spin reverses every half period, the direction in which it is tipped also reverses. The consequence is that the observable will be an up/down modulation of the decay particles that is out-of-phase with the spin orientation by a factor of $\pi/2$.
In practice, the most precise method of determining the tilt is a measurement of the average slope of the decays---upward vs downward---vs. time.  The trajectories can be precisely determined using a set of tracking chambers.

The BNL experiment found $d_\mu = (-0.1\pm0.9)\times10^{-19}\,e\cdot$cm using a limited subset of the data and measured at just 1 of the 24 detector stations~\cite{Bennett:2008dy}.
Both new \gm\ experiments will be sensitive to an EDM at a level close to $10^{-21} e\cdot$cm, a major improvement. In each case, they will rely on the up/down slope asymmetry using trackers.  Here, the J-PARC experiment, which is an all-tracker detector, should have a greater overall acceptance compared to the FNAL experiment, which will feature at first only 3 tracker stations in the 24 discrete detector positions.

In the parasitic method described above, the rapid precession of the magnetic moment  reverses the upward and downward tipping of the spin owing to a possible EDM on every cycle. This leaves at most a very faint modulating signal at the frequency $\omega_{net} \approx \wa$.  In contrast, dedicated storage-ring measurements of EDMs have been proposed by Farley et al.~\cite{PhysRevLett.93.052001} using the ``frozen spin'' technique. In this method, the \wa\ precession is set to zero in crossed vertical magnetic and radial electric fields for a proper selection of field magnitudes and muon momentum. If $\eta \neq 0$, the particle spin will gradually tip out of plane (upward or downward) precessing about the radial electric field, greatly enhancing the signal.  A compact muon EDM experiment designed on this principle has been suggested by Adelmann et al~\cite{Adelmann:2010zz}.

The net spin precession in Eqn.~\ref{eq:omega}, when $\eta = 0$, is frozen when
\begin{equation}
   -{q \over m }\left[ \amu \vec B - \left( \amu - {1 \over \gamma^2 - 1}\right)
   \frac{\vec \beta  \times \vec E_r}{c}    \right] = 0,
   \label{eq:omega-frozen}
\end{equation}
which occurs when the external radial electric field $E_r$ has the magnitude
\begin{equation}
E_r = \frac{\amu B c \beta}{1-(1+\amu)\beta^2)} \approx \amu B c \beta \gamma^2.
\label{eq:E-frozen}
\end{equation}
Using the realistic set of parameters---$E_r = 640$\,kV/m, $p_\mu = 125$\,MeV/$c$, $B = 1\,$T---the authors of \cite{Adelmann:2010zz} predict a sensitivity of $\delta d_\mu \approx 7 \times 10^{-23} e \cdot$\,cm with 1 year of running at PSI on an available muon beamline. The challenges of injection into this unique and compact device, ($R_{cyclotron}$ = 42\,cm), and the potential systematics that can lead to spin tipping from effects unrelated to an EDM are discussed in their paper, and also in Ref.~\cite{PhysRevLett.93.052001}.  While remaining many orders of magnitude behind the linearly-scaled $d_e$ established limits, it is still an idea worth keeping alive. If we are to fully understand any non-zero EDM, we will need many probes to decouple the various interpretations from fundamental CPV sources.  This small-scale experiment would serve well as a general demonstration of the storage-ring based EDM proposals that have been extended to focus on the deuteron and the proton, with promises in those cases of very impressive limits.




%% file: Muonium_final.tex

\section{Muonium hyperfine structure}
\label{sc:muonium}

\subsection{Experimental approaches to muonium spectroscopy}

Muonium ($\mu^+ e^-$) is the electromagnetic bound state
of a positive muon and a negative
electron. It is a purely-leptonic, hydrogen-like atom
that unlike either ordinary hydrogen or muonic hydrogen
is completely free  from the complications associated with the
proton's finite size and its electromagnetic sub-structure.

\begin{figure}
\begin{centering}
  \includegraphics[width=1.0\columnwidth]{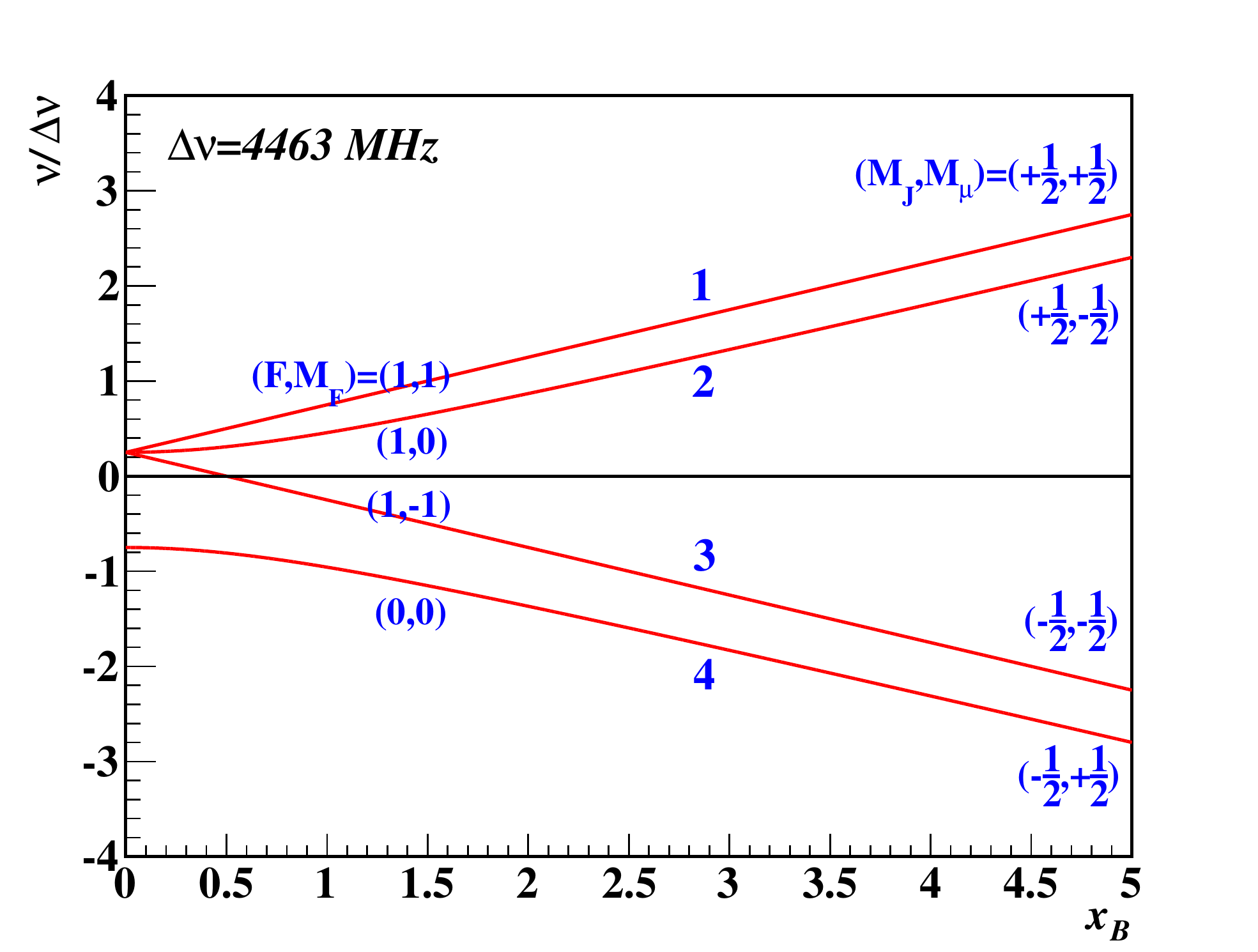}
  \caption{ \label{fg:BreitRabiDiagram} Breit-Rabi diagram of
the energy levels of the muonium $1S$ hyperfine states
versus the field strength parameter
$x_B = (g_J \mu^e_B + g'_{\mu} \mu^{\mu}_B) H / h \Delta \nu$ where
$g_J$, $g'_{\mu}$ are the electron and muon gyromagnetic ratios in
muonium, $\mu^e_B$, $\mu^{\mu}_B$ are the electron and muon Bohr magnetons,
$\Delta \nu$ is the hyperfine interval, and H is the field strength.
The labels 1...4 denote the  two Zeeman transitions ($\nu_{12}$ and $\nu_{34}$)
that are measured in the LAMPF and MuSEUM hyperfine structure
experiments.}
\end{centering}
\end{figure}

Muons and electrons are spin-1/2 particles
and consequently the muonium $1S$ ground state has a
hyperfine structure that comprises a spin $F = 1$ triplet state
with three magnetic substates $M_F = -1, 0, +1$ and a spin $F = 0$
singlet state with one magnetic substate $M_F = 0$.
The interaction
between the magnetic dipole moments
of the muon and the electron
causes an energy splitting
between the  $F = 0,1$ hyperfine states
of about 18~$\mu$eV (4.5~GHz).
As shown in Fig.\ \ref{fg:BreitRabiDiagram}---in the presence
of a static magnetic field---the Zeeman effect causes a further splitting
of the hyperfine states, with the Breit-Rabi equation describing the
energy levels versus field strength (for example
see Refs.\ \cite{Cohen:1977,Feynman:1963uxa}).
In the weak-field limit the magnetic interaction
between the muon-electron magnetic moments dominates
and the aforementioned $( F, M_F )$ are good quantum numbers.
In the strong-field limit the magnetic interaction
with the applied magnetic field dominates
and the electron and muon spin projections $( M_J, M_{\mu} )$
are good quantum numbers.

Muonium was identified in 1960
by the groups of Hughes at Columbia
and Telegdi at Chicago \cite{Hughes:1960zz,PhysRevA.1.595}.
The experiments detected the characteristic signal of muonium precession
in a weak field by recording the high-energy positrons
emitted from polarized muon stops in argon gas.\footnote{In such
``weak field'' experiments,
using polarized muons 
and unpolarized electrons, 
the different ($F, M_{F}$)-states
are populated in proportions
$(1,+1) = 1/2$, $(1,0) = 1/4$,
$(1,-1) = 0$, and $(0,0) = 1/4$.
Such experiments thereby observe
the precession of the
muonium atoms formed in the $( F, M_{F}) = (1, +1)$ hyperfine state.}
In these circumstances
the muonium atoms
were directly formed
in their 1S ground state
by electron capture
from argon atoms.
Since this work,
the formation of muonium has been observed in many other materials,
as well as produced in vacuum through thermal emission
from hot metal foils and fine silica powders.

After the discovery of muonium
much attention was focused on the hyperfine structure
of its 1S ground state.
The first estimates
of the hyperfine splitting
were derived from measurements
of the muonium polarization
versus the applied field.\footnote{The muonium polarization
is a function of the comparative sizes
of the interaction energy of the applied field
and the hyperfine splitting of the 1S ground state.}
Soon afterwards precision experiments
involving microwave resonance techniques
were used to induce transitions between different hyperfine states
and thereby determine the hyperfine structure.
The hyperfine transitions were detected
through the associated muon spin-flip
and the corresponding change
in the decay positron angular distribution.

In low-field resonance experiments,
the transition frequency directly determines
the hyperfine interval, $\Delta \nu$.
In high-field resonance experiments,
the determination of the hyperfine interval
from the measurement of a single transition frequency
requires the use of the Breit-Rabi equation
and knowledge of the muon magnetic moments.
This issue motivated the development by DeVoe {\it et al.}\ \cite{DeVoe:1971ph}
of a so-called ``double-resonance'' technique
involving the high-field measurement of
two hyperfine transition frequencies using
a single microwave cavity.
The technique
allows the concurrent determination
of both the hyperfine interval
and the muon magnetic moment.

More recently, muonium hyperfine spectroscopy
experiments have employed setups that alternate
between two microwave fields
to obtain the two hyperfine frequencies.
They yield the best determinations of
both the muon-to-proton magnetic moment ratio  $\mu_{\mu} / \mu_{p}$
and the muon-to-electron mass ratio  $m_{\mu} / m_{e}$---two
fundamental constants of great importance
to precision spectroscopy of muonic atoms.
The ratio $\mu_{\mu} / \mu_{p}$
is also crucial to the extraction
of the muon anomalous magnetic moment, $a_\mu$,
in the muon \gm\ experiments (see Sec.~\ref{sc:muondipole}).


Note that the hyperfine interval
is both measurable and also calculable with
extraordinary precision (see Ref.\ \cite{Mohr:2012tt} for
details of theoretical calculations). With input
of other fundamental constants---most importantly
the fine structure constant and the
Rydberg constant---the comparison between measured and calculated values
of the hyperfine interval is
considered a definitive test
of QED theory in bound-state systems.\footnote{Although
not discussed in detail here,
the $1S$-$2S$ interval in muonium is
also measured. The experiment \cite{Meyer:1999cx}---utilizing
Doppler-free, two-photon, pulse laser spectroscopy---yielded
a 4~ppb determination of $\Delta\nu_{1S2S}$ in good agreement
with theory. Using the combination of the results from the
hyperfine experiment and the  $1S$-$2S$ experiment, the authors
obtained a verification of charge equality between muons and
electrons of $1 + q_{\mu^+}/q_{e^-} = ( -1.1 \pm 2.1 ) \times 10^{-9}$.}



\subsection{LAMPF hyperfine structure experiment}

\begin{figure}
\begin{centering}
  \includegraphics[width=0.8\columnwidth]{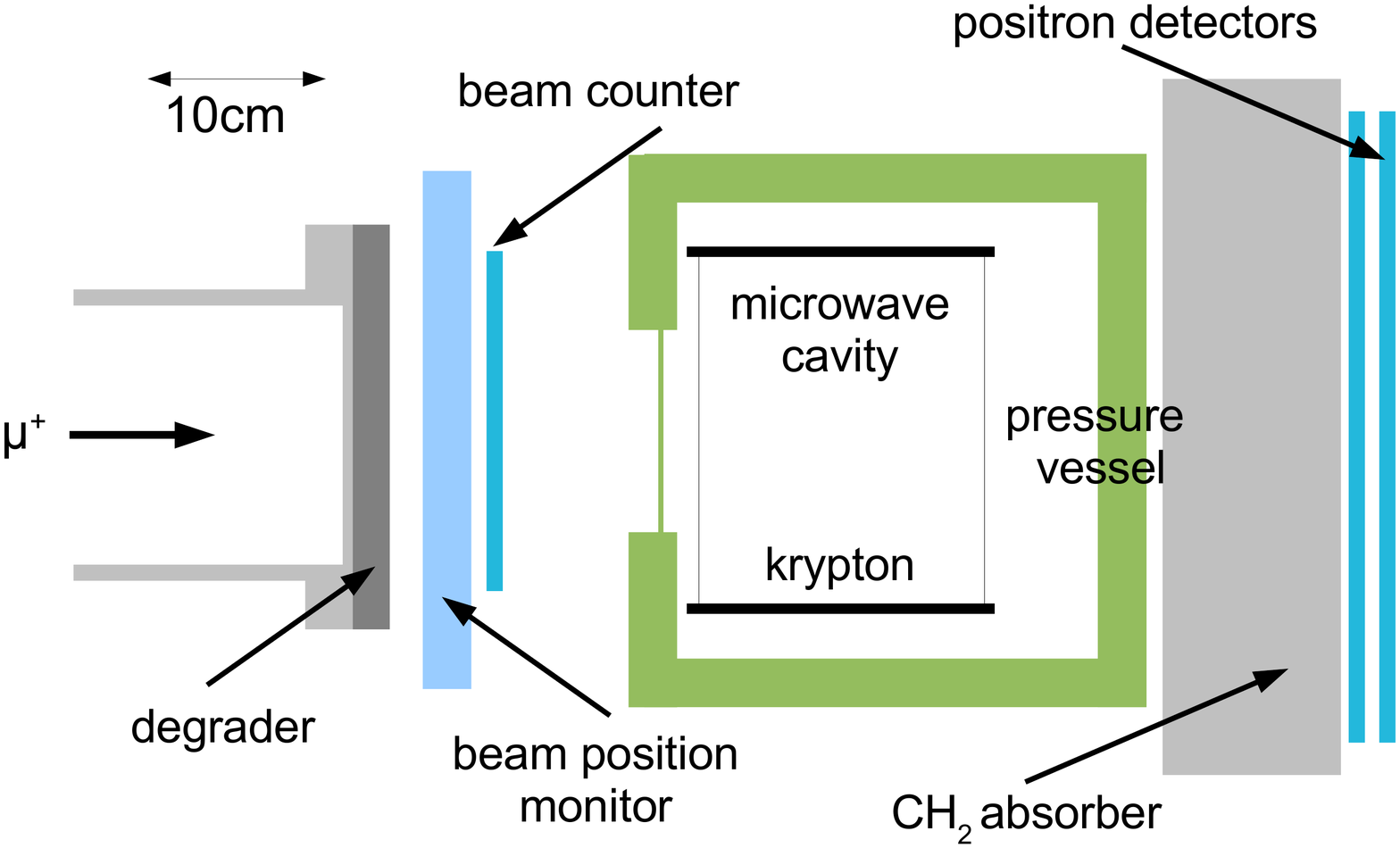}
  \caption{ \label{fg:HFSexpt} Schematic diagram of the LAMPF HFS experiment
\cite{Liu:1999iz} indicating the muon beam, beam counter and profile monitor,
krypton target, microwave cavity, pressure vessel and
downstream high-energy positron detector system.
Figure courtesy Liu {\it et al.}}
\end{centering}
\end{figure}

The most recent measurement of the muonium 1S ground state hyperfine structure
was  conducted by Liu {\it et al.}\ \cite{Liu:1999iz} at LAMPF.
The experiment measured the frequencies
of the two high-field, muon spin-flip transitions
$( M_J$, $M_{\mu} )$ = $( +1/2, +1/2) \leftrightarrow ( +1/2, -1/2) $
and $( -1/2, -1/2) \leftrightarrow ( -1/2, +1/2)$,
denoted respectively as $\nu_{12}$ and $\nu_{34}$
in Fig.\ \ref{fg:BreitRabiDiagram}.
The experiment employed the double-resonance technique
with a novel line-narrowing approach
using a custom time-structured muon beam.
The setup---including the muon beam, gas target, microwave cavity and positron
detector---is depicted in Fig.\ \ref{fg:HFSexpt}.

The experiment used a high rate,
100\%-longitudinally polarized, surface muon beam
derived from the 120~Hz repetition-rate, 650~$\mu$s pulse-period,
LAMPF primary proton beam.
An electrostatic kicker was used
to produce a cycle of 4~$\mu$s beam-on accumulation  periods
followed by 10~$\mu$s beam-off measurement periods.
The beam extinction between accumulation periods
was roughly 99\%.

The incident muons entered a large-bore, high-uniformity, 1.7~T
magnet containing a copper microwave cavity filled with pure krypton gas.
Muon stops formed polarized, ground state muonium--the
$( M_J$, $M_{\mu} )$ = $( 1/2, -1/2)$  and $( -1/2, -1/2)$
states---by electron capture from krypton atoms.
A double-layered scintillator telescope recorded
the high-energy decay positrons
that were emitted downstream
of the stopping target.
A combination of plastic scintillators
and wire chambers were used for beam monitoring.

The microwave cavity was designed to resonate
at both the $\nu_{12}$ transition frequency of
1897.5\,MHz and the $\nu_{34}$ frequency of
2565.8\,MHz. NMR magnetometry using
multiple fixed and movable probes
was used to monitor the magnetic field.

When precisely tuned to $\nu_{12}$ or $\nu_{34}$
the microwave field induces muon spin-flip transitions
and thereby changes the angular distribution of decay positrons.
In earlier measurements all positrons are detected, both those
from ``fast decays'' where the muon interacts only briefly
with the microwave field and those
from ``slow decays'' where the muon interacts at length
with the microwave field.
Liu {\it et al.}\ utilized
the 4~$\mu$s beam-on, 10~$\mu$s beam-off, time structure
to only observe the decay positrons from
long-lived muonium with extended microwave interactions.
This procedure narrowed the
resonance lineshape
by roughly a factor of three.

The positron data was collected in super-cycles
of ten beam pulses with the microwave cavity alternately switched between on and off
and the microwave frequency alternately tuned to $\nu_{12}$ and $\nu_{34}$.
The positron signal for deriving the transition frequencies
was defined as
\begin{equation*}
S( \nu , H ) = ( N_{on} / N_{off} - 1)
\end{equation*}
where $N_{on}$ ($N_{off}$) represents the positron counting rates
with the microwave field on (off)  and  $H$ and $\nu$ are the
static field strength and the microwave frequency, respectively.
Near the Zeeman resonances the spin-flip transitions
caused a large increase in $S( \nu , H )$.

Resonance curves were collected by (i) sweeping the
static magnetic field at a fixed microwave frequency and
(ii) sweeping the microwave frequency at
a fixed static magnetic field.
The resonance curves $S( \nu , H )$ were then fit to lineshapes
to extract the transition frequencies.
The fitted lineshapes incorporated the
muon stopping distribution,
static field distribution,
microwave power distribution,
and positron detection efficiency.
Data were collected at two gas pressures
and extrapolated to zero pressure;
the procedure accounted for a slight shift
of the resonance frequency
due to muonium-atom collisions
and the resulting wavefunction distortions.


The measured resonance curves determined the two Zeeman frequencies to
precisions of 17-18~ppb (the statistical errors
on the $S( \nu , H )$ resonance curves  were the dominant experimental uncertainties).
Using the Breit-Rabi Eqn.\ and the values for $\nu_{12}$ and $\nu_{34}$
a hyperfine interval of
\begin{equation*}
\Delta \nu = 4~463~302~765(53)~{\rm Hz} ~~~ (12~{\rm ppb})
\end{equation*}
and a
magnetic moment ratio of
\begin{equation*}
\mu_{\mu} / \mu_p =  3.183~345~13(39) ~~~ (120~{\rm ppb})
\end{equation*}
were obtained.
The results were a three-fold improvement over the earlier experimental work.

As mentioned earlier, an improved value for the
mass ratio $m_{\mu} / m_e $ can be obtained from the measured value
for the ratio $\mu_{\mu} / \mu_p$ with the input of the
muon anomalous magnetic moment $a_{\mu}$ and the proton-to-electron
magnetic moment ratio $\mu_p / \mu_B$.
The precise measurement of $\mu_{\mu} / \mu_p$ thus yielded a precise
determination of the mass ratio
\begin{equation*}
m_{\mu} / m_e  =  206.768~277(24) ~~~ (120~{\rm ppb})
\end{equation*}
The hyperfine structure of $1S$ muonium thus renders
the best determinations of both the magnetic moment ratio $\mu_{\mu} / \mu_p$
and the lepton mass ratio $m_{\mu} / m_e$.

The current status of theoretical calculations of $\Delta \nu$
is given in Ref.\ \cite{Mohr:2012tt}. 
Although the largest theoretical uncertainties arise from recoil correction terms,
the overall uncertainty in the $\Delta \nu$ prediction
is the aforementioned experimental knowledge of the mass ratio $m_{\mu} / m_e$.\footnote{The
corrections to the hyperfine splitting from hadronic vacuum polarization and $Z^0$ exchange
are much smaller than the uncertainty arising from the mass ratio $m_{\mu} / m_e$.}
The corresponding calculated and measured values of the hyperfine interval
are in good agreement within their respective uncertainties
of 272~Hz (61~ppb) and 53~Hz (12~ppb). This agreement is
considered an important verification of
bound-state QED calculations; the precision
being much greater than analogous comparisons in ordinary hydrogen and muonic hydrogen.

Alternatively---by equating the theoretical expression
and measured value for the hyperfine interval $\Delta \nu$ and regarding
$m_{\mu} / m_e$ as a free parameter---an indirect determination of the
muon-to-electron mass ratio and the muon-to-proton magnetic moment
ratio
\begin{equation}
m_{\mu} / m_e = 206.768~2843(52) ~~~ [{\rm 25~ppb}]
\end{equation}
\begin{equation}
 \mu_{\mu} / \mu_p = 3.183~345~107 (84) ~~~ [{\rm 26~ppb}]
\end{equation}
can be obtained \cite{Mohr:2012tt}. The result for $\mu_{\mu} / \mu_e$ is
important in the determination of the muon anomalous
magnetic moment $a_{\mu}$ (see Sec.\ \ref{ssc:magneticmoment} and Ref.\ \cite{Jungmann:2004sa}).


\subsection{MuSEUM hyperfine structure experiment}
\label{sc:museum}
\begin{figure}
\begin{centering}
  \includegraphics[width=1.0\columnwidth]{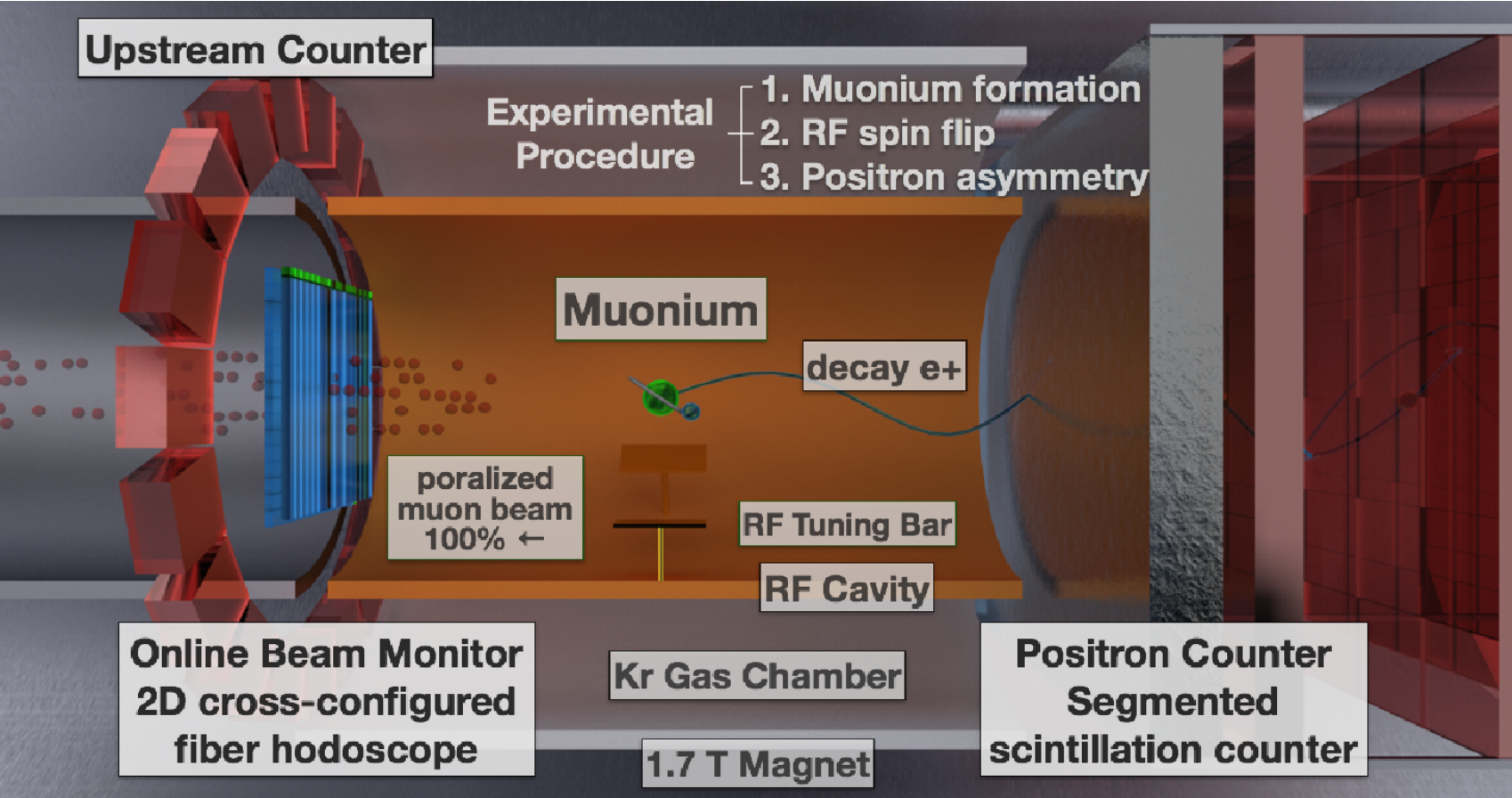}
  \caption{ \label{fg:MuSEUM} Schematic diagram of the J-PARC
muonium experiment showing the pulsed muon beam, beam profile monitor,
krypton gas chamber and microwave cavity, and the absorber
and segmented detector for high energy positron detection.
Figure courtesy MuSEUM collaboration.}
\end{centering}
\end{figure}

An improved hyperfine spectroscopy experiment (MuSEUM)
is under development at J-PARC \cite{Tanaka:2014mja}.
The experiment will employ
the same basic approach as the LAMPF HFS experiment
with the measurement of the same Zeeman frequencies
in a high magnetic field via microwave excitation
of muon spin-flip transitions. The experimental setup is
shown in Fig.\ \ref{fg:MuSEUM}.

As mentioned the LAMPF experiment was limited by statistical uncertainties.
The J-PARC  experiment plans  to increase by one order-of-magnitude the muon beam intensity
and increase by two orders-of-magnitude the recorded decay positrons.
To achieve these goals the experiment will combine
the J-PARC H-line beamline designed for rates
of $1 \times 10^8$~$\mu^+$/s and
a high-rate decay-positron detection system.

The MuSEUM setup will incorporate
a low-mass, high-rate, 2D-imaging, scintillating fiber hodoscope
to enable pulse-by-pulse measurement of beam profiles.
The experiment will use a 1.7~T superconducting magnet with 1~ppm
homogeneity and 50~ppb NMR magnetometry.
A longer microwave cavity will increase the stopping fraction
and reduce the gas pressure correction.
A thin scintillator viewed by an image intensifier and CCD camera
will determine the muon beam profile.
A finely segmented, high rate, scintillator array
with SiPM readout will provide the measurement
of the downstream-going, high energy, decay positrons.


During 2014 the MuSEUM collaboration have conducted tests
of detector sub-systems and performed measurements of beam properties.
The beginning of data taking is anticipated for 2015.

%% file: MuonProtonRadius_final.tex
\section{Muonic Lamb shift}
\label{sc:mup}

At first blush, one might wonder why we include a discussion of the proton charge radius in a muon physics review.  Indeed, the inclusion makes sense here because it is a precision muon experiment that stirred things up with a result now known as the ``Proton Radius Puzzle''~\cite{Pohl:2013yb}.  The CREMA collaboration at PSI reported a very precise measurement of the muonic hydrogen Lamb shift $L_{{\rm 1S}}$; i.e., the 2S$_{1/2} \rightarrow$ 2P$_{1/2}$ energy level difference in the $\mu p$ atom. The S energy levels are sensitive to the proton finite size, owing to their spherical symmetry and consequent overlap with the distributed charge distribution.

The motivation for the experiment was based on the need for a more precise determination of the proton charge radius $r_p$.  Along with the Rydberg constant $R_\infty$, it is one of the two required inputs to calculate hydrogen energy levels using QED, where the S-state energy is given approximately by
\begin{equation}
E(n{\rm S}) \simeq \frac{R_\infty}{n^2} + \frac{L_{{\rm 1S}}}{n^3},
\end{equation}
with $n$ the usual principal quantum number. The connection to $r_p$ can be obtained from the energy shift of an S-state level by
\begin{equation}
\label{eq:lambshift}
\Delta E = \frac{2}{3}\pi \alpha \mid \Psi_{\rm S}(0) \mid^2 r_p^2.
\end{equation}
with $\Psi_{\rm S}(0)$ the electron wavefunction at the origin.
The experimental situation in hydrogen spectroscopy had achieved a precision great enough such that improved knowledge of $r_p$ in computing the finite-size effect was limiting~\cite{Pohl:2013yb,Carlson:2015jba}.

In muonic-hydrogen, the Bohr radius is $\sim186$ times smaller than the corresponding one in ordinary hydrogen, which implies a greater overlap with the nucleus by a factor proportional to the cube of the radii, or $6.4\times 10^6$.  Thus, the sensitivity to the finite-size effect is greatly enhanced.  The logical sequence to an improved QED test is as follows:
\begin{enumerate}
\item Measure the energy levels precisely in the $\mu p$ system.
\item Extract the proton charge radius $r_p$ from the shift in S-state levels relative to the essentially unaffected P states.
\item Use the improved knowledge of $r_p$ to better compare the measured $e p$ atomic energy levels to QED predictions.
\end{enumerate}
At least, that was the idea.

However, the results of the $\mu p$ measurements indicated a 4\% smaller charge radius, with a 0.6\% uncertainty, compared to what had been commonly assumed.  The standard methods had been low-energy $e-p$ scattering and ordinary hydrogen spectroscopy, where the finite size effect enters and one might {\em assume}  QED to obtain the finite-size level shift, rather than {\em test} QED by externally knowing the finite size.  The discrepancy between the muonic and electronic methods is significant---7 standard deviations---such that any hope of simply using the independent $r_p$ extracted from muonic atoms is problematic. Consider the numeric values. The previous CODATA recommended value was $r_p = 0.8775(51)$\,fm~\cite{Mohr:2012tt}, while the muon result alone gives $r_p = 0.84087(39)$\,fm~\cite{Pohl:2010zza,Antognini:1900ns}; note the 13 times smaller uncertainty in the muon measurement.

Besides the statistical significance, it has been generally agreed that the method employed in the muonic hydrogen measurement is nearly irrefutable; it involves laser spectroscopy with accurately calibrated absorption lines.  The checking and double checking of the extrapolation of $r_p$ from $\Delta E_{{\rm 2S-2P}}$ has revealed no error, or even much wiggle room of uncertainty.  Perhaps the muon behaves differently than an electron.  Perhaps it has a more complex sensitivity to the proton charge distribution, which would totally violate our earlier assertions of lepton universality. These ideas and other exotic suggestions have been discussed, and often dismissed or ruled out, in a vast literature that is summarized in Refs.~\cite{Pohl:2013yb,Carlson:2015jba}.

If there is no problem with the muon measurement, and not finding a credible exotic origin for the difference, then possibly there is a problem with {\em both} ordinary hydrogen spectroscopy {\em and} the low-energy elastic electron-proton scattering measurements. These latter two methods agree with one another on $r_p$, albeit with much larger error bars than the muonic Lamb shift measurement.  Nevertheless, investigations have been raised about each method in searching for a resolution to the puzzle.

\subsection{Proton radius from electron scattering and hydrogen spectroscopy}
While we refrain from a deep departure into $e-p$ scattering formalism, the basic assumptions follow; more complete reviews with details, history, and uncertainty discussions exist~\cite{Pohl:2013yb,Carlson:2015jba}.  The relativistic electron-proton scattering cross section $d\sigma/d\Omega$ is usually expressed in terms of a combination of the proton electric and magnetic form factors, $G_E$ and $G_M$.  These both depend separately on $Q^2$, the four momentum transfer squared. One can map out the cross section over a series of energies and angles, and extract $G_E(Q^2)$ and $G_M(Q^2)$ using a so-called Rosenbluth separation~\cite{Rosenbluth:1950yq}.  This workhorse technique should serve well here, although one of the major findings in the JLab program has been the deviation of the ratio of $G_E/G_M$ vs. $Q^2$ at high $Q^2 \geq 3$\,GeV$^2$ from the Rosenbluth method compared to a modern recoil polarization technique. Generally, the discrepancy is attributed to the non-inclusion of important two-photon exchanges in the Rosenbluth method that turn out to be very important there, but do not impact the recoil method~\cite{Adikaram:2014ykv,Milner:2013daa}.  We mention this only to illustrate that surprises can lurk in extrapolations that seem otherwise straightforward.

The definition of the charge radius of the proton from $e-p$ scattering is
\begin{equation}
r_p^2 \equiv  -6 \frac{dG_E}{dQ^2}\mid_{Q^2 = 0}.
\end{equation}
One must extrapolate the $G_E$ vs. $Q^2$ results to the unmeasurable intercept at zero momentum transfer. Important to the procedure is the application of radiative corrections, which must be under control at the sub-percent level in the overall error.  In the most recent results from Mainz~\cite{Bernauer:2010wm}, the A1 collaboration obtains the electron scattering (ES) proton charge radius:  $r_p ({\rm ES}) = 0.879(8)$\,fm.

Turning next to ordinary hydrogen spectroscopy, one recalls that only the S orbitals are affected by the proton finite size because they overlap with the charge distribution of the proton, see Eq.~\ref{eq:lambshift}. This means any transitions to S levels that are measured provide input to $r_p$.  Carlson nicely explains the problem of correlations in the data: {\em the high precision value of the Rydberg constant is obtained from the very same atomic energy level experiments that measure the proton radius}~\cite{Carlson:2015jba}.  Percent level measurements of several transitions yield results that are sensitive to the finite size, but are by no means competitive to the precise measurements in the muon system.  Collectively the charge radius from hydrogen spectroscopy (HS) is: $r_p ({\rm HS}) = 0.8758(77)$\,fm~\cite{Mohr:2012tt}.

\subsection{Muonic Lamb shift experiment}
\begin{figure}
  \includegraphics[width=\columnwidth]{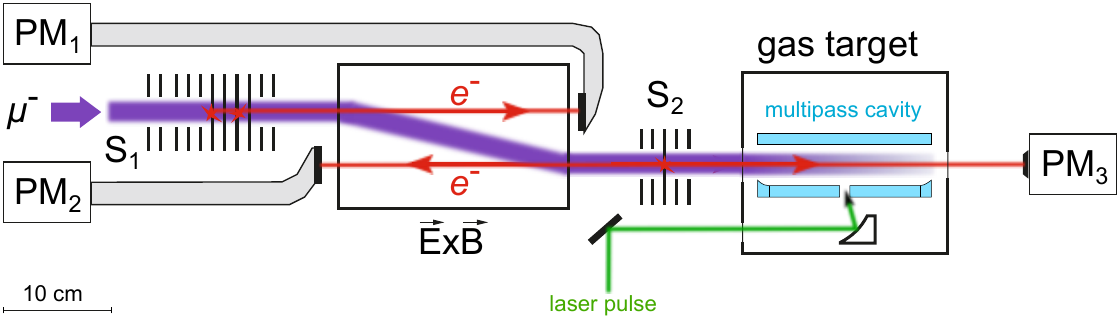}
  \caption{\label{fg:CREMA}
The CREMA experimental layout.  The low-energy muon beam enters from the left and is detected by the emission of electrons in a thin set of carbon foils that then excite scintillators.  The muon passes through a $\vec E \times \vec B$ velocity selector to separate it from the electrons.  It then enters an ultra-low pressure hydrogen gas target. The PMT signals indicate the $t_0$ time and signal the laser to fire. The 5\,ns long pulse is reflected in the multipass cavity giving an effective overlap time with the muonic atoms of about 75\,ns. If the delayed 2P state is formed, large-area avalanche photodiodes placed near the target are positioned to record the characteristic 1.9\,keV x\,rays.  Figure courtesy R.\,Pohl.}
\end{figure}

To study the muonic-hydrogen energy levels, one first needs to form the atom itself.  In practice, this was done in the CREMA experiment by beginning with a very low-energy negative muon beam at PSI, derived from pions spiraling toward the center of a so-called cyclotron trap.  About 30\% of the pions decay in the $\pi^- \rightarrow \mu^- + \bar{\nu}_\mu$ process, emitting negative muons that are further decelerated to keV energies by multiply passing through a metalized foil at -20\,kV.  Once confined, they can leave along the axis of the trap into a toroidal magnetic momentum filter and then onward to a target region that contains pure hydrogen gas at low pressure, see Fig.~\ref{fg:CREMA}.  A muon entering the target will have passed through thin stacks of carbon foils.  Low-energy electrons can be ejected, and subsequently detected in thin scintillators viewed by photomultiplier tubes.  The signals so produced serve as a time marker of the arriving muon. The muon slowed in the process by normal $dE/dx$ energy loss has a probability to atomically capture in an excited $n$ level---typically 14---and begin an ordinary cascade to the ground state.  For optimized conditions of pressure, $\approx 1\%$ of the atoms will result in a muon cascade terminating in the 2S metastable state.  This is the required starting point for a Lamb-shift measurement, and it was by no means an obvious or easy situation to prepare.  For the CREMA conditions of a 1\,mbar pressure hydrogen target, the metastable 2S state will have a lifetime of about $1\,\mu$s before undergoing collisional de-excitation~\cite{Pohl:2013yb}. It is in that short window that the rest of the experiment must then work.

\begin{figure}
  \includegraphics[width=\columnwidth]{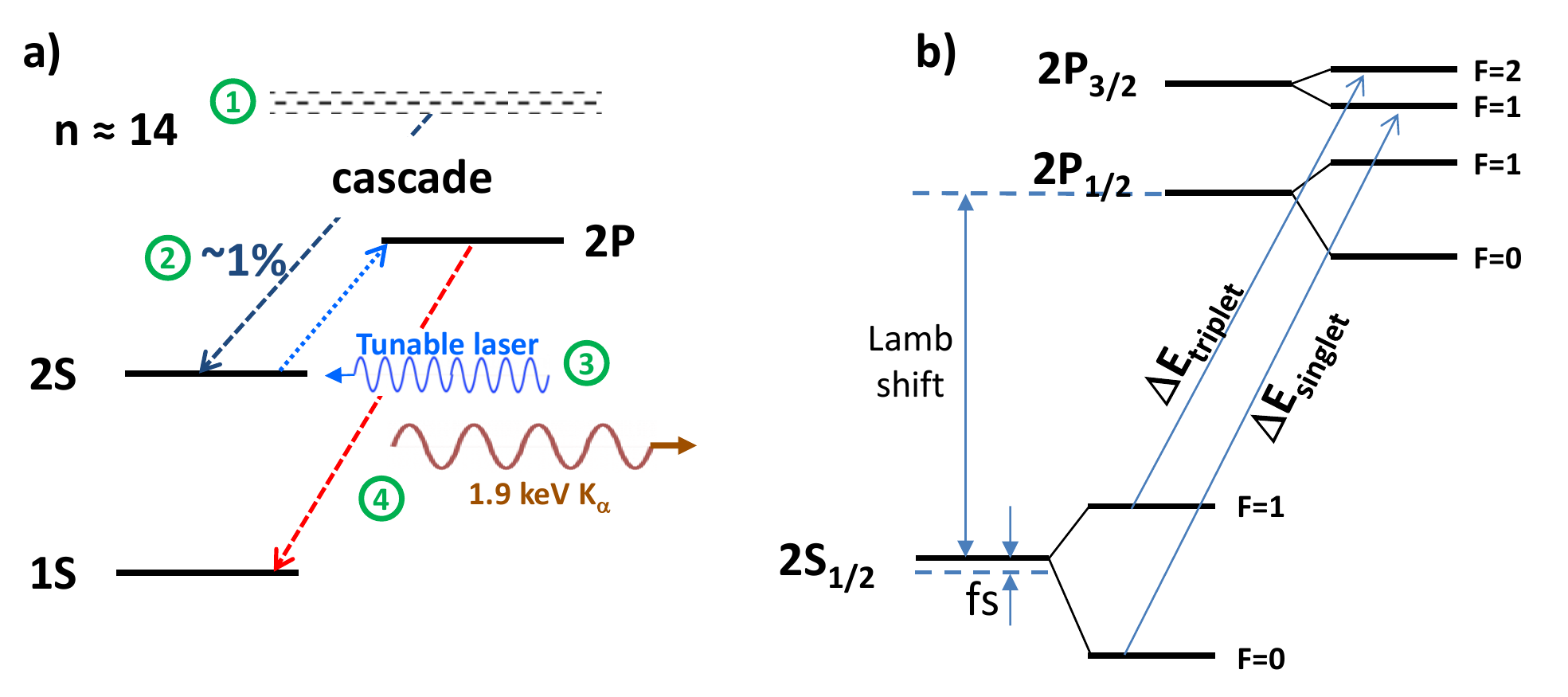}
  \caption{\label{fg:Cascade}
a)  The four step schematic of how the experiment works. 1) The muon arrives in the target and captures in a high $n$ shell. 2) Muon cascade, with $\sim 1\%$ stopping in the metastable 2S state.  3) A tunable laser is fired $0.9\,\mu$s after each muon arrives in the target. 4) The 2S $\rightarrow$ 2P transition is induced if the laser frequency is on resonance, and the 2P state will decay rapidly to the 1S ground state, emitting a $K_\alpha$ x\,ray that can be detected, serving to tag the process.  b) Blowup of the level scheme showing the fine-structure splitting of the P levels and the hyperfine structure of S and P levels.  The two measured and published transitions are indicated.}
\end{figure}

Once formed, the next challenge is to induce the 2S $\rightarrow$ 2P level transition by shining a properly tuned laser on the atoms.  If the laser frequency corresponds to $\Delta E_{{\rm 2S-2P}}$, the muon will transition to the appropriate 2P level, where it will then rapidly cascade to the 1S level, emitting a characteristic 1.9\,keV x\,ray.
These four steps are shown in the left panel of Fig.~\ref{fg:Cascade}. The idea of the experiment is to form the atom and then fire the laser $0.9\,\mu$s later, simultaneously opening up a 75\,ns wide gate to observe whether a 1.9\,keV x\,ray has been emitted.  The laser frequency is tuned in discrete steps to scan the anticipated energy region around the expected 2S $\rightarrow$ 2P transition energy.  The recorded number of x\,rays corresponding to the 2P $\rightarrow$ 1S x\,ray vs. time will have two features.  First, a large prompt peak will be present in 99\% of the cases owing to the atom following a normal cascade to the 1S ground state. A second, delayed peak, at $\sim 100$ times lower intensity will be present {\em if and only if} the laser frequency is correctly tuned to the 2S $\rightarrow$ 2P resonance; otherwise, the second peak is absent.  By repeating the experiment and slowly sweeping the laser frequency, one obtains a very precise measurement of $\Delta E_{{\rm 2S-2P}}$.

In practice, the P orbitals are split by atomic fine structure into the P$_{1/2}$ and P$_{3/2}$ levels, and both S and P levels are further split by the hyperfine interaction.  The level scheme is shown on the right panel of Fig.~\ref{fg:Cascade}, where the Lamb shift is defined as shown, and the effect of the finite-size (fs) correction on the S shell is highlighted.  The two measured transition energies, $\Delta E_T$ (triplet) and  $\Delta E_S$ (singlet) can be used in linear combinations that, together with known QED-based corrections not dependent on $r_p$, yield both the Lamb shift and the hyperfine splitting.  In turn, one can not only deduce the discussed charge radius, but also the Zemach radius, $r_Z$, which is essentially a measure of the magnetic distribution inside the proton.  While the extracted $r_p$ is significantly more precise than other methods, $r_Z$ is not. Its value of $r_Z = 1.082(37)$\,fm is compatible with other methods and its uncertainty is many times larger. As quoted above, we obtain here from muon spectroscopy (MS): $r_p (MS) = 0.84087(39)$\,fm~\cite{Pohl:2010zza,Antognini:1900ns}

\subsection{Present and future work involving muons}

The CREMA collaboration has completed, but not yet published, additional Lamb-shift measurements in deuterium and helium.  Three transitions in deuterium have been accurately measured, with preliminary interpretations that suggest consistency with the muonic hydrogen result; however, theoretical work continues so it is premature to draw firm conclusions. They have also completed Lamb-shift measurements on $^3$He and $^4$He systems in 2014, which are undergoing analysis.

Suppose we do not find a ready solution.  Then what? A proposal by the MUSE collaboration~\cite{Gilman:2013eiv} to help add a different data set to this discussion has been approved at PSI and the design of the experiment is in progress. The idea is to measure low-energy $\mu - p$ scattering and, parasitically, $e - p$ scattering in a large-angle, open-geometry spectrometer.  This novel idea would then complete the set of determinations of $r_p$ from electron and muon scattering and electron and muon spectroscopy. If the muon is somehow different from the electron, or if scattering and spectroscopy methods differ, then data of this type can help resolve the puzzle.

At the present time, the proposed plan aims at uncertainties roughly at the level of current $e - p$ scattering measurements. Systematic uncertainties should be controlled by using both $\mu^+$ and $\mu^-$ beams, with their accompanying $e^+$ and $e^-$ components.  Incoming $\pi, \mu$ and $e$ particles in the secondary beamline are tagged on an event-by-event basis using time-of-flight with respect to the RF structure.  Because of the muon's higher mass compared to the electron, radiative corrections in $\mu - p$ scattering are relatively small and therefore under better control than for electrons, an advantage here for the muon experiment. Of course the difficulty of carrying out a precision scattering experiment utilizing a secondary decay beam introduces a complexity that will certainly challenge the team.

%% file: MuonCapture_final.tex
\section{Nuclear muon capture}
\label{sc:muz}

\subsection{Basic features of muon capture}

Muon capture and beta decay are
close cousins. Both processes
\begin{equation*}
1.~~~ \mu^- ~+~ [ Z , A ] \rightarrow [ Z-1 , A ] ~+~ \nu_{\mu}
\end{equation*}
\begin{equation*}
2.~~~ [ Z , A ] \rightarrow [ Z\pm1 , A ] ~+~ e^{\mp} ~+~ \nu_{e}
\end{equation*}
involve transmutations of protons into neutrons
or vice-versa through a semi-leptonic weak interaction
with a precisely-known leptonic current.
However the energy release in the two reactions---set by the muon mass in the muon process
and the nuclear mass difference in the beta process---are quite different.
Consequently, the two processes can illuminate different features
of the underlying weak nucleonic and nuclear interactions.

Muon capture occurs from the 1S ground state of a muonic atom;
such atoms are formed when muons are stopped in matter.
In light nuclei, where the overlap of the muon orbital with the
nuclear volume is relatively small, the capture rate
is small compared to muon decay.
In heavier nuclei, where the overlap 
is much larger, the capture rate is large compared to muon decay.\footnote{The
Z$^4$-law for muon capture \cite{Wheeler:1949zz}
states the capture rate is proportional to the fourth power of
the effective charge $Z$ of the atomic nucleus.}
For muonic hydrogen and muonic deuterium about 0.1\% of muons
undergo capture.



When muonic atoms are formed on non-zero spin nuclei ($I \neq 0$)
the 1S ground state is
split into two distinct hyperfine states
with total angular momenta of
$F = I + 1/2$ and $F = I - 1/2$.
The possibility of muon capture
from the singlet and triplet hyperfine states in hydrogen,
and doublet and quartet  hyperfine states in deuterium,
is responsible for engendering
capture on hydrogen isotopes
with additional richness and additional complexity.

The first observation of muon capture on hydrogen
was reported in 1962 by Hildebrand \cite{PhysRevLett.8.34} using a hydrogen bubble chamber.
This experiment---together with
other early experiments using bubble chambers
and liquid scintillator
detectors---were important as evidence in support
of the nascent $V$-$A$ theory of the weak interaction \cite{Sudarshan:1958vf,Feynman:1958ty}.
Many other muon capture experiments
have since been conducted.

Muon capture on hydrogen isotopes offers a unique opportunity
to determine elusive components of weak nuclear interactions---the
induced pseudoscalar coupling of the proton and the two-body axial current
of the deuteron. Herein we describe the recent progress
in precision $\mu p$ and $\mu d$ experiments
that address such elementary features of weak interactions.

\subsection{Muon capture on hydrogen, $\mu p \rightarrow n \nu$ }

Muon capture is generally treated
as a current-current weak interaction
where the leptonic current and the nucleonic current
have the familiar parity violating $V$-$A$ structures.
The leptonic current has the
simple \makebox{$\gamma_\mu (1-\gamma_5)$} form.
The nucleonic current---because of
its quark constituents and their strong interactions---is more complicated.

The most general form of the nucleonic $V$-$A$ current is
\begin{eqnarray}
\label{e: general current}
&+g_v \gamma^\mu+\frac{i g_m}{2 m_N}\sigma^{\mu\nu}
q_{\nu}+\frac{g_s}{m_\mu} q^\mu & \nonumber \\
&-g_a \gamma^\mu
\gamma_5-\frac{g_p}{m_\mu}q^\mu \gamma_5-\frac{i g_t}{2
m_N}\sigma^{\mu\nu} q_\nu \gamma_5 &
\end{eqnarray}
where $\gamma^{\mu}$ are the Dirac matrices,
$q = p_n - p_p$ is the momentum transfer
and $m_\mu$ and $m_N$ are the muon and nucleon mass.
The nucleonic current contains six ``coupling constants''
that are functions of the momentum transfer-squared $q^2$.
The couplings $g_v$, $g_m$, and $g_s$ are the vector, weak magnetism
and induced scalar couplings of the hadronic vector current, $V$.
The couplings $g_a$, $g_p$, and $g_t$ are the axial, induced pseudoscalar
and induced tensor couplings of the hadronic axial current, $A$.

The terms involving $g_v$, $g_a$, $g_m$ and $g_p$ are called first-class currents
while the terms involving $g_s$ and $g_t$ are called second-class currents \cite{Weinberg:1958ut}.
This distinction arises as  first-class currents and second-class currents 
have opposite transformation properties under G-parity---an operation that links the transmutation
of protons into neutrons with the transmutation of neutrons into protons.
Consequently, the second-class contributions to the leading first-class currents
only arise through G-parity breaking effects
({\it e.g.} the u-d quark mass difference and the electromagnetic corrections).
No experimental evidence for second-class currents exists
(for recent discussions see Ref.\ \cite{Hardy:2008gy} regarding $g_s$
and Ref.\ \cite{Minamisono:2011zz} regarding $g_t$).

In the conserved vector current hypothesis (CVC) \cite{Feynman:1958ty}
the weak vector current and isovector electromagnetic current
are the components of a conserved vector-isovector current.
The hypothesis relates the $g_v$ and $g_m$ terms of the weak vector current
to the charge and magnetism terms of the isovector electromagnetic current.
It predicts  a ``weak magnetism'' analogous to
magnetic effects in electromagnetic interactions
and yields $g_v = 1.0$ and $g_m = 3.706$ at $q^2 = 0$.
The roots of CVC  were the closeness (1-2\%)
of the constant $G_F$ determined in muon decay
and the constant $G_F$$g_v$ determined in beta decay.
This observation resembles the equality of
the electric charge of the electron
and the proton. Apparently, the bare weak vector charge
of the proton, like the bare electric charge
of the proton, is protected from renormalization
via emission and absorption of virtual pions,
by a conservation law \cite{GellMann:1958zz}.
Nowadays, the conserved vector-isovector current
is an integral part of our understanding
of the nucleon's quark structure.

Concerning the axial current, the axial coupling is well-determined
from measurements of neutron beta decay that yield $g_a = 1.2723 \pm 0.0023$ \cite{Agashe:2014kda}.
In contrast with $g_v$ the value of $g_a$
is (slightly) modified by strong interactions
and we speak of a partially conserved axial current
in place of an exactly conserved vector current.
This partially conserved axial current reflects
an underlying approximately conserved chiral symmetry of
strong interactions \cite{GellMann:1960np}.

The remaining term is the induced pseudoscalar coupling $g_p$;
an interaction that plays
a significant role in muon capture
but not in beta decay. For fifty years
the value of $g_p$ has been uncertain
and the predictions for $g_p$ have been untested.
The interest in $g_p$ stems from more than just its status
as the poorly-known piece of the weak nucleonic current.
A precise value \cite{Bernard:1994wn}
\begin{equation}
\label{eq:CHPTgp}
g_p = 8.44 \pm 0.23
\end{equation}
is predicted by quantum chromodynamics---a prediction that is closely connected
to spontaneous symmetry breaking in low energy QCD
and the dynamical origins of the hadronic masses.

To understand the low-energy realization of  chiral symmetry
it is helpful to consider the hypothetical limit of massless u and d quarks.
For $m_u = m_d = 0$ QCD possess an exact SU(2)$_L$$\times$SU(2)$_R$
chiral symmetry, {\it i.e.}, there are separate copies of isospin symmetry
for the left-handed quarks and the right-handed quarks.
This symmetry generates two conserved currents, a vector
current  corresponding to the sum of left-
and right-handed quark currents and a conserved axial current
corresponding to the difference of left- and
right-handed quark currents. At low energies this chiral symmetry
is spontaneously broken through QCD interactions.
As a result the hadrons acquire mass
and the pion appears as the Goldstone boson
of the broken symmetry.
Still the underlying currents remain conserved currents
and  thereby dictate a precise relation between $g_a$ and $g_p$.

Of course, up and down quarks are not exactly massless and consequently
chiral symmetry and axial current conservation are also not exact.
This small explicit breaking of chiral symmetry modifies
the relation between $g_a$ and $g_p$ but still---through
older current algebra techniques or newer
chiral perturbation theory---a precise prediction for $g_p$
results. The value of $g_p$ is thereby tied
to our modern understanding of the strong interaction
that incorporates its approximate chiral symmetry and
partial axial current conservation as well
as the dynamical origins of the hadronic masses
(for further details see Ref.\ \cite{gorringe:2002xx,Pastore:2014iga}).

\subsubsection{Muon chemistry in pure hydrogen}

\begin{figure}
\begin{centering}
  \includegraphics[width=1.0\columnwidth]{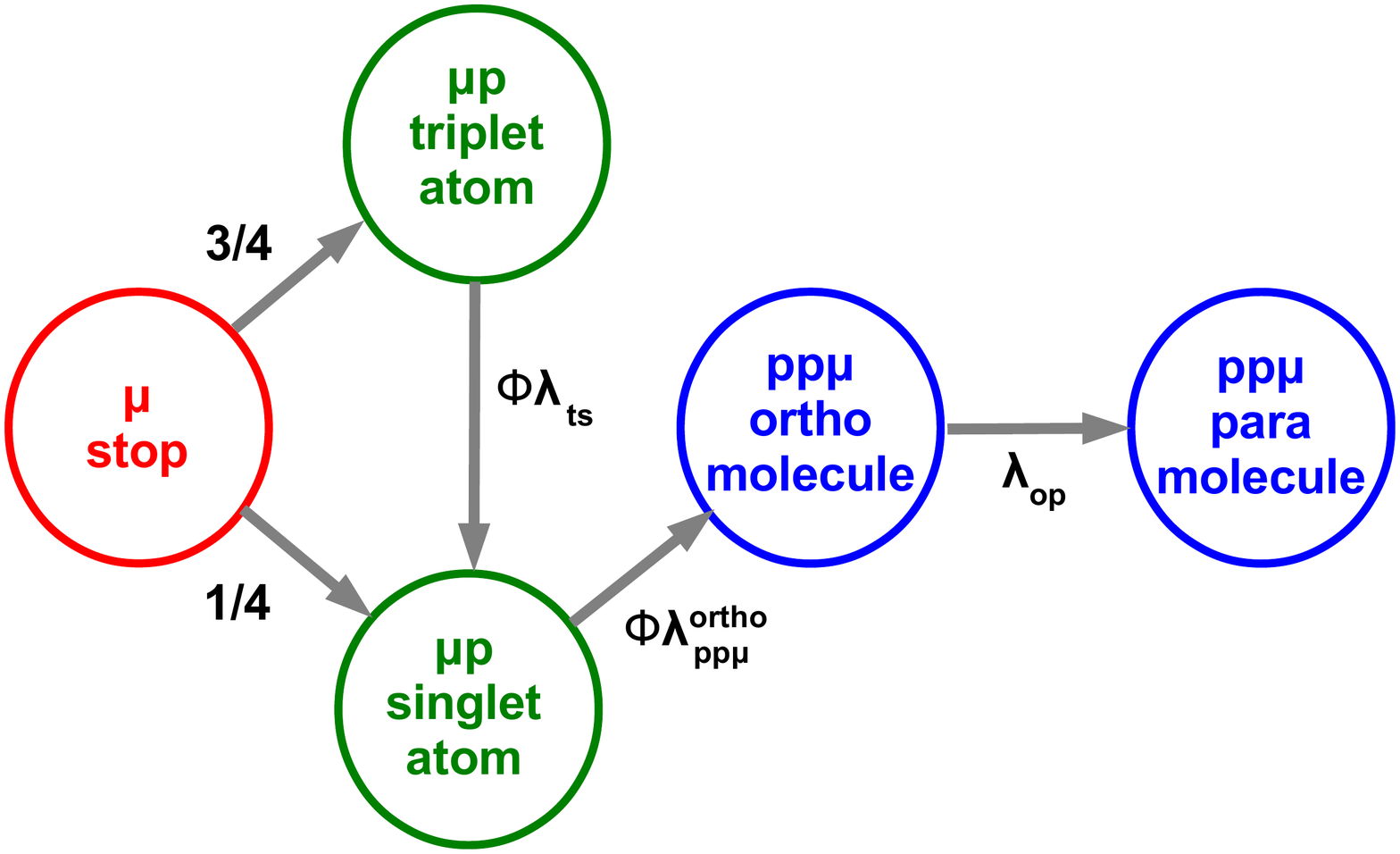}
  \caption{ \label{fg:mupchemistry} Schematic diagram showing
the important atomic and molecular states and transition rates for
muon stops in isotopically pure hydrogen. The $\mu p$ atoms
are initially formed in a statistical mix of triplet atoms (3/4)
and singlet atoms (1/4). $\phi \lambda_{ts}$ is
the density-dependent triplet-to-singlet transition rate,
$\phi \lambda^{ortho}_{pp \mu}$ is the density-dependent ortho-molecular
formation rate, and $\lambda_{op}$ is the density-independent ortho-to-para
transition rate.}
\end{centering}
\end{figure}

Although the theoretical relation
between the $\mu p \rightarrow n \nu$ capture rate
and the weak coupling constants
is quite straightforward---a complication exists.
The $\mu p$ atoms that form when
muons are stopped in hydrogen
are small and  neutral.
Consequently, they scatter off and react
with the surrounding H$_2$ molecules,
thus causing the $F = 0, 1$ hyperfine populations
to evolve with time.
This evolution depends on the thermalization
of the `hot' $\mu p$ atoms in the H$_2$ environment
as well as chemical reactions that form muonic molecules
(the bound-states of a single negative muon and two hydrogen nuclei).
A detailed knowledge of the relevant atomic and molecular
processes---as shown in Fig.\ \ref{fg:mupchemistry}---is therefore needed
to extract $g_p$ from experimental data.

%


The $\mu$p atom is initially formed in an excited state with a
principal quantum number $n \sim 14$.
The excited atom rapidly de-excites through
combinations of Auger emission, 
radiative decays 
and Coulomb de-excitation. 
On reaching the 1S ground state
the $\mu p$-atoms have kinetic energies
of typically  1~eV (for details see \cite{Badertscher:1996ka})
and a statistical population of the hyperfine states
(3/4 triplet atoms and 1/4 singlet atoms).

These energetic atoms are rapidly thermalized by
elastic  and spin-flip collisions
with the atomic nuclei of the surrounding H$_2$ molecules.
When their energies fall below the
0.18~eV $\mu p$ hyperfine splitting,
the singlet-to-triplet transitions are energetically forbidden
and  triplet-to-singlet transitions depopulate the higher-lying triplet state.
The triplet lifetime is about $0.1$~ns
in liquid H$_2$ and about $10$~ns in 10 bar H$_2$ gas.

At sufficient densities $pp\mu $ molecules form.
Like ordinary H$_2$ molecules,  there exists both para $\mu$-molecular hydrogen
with a total nuclear spin $I = 0$ and ortho $\mu$-molecular hydrogen
with a total nuclear spin $I = 1$.
The para molecule is the true ground state of the $pp \mu$ molecule.
Importantly, the two molecules  have different $\mu p$-spin decompositions;
the para-molecule being 3:1 triplet-to-singlet
and the ortho-molecule being 1:3 triplet-to-singlet.

The $p p \mu$ molecules are formed
by collisions between $\mu p$ atoms and surrounding H$_2$ molecules
via Auger emission $\mu p + H \rightarrow pp \mu + e$.
The calculated rate of the E1 Auger transition forming ortho-molecules
$\lambda^{ortho}_{pp \mu} \simeq \phi \times 1.8 \times 10^6 s^{-1}$
is much faster than the E0 Auger transition forming
para-molecules $\lambda^{para}_{pp \mu} \simeq \phi \times 0.75
\times 10^4 s^{-1}$ ($\phi$ is the H$_2$ density
normalized to the liquid H$_2$ density $\phi_o = 4.25 \times 106{22}$~atoms/cm$^3$).
A recent measurement \cite{Andreev:2015evt} of the total rate of molecular formation
found $\lambda_{p \mu p} =  \phi \times 2.01 \pm 0.07 \times 10^6 s^{-1}$
in reasonable agreement with theoretical predictions.

Naively, the $\Delta I = 0$ selection rule for E1 transitions
forbids the decay of ortho molecules to para molecules.
However, as recognized by Weinberg, through
relativistic effects that mix ortho- and para-states
the ortho $pp \mu$ molecules do gradually decay into para $pp \mu$ molecules.
The decay rate was computed by Bakalov {\it et al.}\ \cite{Bakalov:1980fm}
to be $\lambda_{op} = 7.1 \pm 1.2 \times 10^4 s^{-1}$
(this rate is independent of density).
Unfortunately, the two published measurements
for $\lambda_{op}$  of $(4.1 \pm 1.4) \times 10^4 s^{-1}$ \cite{Bardin:1981cq}
and $(10.4 \pm 1.4) \times 10^4 s^{-1}$ \cite{Clark:2005as},
are in significant disagreement.


%
%


\subsubsection{Experimental approaches to $\mu p$ capture}

The ``neutron approach''
to studying $\mu p \rightarrow n \nu$ capture
involves stopping muons in hydrogen
and detecting the resulting 5.2~MeV capture neutrons.
Such experiments were conducted
in liquid hydrogen and gaseous hydrogen
and achieved precisions
of roughly 10\% in the effective capture rate
for the relevant $F = 0, 1$ populations.
The neutron method was, however, limited
by the necessary determination of the neutron detection efficiencies.




The ``lifetime approach''
to studying $\mu p \rightarrow n \nu$ capture
avoids directly detecting the reaction products of muon capture.
Rather, it determines the capture rate $\Lambda$
from the difference between the disappearance rates of the $\mu p$ atom
and the positive muon, {\it i.e.}\footnote{For $\mu p$ atoms the disappearance rate $\lambda_{\mu p}$ is the sum of
the $\mu$ decay rate and the $\mu$ capture rate whereas
for positive muons the disappearance rate is simply the $\mu$ decay rate.}
\begin{equation*}
\Lambda = \lambda_{\mu p} - \lambda_{\mu^+}
\end{equation*}
where $\lambda_{\mu p} \equiv 1 / \tau_{\mu p}$ and
$\lambda_{\mu^+} \equiv 1 / \tau_{\mu^+}$ are obtained from the measured time distributions
of the decay electrons and positrons, respectively.
The experiment is difficult as the decay rate
is roughly 1000 times the capture rate and therefore the two
disappearance rates are very similar---thus
requiring extraordinarily precise $\mu p$ and $\mu^+$ lifetime measurements.
The lifetime approach was pioneered
by Bardin {\it et al.}\ \cite{Bardin:1980mi} at Saclay.

A serious concern for both approaches was muon stops
in  $Z > 1$ surrounding  materials.
The detection of capture neutrons or decay electrons
from muon stops in surrounding materials
would alter the time distribution
and distort the measured lifetime.
Therefore the target vessel, etc.,
were typically constructed from high-Z
materials so stopping muons were rapidly absorbed.
Similarly, the small size and neutrality of
$\mu p$ atoms exposes the muon to transfer
to any $Z> 1$ contaminants in the H$_2$ gas.
Again detection of capture neutrons
or decay electrons from gas contaminants
would distort the measured lifetime.




\subsubsection{MuCap experiment}

The MuCap experiment \cite{Andreev:2012fj} was conducted at PSI.
It used  a custom-built, muon-on-demand beam
to increase the sample of decay electrons
while mitigating the effects of muon pileup.
It also used an active target
to verify the stopping of muons in hydrogen
and monitor the effects of gas impurities.

The experiment was performed in H$_2$ gas
of high chemical and isotopic purity
at 10~bar pressure and room temperature.
Under these conditions
the triplet atoms
are short-lived
and the $pp \mu$  molecules
are rarely formed---thus preparing
a nearly-pure sample
of singlet atoms
and enabling an
unambiguous measurement
of the $\mu p$ singlet capture rate $\Lambda_S$.

\begin{figure}
\begin{centering}
  \includegraphics[width=0.9\columnwidth]{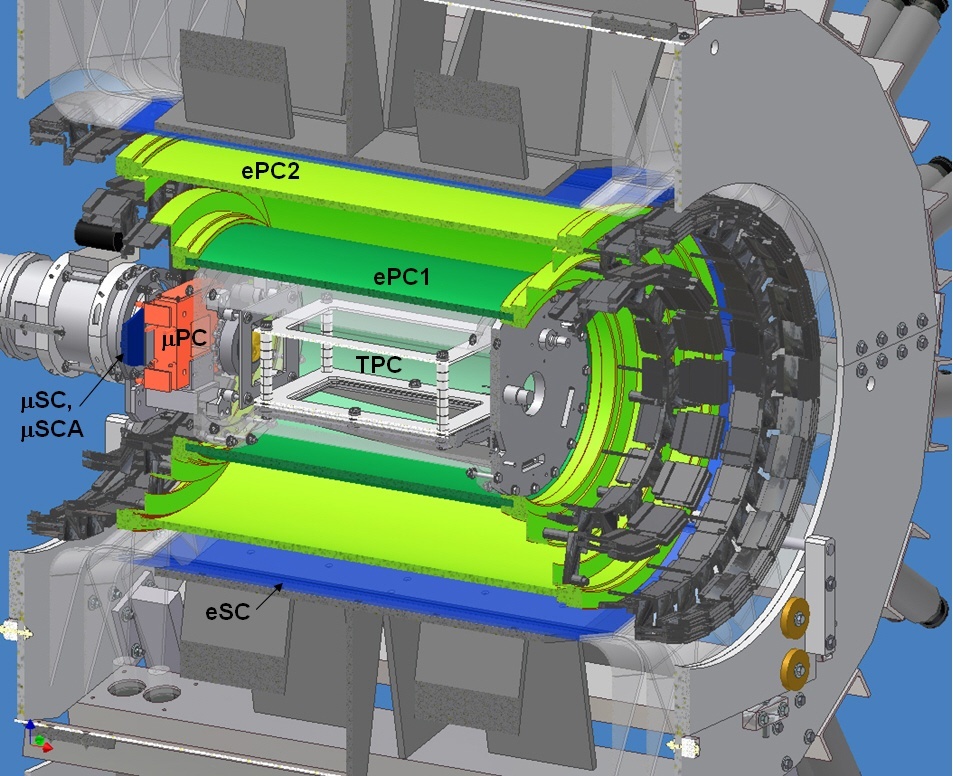}
  \caption{ \label{fg:MuCapSetup} Cutaway diagram of the MuCap experiment
indicating the
muon counters ($\mu$SC, $\mu$PC), H$_2$ time projection chamber (TPC),
and electron counters ($e$PC1, $e$PC2, eSC). The muon counters
determine the muon arrival time, the electron counters determine the
muon decay time, and the H$_2$ TPC validates the muon stopped H$_2$ gas.}
\end{centering}
\end{figure}

The experimental setup is shown in Fig.\ \ref{fg:MuCapSetup}.
A series of incoming muon detectors
that consisted of a plastic scintillator and
a planar multiwire chamber
together determined the arrival time
and provided the pile-up protection
of beam muons.
A series of outgoing electron detectors
that consisted of consecutive layers
of cylindrical multiwire chambers and
segmented plastic scintillators
together determined the times
and trajectories of the decay electron.
On identifying a muon
the upstream electrostatic kicker was turned on
and thereby the muon beam was turned off.

The incoming muons were stopped in
a hydrogen-filled time projection chamber (TPC).
The TPC comprised a 5.04~liter active volume
with a vertical drift field
and a horizontal readout plane
of perpendicular anode wires and cathode strips.
The analog signals from anode wires and cathode strips
were fed to three discriminator thresholds that
triggered on: incoming muons (denoted $E$$L$),
the Bragg peak of stopping muons (denoted $E$$H$),
and the high ionization of charged products from muon capture on gas impurities (denoted $E$$V$$H$).
The discriminator hits
were recorded by multihit TDCs.

The experiment employed custom isotope separation
and gas purification units.
Isotopically pure H$_2$ gas was prepared
from commercial, isotopically-pure hydrogen
by repeated cycles of fractional distillation.
Chemically pure H$_2$ gas was maintained
by recycling the gas through a purification system
that incorporated a cold trap and micro-porous filters.
The experiment achieved a deuterium contamination
of $<$10~ppb and a water contamination of about 9~ppb.

A crystal oscillator was used
for the timebase of the readout electronics.
The collaboration was blinded
to the exact frequency of the timebase during the
data taking and the data analysis.
Only after completing all the analyses
was the frequency revealed.


The experiment accumulated about $1.2 \times 10^{10}$ negative muon decays
from pure H$_2$ gas, $0.6 \times 10^{10}$ positive muon decays
from pure H$_2$ gas, as well as
decay electrons from impurity-doped gas that permitted the investigation
of muon transfer to gas impurities and formation rates of muonic molecules.

The decay curves  were constructed from the
measured time difference $(t_e~-~t_{\mu})$ between an incoming muon signal in the muon scintillator
and an outgoing electron signal in the electron scintillators.
The TPC was used to validate that
the  muon had stopped in
H$_2$ gas.
The TPC data showed stopping muons as a trail of $EL$ hits
that led to several $EH$ hits
at the stop location.
The algorithm for authenticating a stop
was optimized to handle the
effects of (i) the hard scattering of incident muons
from target protons into surrounding materials,
and (ii) the possible interference between
the incoming muon ionization and the outgoing
electron ionization in the TPC.


The experiment also pioneered the {\em in-situ} measurement
of gas impurities with the TPC.
Muon transfer to gas impurities and
subsequent capture on $Z > 1$ nuclei
was identified by single, delayed $E$$V$$H$ hits
at the stopping location.
From the measured rates and time distributions
of muon stops with subsequent $E$$V$$H$ hits
the necessary corrections due to N$_2$ / H$_2$O impurities
in the pure H$_2$ gas were then derived.

The measured decay curves were fit to determine the muonic hydrogen lifetime.
In principle---due to the muon kinetics and the time evolution
of the $\mu p$ spin-states---the theoretical time distribution is not exactly a
single exponential decay curve. However, in practice
a single exponential was a good fit to the time distribution
and  adequately determined the muon disappearance rate.

Two corrections were necessary to extract the
singlet capture rate $\Lambda_S$
from the difference $\lambda_{\mu p} - \lambda_{\mu^+}$
between the $\mu^{\pm}$ disappearance rates.
One correction accounted for the small population
of $pp \mu$  molecules with singlet atoms.
Another correction accounted for the slight difference in the
decay rate of a bound muon versus a free muon.
After these corrections of about $18$~s$^{-1}$ and $12$~s$^{-1}$
respectively the final result of
\begin{equation*}
\Lambda_S = 715.6 \pm 5.4 (stat) \pm 5.1 (syst)~s^{-1}
\end{equation*}
was obtained \cite{Andreev:2015evt}.
Unlike many earlier experiments, the MuCap result is
essentially free from ambiguities associated
with muonic molecule formation.


%
\begin{figure}
\begin{centering}
  \includegraphics[angle=90,width=1.0\columnwidth]{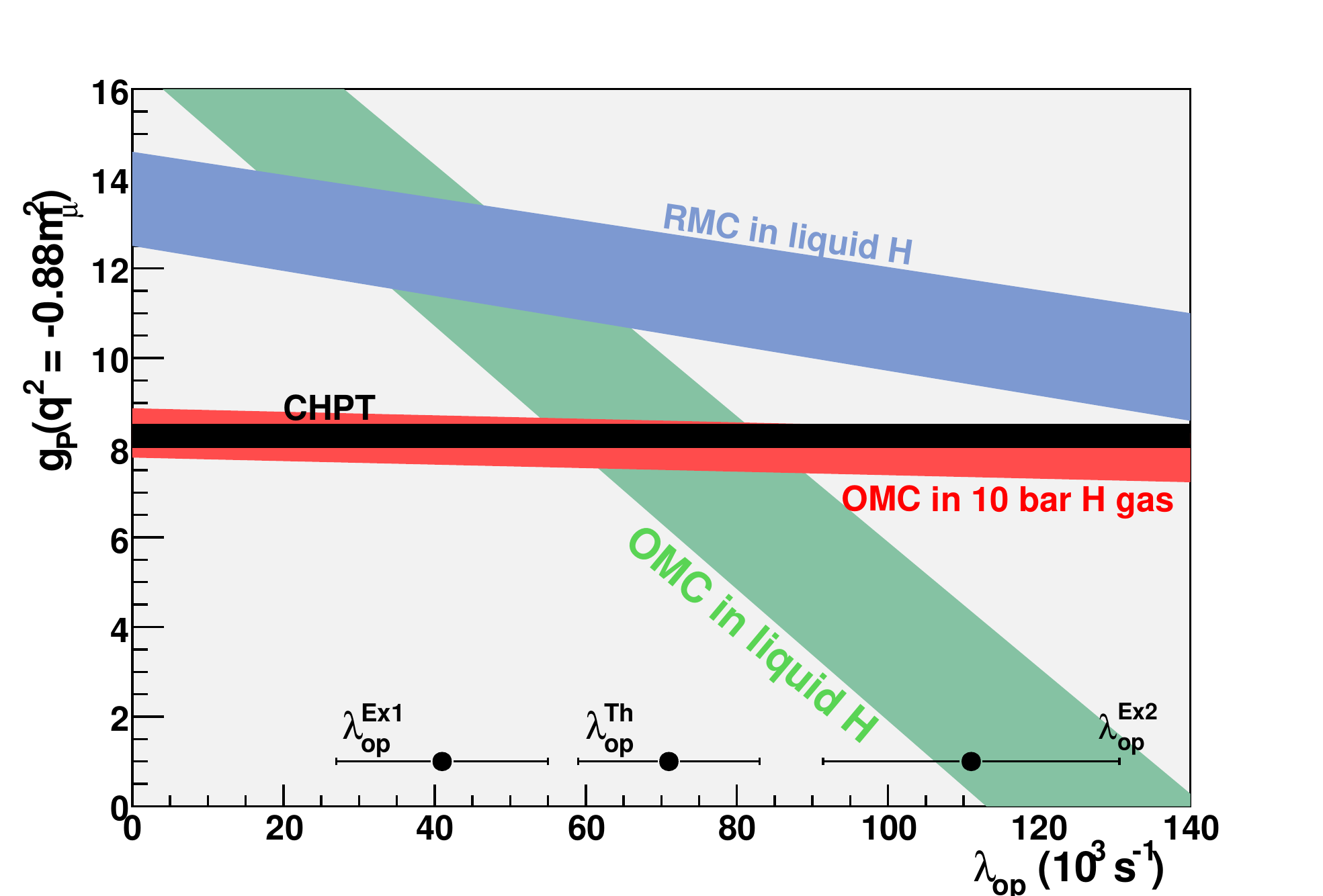}
  \caption{ \label{fg:gp_v_lambdaop} The induced pseudoscalar coupling $g_p$
versus the ortho-to-para molecular transition rate $\lambda_{op}$.
It shows the recent result of the MuCap experiment derived
from ordinary muon capture in 10~bar gas  and earlier results
of Ref.\ \cite{Bardin:1980mi} for ordinary muon capture in liquid hydrogen
and Ref.\ \cite{Wright:1998gi} for radiative muon capture in liquid hydrogen.
The data points indicate the experimental
results \cite{Bardin:1981cq,Clark:2005as} and the theoretical
calculation \cite{Bakalov:1980fm}
of the ortho-to-para transition rate. The MuCap result from
10 bar H$_2$ gas is essentially
free of ambiguities associated with $\lambda_{op}$.}
\end{centering}
\end{figure}

Using the latest theoretical calculations of $\mu p \rightarrow n \nu$ capture \cite{Bernard:1994wn}
and incorporating radiative corrections \cite{Czarnecki:2007th}, the MuCap measurement of singlet capture
thereby determines the coupling $g_p$.
Using $g_a = 1.2701 \pm 0.0025$
for the axial coupling \cite{Beringer:1900zz},
the MuCap result for $\Lambda_S$ implies
a value of
\begin{equation*}
g_p = 8.06 \pm 0.48 \pm 0.28
\end{equation*}
for the induced pseudoscalar coupling
(the uncertainties are associated
with the MuCap measurement and the
$\chi$PT calculation of the capture rate, respectively).


The MuCap result for $g_p$ is in good agreement with the
original predictions of current algebra and the modern predictions
of chiral perturbation theory (Eqn. \ref{eq:CHPTgp}).
As shown in Fig.\ \ref{fg:gp_v_lambdaop}---the result
for $g_p$ is also essentially free from the ambiquities
associated with the limited knowledge of the ortho-to-para
molecular transition rate that afflicted earlier
experiments in liquid hydrogen.
The result verifies our modern understanding
of approximate chiral symmetry
and partial axial current conservation
in QCD. Such concepts are the foundations of our understanding
of the origins of the hadron masses and the pion's significance
as the Goldstone boson of a broken symmetry.

\subsection{Muon capture on deuterium, $\mu d \rightarrow n n \nu$ }




At a basic level the atomic nucleus
is more than just an assembly of neutrons and protons.
It incorporates such
non-nucleonic degrees-of-freedoms
as virtual pions and delta particles.
With  $g_v$, $g_m$, $g_a$, and $g_p$ all well-measured,
the weak interaction offers a precise probe
for exploring such exotic constituents of nuclear matter.

Muon capture on deuterium is nature's bridge
between weak nucleonic and nuclear interactions.
As such it parallels the role
of radiative capture $n p \rightarrow d \gamma$ on hydrogen and
photo-disintegration   $\gamma d \rightarrow n p$ of deuterium
for electromagnetic processes.
The $n p \rightarrow d \gamma$ reaction
provided the first unequivocal evidence
for non-nucleonic degrees-of-freedom
in electromagnetic interactions.
These non-nucleonic effects were surprisingly large
with pion currents contributing roughly 10\%
of thermal neutron capture.

The interest in exchange currents in weak interactions
is more than theoretical.
The $\mu d \rightarrow n n \nu$ reaction
is closely related to other $A = 2$ weak processes
including $pp \rightarrow d e \nu$ thermonuclear fusion in stars
and $\nu$d interactions in heavy-water neutrino detectors.
It represents the only $A = 2$ weak interaction
that is measurable and calculable to high precision.
As such, the reaction is crucial to quantitatively understanding
the non-nucleonic contributions to weak nuclear interactions
and their influence on such processes
as  big-bang nucleosynthesis and stellar evolution
as well as ordinary and double $\beta$-decay.

The $\mu d \rightarrow n n \nu$ process
is dominantly an allowed Gamow-Teller transition
from the $^{3}$S$_{1}$ deuteron ground state
to the $^{1}$S$_{1}$ $nn$ continuum state.
Due to the $V$-$A$ structure of the weak interaction
the deuterium capture rate
from doublet ($F = 1/2$) atoms
is much larger than
quartet ($F = 3/2$) atoms.
Given our excellent knowledge
of the nucleon weak couplings
and the deuteron nucleonic wavefunction,
the major uncertainty
in $\mu d$ capture
is the poorly-known contribution
of the two-body axial current.
A precision measurement of $\mu d$ capture
can resolve this two-body current
and thereby advance
our theoretical understanding
of many weak nuclear processes.

Until fairly recently the theoretical work on $\mu$d capture
was based on phenomenological potential models of nucleon-nucleon interactions.
Using phenomenological potentials, sophisticated calculations that incorporated
detailed initial- and final-state nucleonic wavefunctions augmented
by simplified models of non-nucleonic contributions were
performed by Tatara {\it et al.}\ \cite{Tatara:1990eb},
Doi {\it et al.}\ \cite{Doi:1991ns} and Adam
{\it et al.}\ \cite{Adam:1990kf}. These calculations gave
rates $\Lambda_D$ for doublet capture of typically 390-400~$s^{-1}$.
The calculations suggested a two-body axial contribution
of roughly 5\%  that mostly originates
from delta excitation via pion exchange between nucleons.

Chiral effective field theory ($\chi$EFT) was first developed for systems of pions,
then extended to systems involving a single nucleon, and eventually applied
to few-body nuclear systems. Its development has profoundly altered
the theoretical treatment of weak interactions on few-body nuclei.
$\chi$EFT established a rigorous, unified framework
for calculations that obeys the underlying symmetries of quantum chromodynamics
while utilizing the pions and nucleons as low-energy degrees-of-freedom.
It is based on a systematic expansion in small
parameters---the momentum transfer, pion mass and nuclear binding
energy---where leading-order terms are computed and higher-order terms are neglected.
Each of the calculated terms involves a low-energy constant that must
be determined from data.

In $\chi$EFT, a single low energy constant determines
the two-body axial current contribution in weak nuclear processes
(this low energy constant is denoted as \^{d}$_R$ or  L$_{1A}$ in different
versions of effective theories).
The $\mu d \rightarrow n n \nu$ process offers an unmatched opportunity
for determining this constant to a precision comparable
to the recent caculations.

Over recent years a number of  calculations  of $\Lambda_D$
have been performed with increasing sophistication
in the $EFT$ framework.
The two most recent calculations, which consistently
treat the nuclear wavefunctions and the weak
operators, were conducted by Marcucci {\it et al.}\ \cite{Marcucci:2011jm}, yielding
399$\pm$3~$s^{-1}$, and by Adam {\it et al.}\ \cite{Adam:2011mr},
yielding 383.8-392.4~$s^{-1}$.

\subsubsection{Muon chemistry in pure deuterium}

\begin{figure}
\begin{centering}
  \includegraphics[width=1.0\columnwidth]{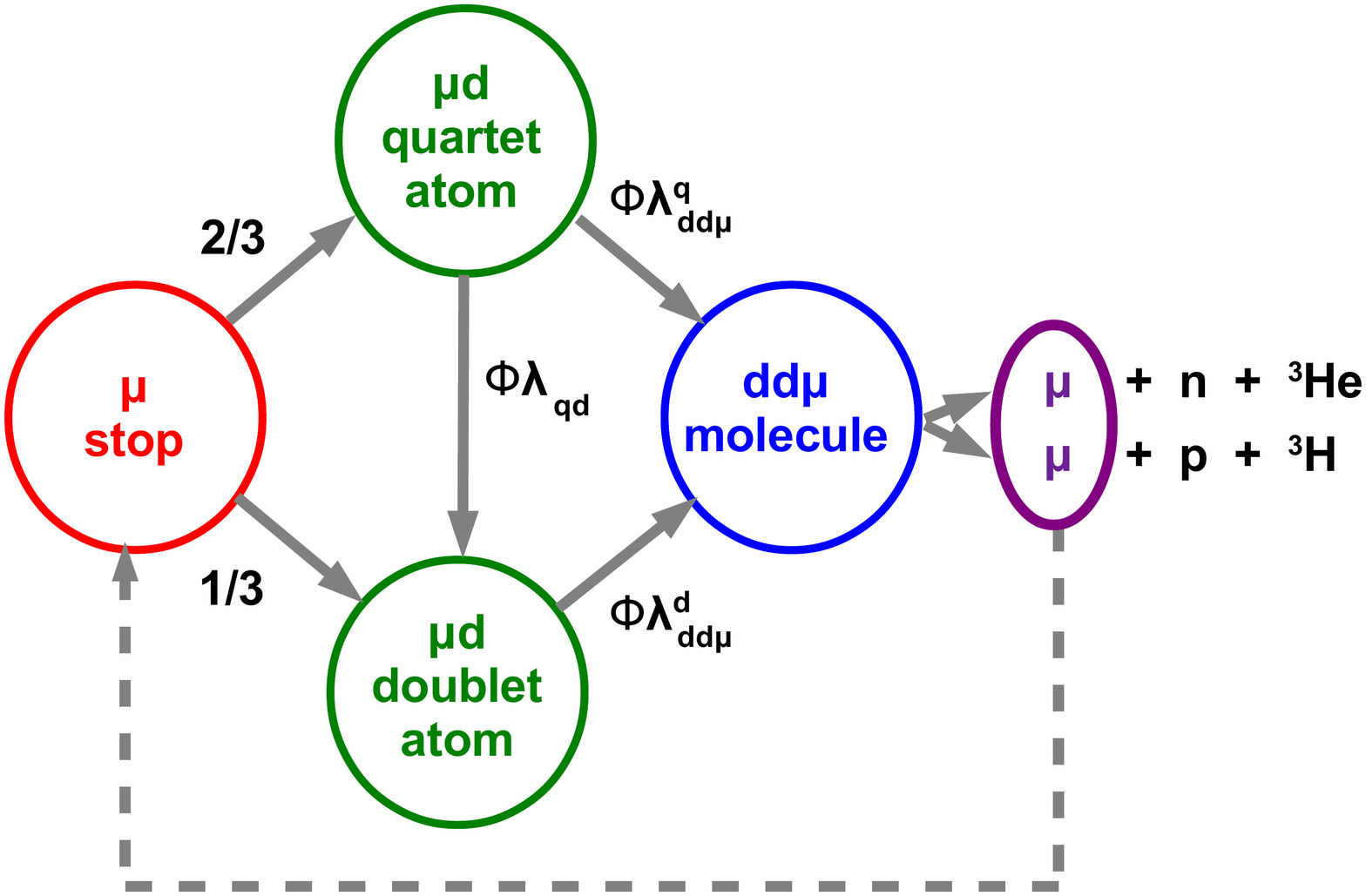}
  \caption{ \label{fg:mudchemistry} Schematic diagram showing
the important atomic and molecular states and transition rates for
muon stops in isotopically pure deuterium. The $\mu p$ atoms
are initially formed in a statistical mix of quartet atoms (2/3)
and doublet atoms (1/3). $\phi \lambda_{qd}$ is
the density-dependent quartet-to-doublet transition rate
and $\phi \lambda^{q}_{dd \mu}$ and $\phi \lambda^{d}_{dd \mu}$ are
the density-dependent molecular formation rates from the quartet and
doublet states. Also shown is the muon recycling
following $dd \mu$ molecule formation and
$\mu$-catalyzed fusion.}
\end{centering}
\end{figure}

Just as atomic and molecular processes can complicate the
interpretation of $\mu p$ capture data, such atomic and
molecular processes also complicate the interpretation
of $\mu d$ capture data. A detailed knowledge of relevant
atomic and molecular processes---as shown in Fig.\ \ref{fg:mudchemistry}---is
therefore required.

The $\mu d$ atoms are formed in
excited states that rapidly de-excite
to the 1S ground state by Auger emission,
radiative decays and Coulomb collisions, thus yielding
a statistical mix of ``hot'' doublet and quartet atoms.
In many respects the chemical reactions
of $\mu d$ atoms are very similar to $\mu p$ atoms.
Both $\mu d$ and $\mu p$ are tiny, neutral
atoms that easily penetrate the electronic clouds
of surrounding molecules to scatter off and react with
atomic nuclei.
Like $\mu p$ atoms, the $\mu d$ atoms
are thermalized by elastic
and spin-flip collisions with surrounding nuclei.
When the $\mu d$ energy falls below
the 0.043 eV hyperfine splitting,
the spin-flip collisions
then depopulate the quartet atoms
in favor of doublet atoms.
However, the cross sections
are considerably smaller for $\mu d + d$ scattering
than  $\mu p + p$ scattering and consequently
the quartet $\mu d$ atoms in D$_2$ are longer-lived
than triplet $\mu p$ atoms in H$_2$
(for details see \cite{Kammel:1984vu}).

One new feature of $\mu d$ chemistry is
temperature-dependent resonant formation of
$dd \mu$ molecules---a process by which the $dd \mu$ binding
energy is absorbed by D$_2$ vibro-rotational modes.
For example, at cryogenic temperatures,
while $dd \mu$ formation by doublet $\mu d$ atoms
involves a rather slow, non-resonant process, the $dd \mu$
formation by quartet $\mu d$ atoms involves a fast, resonant process.

Another new feature of $\mu d$ chemistry is muon catalyzed fusion \cite{Breunlich:1989vg}.
The possibility of muon catalysis of nuclear reactions
was first proposed  by Frank \cite{Frank:1947} in 1947
and later considered by  Gerstein, Sakharov and Zeldovich
in the early 1950s as a possible energy source.
Its first observation was entirely
accidental---Alvarev {\it et al.}\ \cite{Alvarez:1957un} identifying the puzzling tracks
following muon stops in bubble chambers as muons released
following catalyzed fusion.

This release of muons from $dd \mu$ molecules
is an additional dimension of $\mu d$ chemistry.
In $dd \mu$ molecules the fusion reactions  are
\begin{equation*}
1.~~~ dd \rightarrow n~^3{\rm He}
\end{equation*}
\begin{equation*}
2.~~~ dd \rightarrow p~^3{\rm H} ,
\end{equation*}
the former yielding an intense source
of mono-energetic neutrons in $\mu d$ experiments.
On forming a $dd \mu$ molecule, the
fusion reaction occurs essentially instantaneously.
The reactions generally release muons
but on occasion they will stick to the charged products
of the fusion reaction.




\subsubsection{MuSun experiment}

The $\mu$d doublet capture
in pure deuterium was previously measured
using the lifetime technique and a liquid D$_2$ target
yielding $\Lambda_d = 470 \pm 29$~s$^{-1}$ \cite{Bardin:1986}
and using the neutron technique and a gaseous D$_2$ target
yielding $\Lambda_d = 409 \pm 40$~s$^{-1}$ \cite{Cargnelli:1989}.
These experiments were conducted more than twenty five years ago.

The MuSun experiment \cite{Andreev:2010wd} is using
the lifetime technique to measure the $\mu d$ doublet capture rate $\Lambda_D$
to about 1.5\% and thereby improve by roughly five-fold the
current knowledge of two-body axial current contributions
to $A = 2$ weak nuclear processes.
The approach requires
the preparation of a nearly-pure sample of doublet atoms,
a 10 ppm measurement of the $\mu d$ atom lifetime,
and careful monitoring of isotopic and chemical impurities in deuterium.
The experiment builds on the development and the innovations
in the MuCap experiment.

\begin{figure}
\begin{centering}
  \includegraphics[width=0.9\columnwidth]{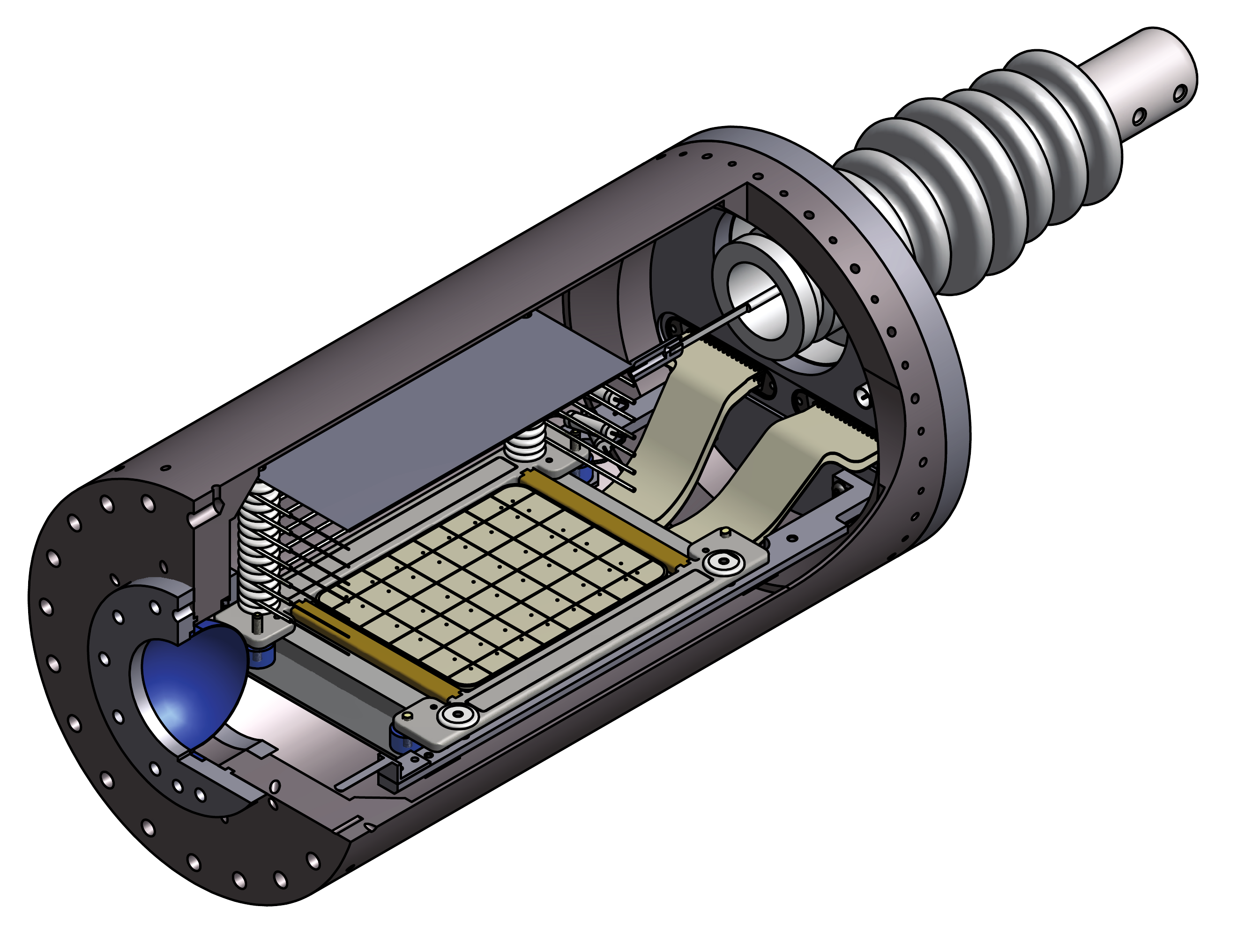}
  \caption{ \label{fg:MuSunTPC} Cutaway diagram of the MuSun low temperature, high pressure,
 D$_2$ time projection chamber. It shows the cryogenic pressure vessel,
the beryllium beam entrance window and liquid neon cooling system, as well as the horizontal cathode plane, $6 \times 8$ segmented anode plane
and field-shaping wires.}
\end{centering}
\end{figure}

As shown in Fig.\ \ref{fg:MuSunTPC}, the MuSun experiment
is using a novel cryogenic D$_2$ time projection chamber.
The temperature of 34~K and pressure of 5-6~bar
were chosen to prepare an optimal population of nearly-pure doublet atoms.
At this temperature and pressure
the quartet atoms rather quickly decay to doublet atoms
but regeneration of quartet atoms by muon recycling
following $dd \mu$ molecule formation and $dd$-fusion
is quite small.
The time projection chamber also provides the stop definition
for incoming muons and monitoring of $\mu d$ kinetics including
both the formation of muonic molecules
and the transfer to gas impurities.

%
%


The MuSun setup consists of an incoming muon counter package,
outgoing electron counter package, the cryogenic D$_2$ time projection
chamber and a liquid scintillator neutron detector array.
The muon and electron counter packages are conventional arrangements
of plastic scintillators and proportional chambers
that were originally constructed for the MuCap experiment.

The high pressure, low temperature, D$_2$ TPC works as follows.
Ionization is collected via a vertical drift field and
a $6$$\times$8 segmented, horizontal anode plane, then readout
via cryogenic pre-amplifiers and
8-bit, 25~MHz waveform digitizers.
The TPC and associated electronics
were designed for
good energy resolution ($\sim$10 keV)
to thereby enable the
clean identifcation of
fusion products
and muon capture
on gas contaminants.

As in MuCap the  muon  and electron plastic scintillators determine the time interval
between the incoming muon and the outgoing electron in order to construct
the decay curve and extract the $\mu d$ lifetime.
A stop definition that is derived from the signals in the anode pads
is used to validate the entries in the time distribution.

Processes that involve either muon transfer to chemical impurities
or molecular formation on isotopic impurities impose stringent limits
on the possible chemical and isotopic contamination of the D$_2$ gas.
Chemical impurities ({\it e.g.}\ air, water) are worrisome
due to large $\mu$d$\rightarrow$$\mu$Z transfer rates
and require concentrations of N$_2$ of $\leq$1~ppb and O$_2$ of $\leq$3~ppb.
Isotopic impurities  are worrisome due to
$pd \mu$ molecule formation and require concentrations
of ordinary hydrogen of $\leq$10~ppm.
To achieve such purities the D$_2$  gas was prepared in-situ by a
custom isotope separation unit and continuously cleaned of
chemical impurities by a custom gas recycling unit.

\begin{figure}
\begin{centering}
  \includegraphics[width=0.9\columnwidth]{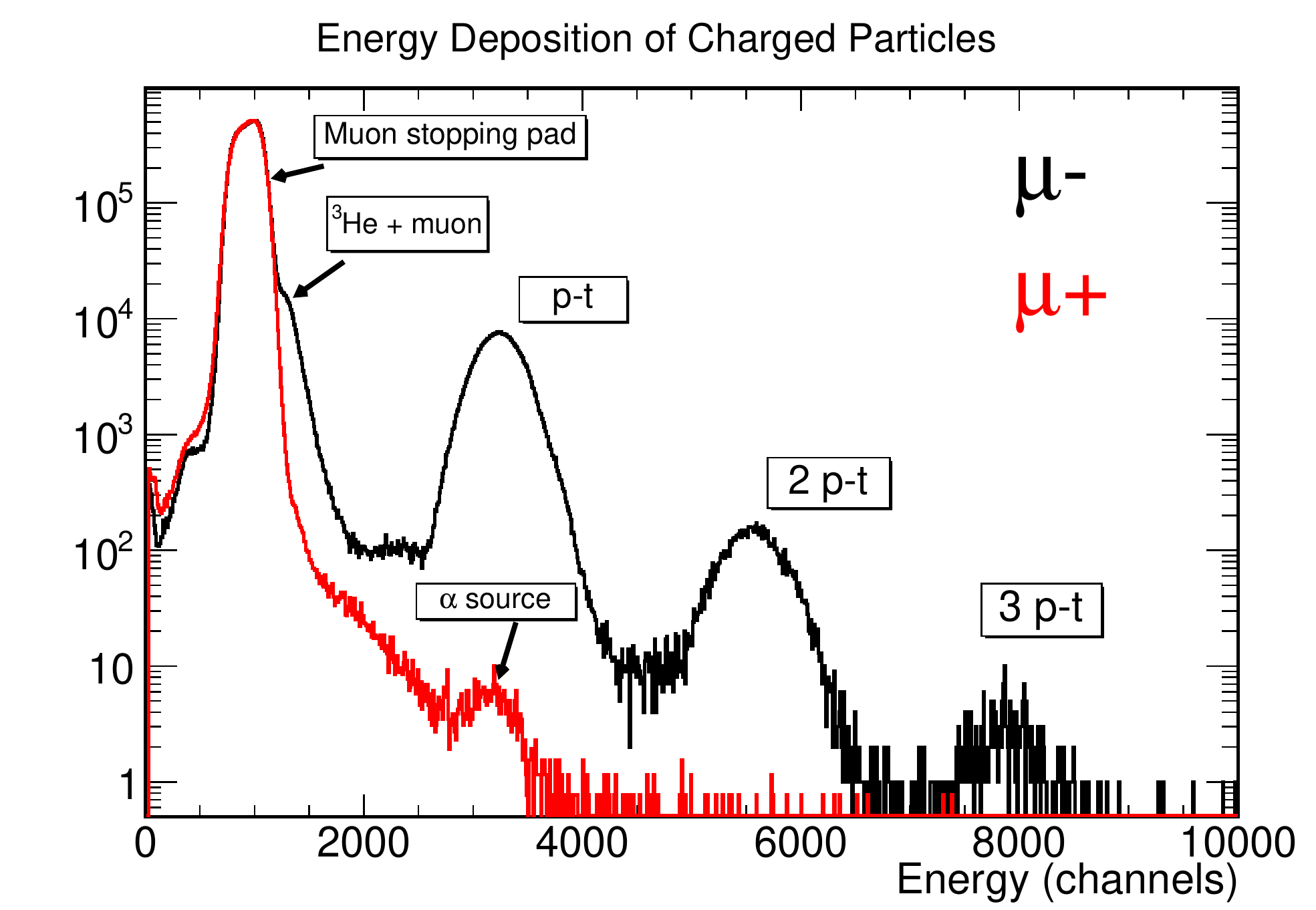}
  \caption{ \label{fg:MuSunEnergy} Measured energy deposition of stopping negative and positive muons
in the MuSun cryogenic TPC. The energy is defined
as the summed signal from the muon stopping pad and its eight neighboring pads. The positive muon
energy distribution
shows a large peak that corresponds to the stopping muons and a
small peak that corresponds to an {\em in-situ} alpha calibration source.
The negative muon
 energy distribution additionally shows the effects of
muon-catalyzed fusion and the additional energy deposition associated with the charged-particle
products of the n~$^3$He and p~$^3$H fusion channels.
In the particular case of the p~$^3$H channel, the occurrence of one, two and three
fusions following a single muon stop are clearly discernable.}
\end{centering}
\end{figure}

The MuSun experiment is well underway at PSI.
Fig.\ \ref{fg:MuSunEnergy} shows the
measured energy deposition of
stopping positive and negative muons
in the cryogenic TPC.
For negative muons the figure indicates the occurrence
of one, two and three
muon-catalyzed fusions are clearly detected.
Production data taking was begun in 2013 and will continue through 2015.

%% file: Summary_final.tex
\section{Summary and Outlook}

We have described many contemporary projects where muons are being used as probes of fundamental atomic-, nuclear- and particle-physics properties, the structure of the weak interaction, and as sensitive probes of physics beyond the standard model.  But, why muons, say, compared to electrons or tauons?  The answer involves comparing known parameters such as mass, lifetime, and decay modes, as well as practical issues such as source yields, polarizability, and once again, lifetime. Typically, greater mass provides enhanced sensitivity to high-energy-scale physics.  Access through radiative quantum loops generically scales as $(m_l/m_{heavy})^2$, where $m_l$ is the lepton mass and the subscript $heavy$ might correspond to the known $W$, or $Z$, or to an unknown new particle that couples to leptons.
For example, given equally precise measurements of the lepton anomalous magnetic moments, the muon will be more sensitive than an electron by a factor of 43,000~\footnote{The electron anomalous magnetic moment has been measured~\cite{Hanneke:2008tm} 2300 times more precisely compared to the muon---a spectacular achievement, which determines the fine-structure constant---but it does not close the gap.}.  The tau will be 300 times better still, but it suffers from its fleeting lifetime of 0.29\,ps, which impedes this and many other desired studies.

We have seen how the relatively long muon lifetime allows for the formation of muonic atoms, which, with their smaller Bohr radii, can be used to sensitively probe nucleons and nuclei.  The purely leptonic, but hydrogen-like, muonium atom is a probe of QED and fundamental properties, such as the muon mass and the magnetic moment ratio to the proton.  The copious production of polarized muons---which are too light to decay by the strong interaction---essentially allows for a laboratory of weak-interaction studies. Highly polarized muons, and the self-analyzing nature of their decay, are essential ingredients to many experiments.

Modern experiments involving muons are generally {\em nth}-generation efforts, having been designed based on earlier pioneering work.  The new experiments excel in precision and sensitivity reach, often being complemented by equally important theoretical improvements. Many experiments being built now will have the rare characteristic of ``discovery'' sensitivity, and an energy-scale reach complementary to or beyond that of the LHC collider program.

To summarize briefly, we first recall recent accomplishments in this field, and then list future directions with their planned sensitivities.

\subsection{Recent accomplishments}
\begin{itemize}
\item Muon lifetime: $\tau_{\mu^+}$ has been measured to 2.2\,ps, (1\,ppm) by the MuLan experiment\,\cite{Tishchenko:2012ie}.  With 2nd-order electroweak corrections now computed~\cite{vanRitbergen:1999fi}, the fundamental Fermi Constant is determined to be $G_F =
    1.166\, 378\, 7(6)\times 10^{-5}~{{\rm GeV}^{-2}}$ (0.5\,ppm).
\item Decay parameters:  Experiments at TRIUMF~\cite{TWIST:2011aa,Bueno:2011fq} and PSI~\cite{Danneberg:2005xv,Prieels:2014paa} have reduced the uncertainties by up to an order of magnitude on the Michel parameters $\rho$, $\xi$, and $\delta$, and on the transverse and longitudinal polarization of the electron in muon decay.  Global fits~\cite{Gagliardi:2005fg,TWIST:2011aa} have strong implications on possible deviations of the $V$-$A$ structure of the weak interaction. Results to date confirm standard model expectations.
\item Muon anomaly:  The final result of the Brookhaven \gm\ experiment~\cite{Bennett:2006fi}, when compared to steadily improved theoretical evaluations~\cite{Davier:2010nc,Hagiwara:2011af} of the muon anomaly results in a $>3 \sigma$ deviation, possibly indicative of new physics.
\item cLFV in $\mu \rightarrow e\gamma$:  The MEG experiment has set the world record on a test of charged lepton flavor violation finding BR($\mu \rightarrow e\gamma) < 5.7 \times 10^{-13}$~\cite{Adam:2013mnn}. The limits constrain new physics involving loop processes to the $10^3$ TeV energy scale, if one assumes {\em maximal} flavor mixing.
\item Muonium HFS: The LAMPF HFS experiment measured the hyperfine splitting of two transitions involving four levels of the $1S$ muonium atom in the high-field limit.  The results~\cite{Liu:1999iz} represent an important test of bound-state QED and established the important ratios: $\mu_{\mu} / \mu_p$ $=$ $3.183~345~13(39)$ (120~ppb), and $m_{\mu} / m_e$ $=$ $206.768~277(24)$.
\item Muon capture on the proton:  The MuCap experiment~\cite{Andreev:2012fj} determined the $\mu^- + p \rightarrow n + \nu_\mu$ singlet capture rate by measuring the $\mu^-$ lifetime in protium gas TPC and comparing it to $\tau_{\mu^+}$. Including updated radiative corrections~\cite{Czarnecki:2007th}, one obtains $g_p = 8.06 \pm 0.48 \pm 0.28$, the weak-nucleon pseudoscalar coupling of the proton.  The result confirms a fundamental prediction of chiral perturbation theory and concludes a fifty year effort to unambiguously determine the coupling $g_p$.
\item Proton radius:  The precise measurement~\cite{Antognini:1900ns} of two $2S \rightarrow 2P$ (Lamb shift) transitions in muonic hydrogen determines the proton charge radius, $r_p = 0.84087(39)$\,fm~\cite{Antognini:1900ns}.  The results---stunningly---are $7\,\sigma$ smaller than the previous world average, which was based on $e - p$ scattering and ordinary hydrogen spectroscopy. The so-called ``proton radius puzzle'' remains unsolved.

\end{itemize}

\subsection{Near-term projects}
At the time of this review a number of approved projects are actively taking data or are in a construction phase.  These are the ones to watch for results in the coming years; they include:
\begin{itemize}
\item cLFV in $\mu \rightarrow e\gamma$: The MEG experiment upgrade of the calorimeter, tracker, and other systems will improve the energy resolution and timing required to achieve a sensitivity goal of $4 \times 10^{-14}$ in the next 4 years.
\item cLFV in $\mu \rightarrow e $ conversion:  COMET at J-PARC and Mu2e at Fermilab will measure the coherent conversion of a muon to an electron in the field of a nucleus at a single event sensitivity approaching $10^{-17}$, a bold, 4 orders of magnitude improvement. Superconducting solenoids are used for particle production, transport, and final spectrometer functions.
\item cLFV in $\mu \rightarrow eee$: Mu3e at PSI is in a R\&D phase with an approved plan to reach a BR sensitivity of $10^{-15}$.  Central to their success is the development of ultra-thin silicon tracking detectors and, for a later phase, the creation of a next-generation high-intensity muon beamline.
\item Muon anomalous magnetic moment:  J-PARC and Fermilab experiments~\cite{Iinuma:2011zz,Grange:2015bea} are being built to reach sensitivities on $\delta a_\mu$ of $\sim 500$\,ppb and 140\,ppb, respectively. They employ radically different beam delivery and storage techniques, and will consequently confront different systematic errors, an important comparison.
\item Muon EDM:  Both \gm\ experiments can parasitically collect data sensitive to a muon EDM, with 1 to 2 orders of magnitude improvement beyond the current limit, $d_\mu < 10^{-19} e\cdot$cm. Dedicated EDM storage-ring plans, using a frozen-spin technique are promising, but are not yet approved.
\item Muonium HFS:  The MuSEUM experiment will explore the hyperfine structure of the muonium atom using the well-established double-resonant cavity technique for exciting spin-flip transitions. They aim for an order of magnitude increase in formed muonium atoms and two orders of magnitude increase in recorded decays compared to the most recent LAMPF experiment.
\item Proton radius:  CREMA will publish Lamb-shift measurements on the deuteron, helium-3, and helium-4 systems, important data to be compared to their existing hydrogen measurements.  The MUSE experiment plans to measure low-energy $\mu - p$ and $e - p$ scattering at low $Q^2$.  These data will provide an important missing clue in the enduring ``proton radius puzzle.''
\item Muon capture on deuterium:  MuSun will complete a measurement of $\mu d$ capture to $1.5\,\%$ precision, which will provide a clean determination of the low-energy constant arising in the effective-field-theory description of the two nucleon weak axial current. The result is relevant for fundamental astrophysics reactions, such as $pp$ fusion and the neutrino breakup reactions in the SNO experiment.
\end{itemize}

Physicists that might identify themselves with the subfield of  ``Muon Physics'' form a diverse group who happen to share a common and unique probe.  Many of the practical issues cross traditional discipline boundaries. The word ``Precision'' in our title evidently describes many of the results described above, but we naturally stretch its meaning to include ultra-high sensitivity experiments as well, such as those involving rare decays.  We trust that the many exciting projects described in this review will convince the reader of the prolific record of this eclectic Precision Muon Physics community and of the very bright future that lies ahead.

%% file: Acknowledgments.tex
\section*{Acknowledgments}
The authors are grateful to our many friends and colleagues in the worldwide muon physics community who have provided us with considerable materials, figures and feedback, incorporated in this review.  In particular, we have had very useful input and contributions from
A.\, Antognini,
R.\,Bernstein,
A.\,de\,Gouv\^ea,
M.\,Eides,
P.\,Kammel,
D.\,Kawall,
G.\,Marshall,
T.\,Mibe,
J.\,Miller,
T.\,Mori,
R.\,Pohl,
B.L.\,Roberts,
and
D.\,Stockinger.
This effort was supported by the National Science Foundation award PHY-1205792 and the DOE Office of Nuclear Physics award DE-FG02-97ER41020. 